\documentclass[10pt]{article}
\usepackage[latin9]{inputenc}
\usepackage{xcolor}
\usepackage{float}
\usepackage{amsmath}
\usepackage{amsthm}
\usepackage{amssymb}
\usepackage{graphicx}
\usepackage{geometry}
\PassOptionsToPackage{normalem}{ulem}
\usepackage{ulem}
\usepackage[unicode=true,pdfusetitle,
 bookmarks=true,bookmarksnumbered=false,bookmarksopen=false,
 breaklinks=false,pdfborder={0 0 1},backref=false,colorlinks=false]
 {hyperref}

\makeatletter

\providecommand{\tabularnewline}{\\}
\floatstyle{ruled}
\newfloat{algorithm}{tbp}{loa}
\providecommand{\algorithmname}{Algorithm}
\floatname{algorithm}{\protect\algorithmname}
\providecolor{lyxadded}{rgb}{0,0,1}
\providecolor{lyxdeleted}{rgb}{1,0,0}
\DeclareRobustCommand{\mklyxadded}[1]{\textcolor{lyxadded}\bgroup#1\egroup}
\DeclareRobustCommand{\mklyxdeleted}[1]{\textcolor{lyxdeleted}\bgroup\mklyxsout{#1}\egroup}
\DeclareRobustCommand{\mklyxsout}[1]{\ifx\\#1\else\sout{#1}\fi}
\DeclareRobustCommand{\lyxadded}[4][]{\texorpdfstring{\mklyxadded{#4}}{#4}}

\theoremstyle{plain}
\newtheorem{thm}{\protect\theoremname}
\theoremstyle{plain}
\newtheorem{lem}[thm]{\protect\lemmaname}
\ifx\proof\undefined
\newenvironment{proof}[1][\protect\proofname]{\par
	\normalfont\topsep6\p@\@plus6\p@\relax
	\trivlist
	\itemindent\parindent
	\item[\hskip\labelsep\scshape #1]\ignorespaces
}{%
	\endtrivlist\@endpefalse
}
\providecommand{\proofname}{Proof}
\fi
\theoremstyle{definition}
\newtheorem{defn}[thm]{\protect\definitionname}

\usepackage[english]{babel}
\usepackage{graphicx}
\usepackage{refcount}
\usepackage{tikz}
\usetikzlibrary{intersections,calc,quotes}
\tikzset{%
  point/.style={circle,inner sep=1.5pt,minimum size=1.5pt,draw,fill=#1},
  point/.default=red,
}

\newcommand{\drawarc}[5][]{%
  \path (#2); \pgfgetlastxy{\startx}{\starty}%
  \path (#3); \pgfgetlastxy{\centerx}{\centery}%
  \path (#4); \pgfgetlastxy{\endx}{\endy}%
  \pgfmathsetmacro{\angleStart}{atan2(\starty-\centery,\startx-\centerx)}%
  \pgfmathsetmacro{\angleEnd}{atan2(\endy-\centery,\endx-\centerx)}%
  \draw[#1] (#3) ++(\angleStart:#5) arc[start angle=\angleStart, end angle=\angleEnd, radius=#5];%
}
\newcommand{\intersectCoord}[5]{%
  \path[name intersections={of=#1 and #2, name=ix#3#4}];
  \coordinate (ca#5) at (ix#3#4-1);
  \coordinate (cb#5) at (ix#3#4-2);
}

\newcommand{\intersectAndLabel}[6]{%
  \intersectCoord{#1}{#2}{#3}{#4}{#5}
  \node[point,label={[label distance=-1pt]#6:{\textcolor{black}{$#5$}}}] (pa#5) at (ca#5) {};
}
\usepackage{xparse} 
\NewDocumentCommand{\myarc}{O{0.3em}m}{%
    \tikz[baseline=(arcnode.base)] {
        \node[inner sep=0pt, outer sep=0pt] (arcnode) {#2};

        \coordinate (arcstart) at ($(arcnode.west) + (0,#1)$);
        \coordinate (arcend) at ($(arcnode.east) + (0,#1)$);

        \draw[line cap=round] (arcstart) to[bend left=5] (arcend);
    }%
}

\makeatother

\usepackage[style=authoryear,backend=bibtex]{biblatex}
\providecommand{\definitionname}{Definition}
\providecommand{\lemmaname}{Lemma}
\providecommand{\theoremname}{Theorem}

\begin{document}
\title{Algorithms for Nonlinear Mixed-Integer Trilateration}
\author{Ophir Uziel, Efi Fogel, Dan Halperin, and Sivan Toledo\\
Blavatnik School of Computer Science and AI, Tel Aviv University}
\maketitle
\begin{abstract}
For three decades, carrier-phase observations have been used to obtain
the most accurate location estimates using global navigation satellite
systems (GNSS). These estimates are computed by minimizing a nonlinear
mixed-integer least-squares problem. Existing algorithms linearize
the problem, orthogonally project it to eliminate real variables,
and then solve the integer least-square problem. There is now considerable
interest in developing similar localization techniques for terrestrial
and indoor settings. We show that algorithms that linearize first
fail in these settings and we propose several algorithms for computing
the estimates. Some of our algorithms are elimination algorithms that
start by eliminating the non-linear terms in the constraints; others
construct a geometric arrangement that allows us to efficiently enumerate
integer solutions (in polynomial time). We focus on simplified localization
problems in which the measurements are range (distance) measurements
and carrier phase range measurements, with no nuisance parameters.
The simplified problem allows us to focus on the core question of
untangling the nonlinearity and the integer nature of some parameters.
We show using simulations that the new algorithms are effective at
close ranges at which the linearize-first approach fails.
\end{abstract}
Keywords: Integer Ambiguity Resolution, Least-squares Ambiguity Decorrelation
Adjustment (LAMBDA), Nonlinear, GNSS

\section{Introduction}

For three decades, carrier-phase observations have been used to obtain
the most accurate location estimates using global navigation satellite
systems (GNSS)~\parencite{doi:10.1007/BF00863419,GNSSHandbookIntegerAmbiguity,MixedIntegerXu1995,ChangPaige:2003:GPS}.
While originally designed for measuring the signal's time of arrival
(ToA), which varies linearly with the range (distance) between the
emitting satellite and the receiver, it turns out that the receiver
can also measure and track the carrier's phase, which also varies
linearly with the range, but wraps around every whole wavelength.
The errors in the phase measurements are orders of magnitude smaller
than in the ToA estimates, when both are expressed in meters, allowing
in principle more accurate location estimates. However, to resolve
the location of the receiver from carrier-phase measurements, the
receiver needs to resolve the so-called \emph{integer ambiguities},
essentially the number of whole wavelengths between each satellite
and the receiver (the actual formulation also involves nuisance parameters,
such as the initial phases at the satellite and the receiver). A method
called LAMBDA~\parencite{doi:10.1007/BF00863419,LAMBDA} was invented
to efficiently resolve the integer ambiguities. There are also some
variants that have been shown to be effective~\parencite{MixedIntegerXu1995,doi:10.1179/1752270612Y.0000000029,ChangPaige:2003:GPS,MLAMBDA,doi:10.48550/arXiv.1204.1398}.
These methods work well in both single-epoch settings and in multi-epoch
settings in which the receiver is either stationary or moving (in
the latter setting the method is referred to as real-time kinematic
positioning, or RTK).

There is now considerable interest in developing carrier-phase positioning
for terrestrial localization systems, where the distances between
the emitters and the receivers are much smaller than in GNSS, including
in cellular and indoor systems~\parencite{doi:10.1109/GLOBECOM54140.2023.10437902,doi:10.1109/TCOMM.2021.3119072,doi:10.1109/MWC.023.2200633}.
Recent articles on short-range carrier localization that discuss the
integer-ambiguities resolution problem propose using LAMBDA and its
variants, implicitly assuming that these methods are suitable for
this setting~\parencite{doi:10.3390/s21206731,doi:10.1109/GLOBECOM54140.2023.10437902,doi:10.1109/TCOMM.2021.3119072}.

We show in this paper that this assumption is often wrong. LAMBDA
and its variants start by linearizing a nonlinear mixed-integer least-squares
problem around an approximate solution of the continuous (non-integer)
constraints. The nonlinearity is the Euclidean distance between the
emitter and receiver, which is very large in GNSS. Therefore, in GNSS
settings the linearization introduces essentially no error, even if
the approximate solution (usually obtained from ToA measurements)
is not very accurate. In terrestrial and indoor settings, where the
distances between the emitters and receivers are on the order of meters
or kilometers, this is no longer true; we show below that LAMBDA variants
fail at these distances. 

In some cases the linearization is not explicit. Constraints that
relate the unknown parameters to observations and to an approximate
solution can also be derived from geometric principles (see, e.g.~\parencite{ChangPaige:2003:GPS}).
However, these constraints are also approximations that assume that
the distances to the satellites is very large. In general, both approaches
lead to the same constraints.

The main contributions of this paper are two families of algorithms
for resolving the integer ambiguities in trilateration problems that
include both range measurements and carrier-phase range measurements.
Our algorithms tackle the nonlinearity head-on. One family of algorithms
is based on elimination of the nonlinear terms. We use a non-linear
elimination technique, reminiscent of closed-form solutions for code-phase
GNSS localization~\parencite{doi:10.1109/TAES.1985.310538,doi:10.1109/WPNC.2010.5653789}.
The elimination process reduces the original problem to a linear least-squares
problem with discrete parameters that are not integers. We show that
a key component of LAMBDA, namely the Schnorr-Euchner search~\parencite{doi:10.1007/BF01581144},
can be adapted to this case. This generalization of the Schnorr-Euchner
search is another key contribution of this article. When the target
is tracked over two or more epochs, the non-linear elimination includes
two subtraction (differencing) stages. This leads to a generalized
mixed-integer least-squares minimization problem; the discrete parameters
are integers, as in LAMBDA. 

A second family of algorithms, described in Section~\ref{sec:geometric-arrangements},
uses geometric arrangements to resolve the ambiguities. These algorithms
essentially project the integers onto the 2D or 3D space of target
locations. Interestingly, these algorithms run in polynomial time
even in the worst case, even though LAMBDA reduces the mixed-integer
localization problem to a pure integer least-square problem which
is known to be NP hard~\parencite{ArgellEtAl2002,CVPisNPHard,vanEmdeBoasCVP}
and hard to approximate~\parencite{DinurCVP}. The reason for this surprising
result is the projection to 2D or 3D, which allows us to enumerate
all the combinations of integers that correspond to actual locations
of the target in polynomial time. This is another important contribution
of this article.

We show in simulations that our methods are both efficient and effective,
and that they can resolve the integer ambiguities at ranges where
LAMBDA fails due to the linearization error.

We focus on trilateration problems that include range observations
and observations of the remainders of division of the ranges by the
wavelength. These problems are simpler than GNSS localization problems
in that they lack nuisance parameters such as clock bias and initial
phases at the emitters and receiver. (The lack of these nuisance parameters
also means that measurements at a reference receiver whose location
is known are not necessary.)

We chose to focus on these simplified localization problems, which
still retain the Euclidean nonlinearity and the mixture of real and
integer parameters, to enable the widest possible exploration of the
algorithmic design space. We plan to explore localization problems
associated with realistic system models and to extend our algorithms
to these cases in the near future. 

In summary, the main contributions of this article are:
\begin{enumerate}
\item A square-and-difference elimination algorithm that transforms the
mixed-integer trilateration problems defined in Section~\ref{sec:Problem-Statement}
to linear least-squares problems with discrete (but not integer) parameters.
\item An extension of the Schnorr-Euchner search that can solve discrete
linear least-squares problems of this type exactly, possibly in exponential
time (just like the original Schnorr-Euchner search for integer least-squares
problems).
\item A square and double-difference elimination algorithm that transforms
multi-epoch problems, also defined in Section~\ref{sec:Problem-Statement},
into linear mixed-integer least-squares problems, which can be solved
using LAMBDA and its variants.
\item Careful characterization of the uncertainties and errors introduced
in the elimination processes, which allows us to correctly weigh the
constraints in the linear least-square problems. These weights are
critical for the success of the algorithms. 
\item Several geometric algorithms for solving the same trilateration problems.
These algorithms project the problems, including the integer parameters,
into 2 or 3 dimensions. These algorithms run in polynomial time, even
though LAMBDA and its variants, as well as the algorithms from items
1 and 3 in this list, reduce the localization problem into an NP-hard
integer least-squares problem.
\item Computational demonstration, using simulations, that our algorithms
are efficient and effective.
\end{enumerate}
The rest of this article is organized as follows. Section~\ref{sec:Problem-Statement}
defines the estimation problems that we tackle and the concrete values
or ranges of parameters under which we test our algorithms. Sections~\ref{sec:Deriving-Linear-Constraints},
\ref{sec:Assembling-Minimization-Problems}, \ref{sec:Quantifying-Error-Terms},
and~\ref{sec:Generalized-Schnorr-Euchner} together present our square-and-difference
algorithms. More specifically, Section~\ref{sec:Deriving-Linear-Constraints}
explains how we derive constraints from the raw observations. This
section describes all the elimination techniques that we use. Section~\ref{sec:Assembling-Minimization-Problems}
specifies how we assemble these constraints, which are all equality
constraints with error or noise terms, into generalized least-squares
minimization problems. Section~\ref{sec:Quantifying-Error-Terms}
characterizes each of the error terms; these characterizations determine
the weight of each constraint in the minimization problems. Section~\ref{sec:Generalized-Schnorr-Euchner}
presents the generalized Schnorr-Euchner search, which we use to solve
mixed-discrete least squares problems. Section~\ref{sec:geometric-arrangements}
presents our polynomial-time geometric algorithms. Section~\ref{sec:Simulations-and-Evaluation}
describe the performance of both our new algorithm and LAMBDA on simulated
data. Section~\ref{sec:Conclusions} presents our conclusions from
this research.

\section{\label{sec:Problem-Statement}Problem Statement}

We wish to localize a target at Cartesian coordinates $\mathring{\ell}=\begin{bmatrix}x & y & z\end{bmatrix}^{T}$
or $\mathring{\ell}=\begin{bmatrix}x & y\end{bmatrix}^{T}$ from
range or pseudo-range measurements from a set of reference points
$\rho_{i}=\begin{bmatrix}x_{i} & y_{i} & z_{i}\end{bmatrix}^{T}$
or $\rho_{i}=\begin{bmatrix}x_{i} & y_{i}\end{bmatrix}^{T}$, as well
as carrier-phase measurements from the same reference points. The
ring above the letter $\ell$ marks this location as the ground truth.
The simplest setting for such problems consists of a set of range
and carrier-phase observations
\begin{eqnarray}
r_{i} & = & \left\Vert \mathring{\ell}-\rho_{i}\right\Vert _{2}+\epsilon_{i}\label{eq:range-observations}\\
\lambda\varphi_{i}+\lambda\mathring{n}_{i} & = & \left\Vert \mathring{\ell}-\rho_{i}\right\Vert _{2}+\delta_{i}\;,\label{eq:phase-observations}
\end{eqnarray}
where $r_{i}$ is a noisy range (distance) measurement, $\epsilon_{i}$
is the range-measurement error (noise), $\lambda$ is the wavelength
of the carrier wave of the signal emitted by the references (or by
the target, in transmitter localization), $\varphi_{i}\in[0,1)$ is
a noisy phase measurement, $\delta_{i}$ is the phase measurement
error and $\mathring{n}_{i}\in\mathbb{N}$ is an unknown integer that
corresponds roughly to the number of whole wavelengths between $\mathring{\ell}$
and $\rho_{i}$. (Because of the measurement error, $\mathring{n}_{i}=\lfloor(\left\Vert \mathring{\ell}-\rho_{i}\right\Vert -\delta_{i})/\lambda\rfloor$
can be smaller or larger by one than $\lfloor\left\Vert \mathring{\ell}-\rho_{i}\right\Vert /\lambda\rfloor$.)
The unknown integer parameters are often called \emph{integer ambiguities}.
In more realistic problems only pseudo-ranges are measured $\left\Vert \mathring{\ell}-\rho_{i}\right\Vert +c\Delta t$
where $\Delta t$ is a clock error or bias and $c$ is the speed of
propagation) and the phases are measured relative to some unknown
initial phases at both the transmitter and the receiver. There might
also be complications from ionospheric delays, tropospheric delays,
etc.

If the receiver makes observations of the forms~(\ref{eq:range-observations})
and~(\ref{eq:phase-observations}) at multiple points in time, called
\emph{epochs}, and can track the carrier phases continuously, we denote
the observations at time $j$
\begin{eqnarray}
r_{i,j} & = & \left\Vert \mathring{\ell}_{j}-\rho_{i,j}\right\Vert _{2}+\epsilon_{i,j}\label{eq:range-observations-1}\\
\lambda\varphi_{i,j}+\lambda\mathring{n}_{i} & = & \left\Vert \mathring{\ell}_{j}-\rho_{i,j}\right\Vert _{2}+\delta_{i,j}\;.\label{eq:phase-observations-1}
\end{eqnarray}
Note that this notation allows both the target and the references
to change their locations between epochs, but that a single integer
is used in all the phase observations of reference $i$. In some settings
we know that the target and/or the references are stationary; it is
easy to exploit this information, as we shall see below.

The assumption that a but that a single integer $\mathring{n}_{i}$
is used in all the phase observations of reference $i$ depends on
the receiver's ability to track the carrier phase continuously; that
is, that there are no \emph{cycle slips }(the term used in the GNSS
literature to describe a failure to track the carrier phase continuously).

These problems are interesting when $\delta_{i}\ll\epsilon_{i}$.
Typical value of $\lambda$ for radio systems range from about 1~m
to about 10~cm and typical standard deviations for $\delta_{i}$
are on the scale of $\lambda/100$, so about 1~cm to 1~mm. In GNSS
typical standard deviations for $\epsilon_{i}$ are on the scale of
10~m, much higher.

We wish to estimate $\mathring{\ell}$ in the generalized least-squares
sense, which corresponds to maximum likelihood under the assumptions
that the $\epsilon_{i}$s and $\delta_{i}$s are unbiased Gaussian
random variables. We further assume that the $\epsilon_{i}$s and
$\delta_{i}$s are uncorrelated and have known standard deviations
$\sigma(\epsilon_{i})$ and $\sigma(\delta_{i})$. Under these assumptions,
we seek the minimizer 
\begin{equation}
\hat{\ell},\hat{n}=\arg\min_{\ell,n}\left\Vert W\left(M\left(\ell\right)-\begin{bmatrix}0\\
\lambda I
\end{bmatrix}n-\begin{bmatrix}r\\
\lambda\varphi
\end{bmatrix}\right)\right\Vert _{2}\label{eq:nonlinear-generalized-ls}
\end{equation}
where $W$ is a diagonal matrix with $\sigma^{-1}(\epsilon_{i})$
and $\sigma^{-1}(\delta_{i})$ on the diagonal, and $M$ is a function
such that 
\[
M_{i}\left(\ell\right)=M_{m+i}\left(\ell\right)=\left\Vert \ell-\rho_{i}\right\Vert _{2}\;,
\]
where $m$ is the number of references. Note that we use the plain
letters $\ell$ and $n$ to denote hypothetical locations and integer
vectors, and that we decorate $\ell$ and $n$ with a hat to denote
the estimates. The integer vector $\mathring{n}$ is a vector of nuisance
parameters; we normally do not need to estimate it, but algorithms
for estimating $\mathring{\ell}$ usually estimate $\mathring{n}$
it as a side effect.

We can obtain a coarse estimate $\ell_{0}$ of $\mathring{\ell}$
from the range measurements alone and linearize $M$ around $\ell_{0}$,
\[
M\left(\ell\right)\approx M\left(\ell_{0}\right)+\mathrm{J}_{M}\left(\ell-\ell_{0}\right)=M\left(\ell_{0}\right)+\frac{\partial M}{\partial\ell}\left(\ell-\ell_{0}\right)
\]
where $\mathrm{J}_{M}=\partial M/\partial\ell$ is the Jacobian of
$M$ and where the derivatives are evaluated at $\ell_{0}$. In GNSS,
the distances $\left\Vert \mathring{\ell}-\rho_{i}\right\Vert $ are
20,000~km or more, so when $\ell_{0}$ is within tens or even hundreds
of meters of $\ell$, this linear approximation is superb (see Section~\ref{sec:Errors-in-Linearization-of-Ranges}
below), so we can estimate $\mathring{\ell}$ and $\mathring{n}$
by solving the mixed-integer linear least-squares problem 
\begin{equation}
\arg\min_{\Delta\ell,n}\left\Vert W\left(M\left(\ell_{0}\right)+\mathrm{J}_{M}\Delta\ell-\begin{bmatrix}0\\
\lambda I
\end{bmatrix}n-\begin{bmatrix}r\\
\lambda\varphi
\end{bmatrix}\right)\right\Vert _{2}\;,\label{eq:linearized-mixed-integer-ls}
\end{equation}
where $\Delta\ell=\ell-\ell_{0}$. Note that the same $W$ appears
in this expression, because the linearization errors at GNSS ranges
are so small that their effect on the overall errors is negligible.

We are interested in problems in which the ranges are much smaller,
where the linear approximation may not be good enough. In other words,
in GNSS a reasonably good $\ell_{0}$ allows us to ignore the nonlinearity
in $M$. The main and essentially only computational challenge is
the estimation of the integer ambiguities $\mathring{n}$. This is
true even if there are additional nuisance parameters like clock bias,
because the dependence on them is linear.

We note that in multi-epoch problems, $\ell_{0}$ can also be estimated
from both range measurements and carrier-phase measurements, ignoring
the fact that the elements of $n$ are integers. When the number of
references and/or epochs is large, this approach can sometimes generate
a more accurate $\ell_{0}$ than using the range measurements alone,
and hence can allow the linearization-first approach to work at shorter
ranges~\parencite{ChangPaige:2003:GPS}. We focus in this paper on single-epoch
and two-epoch problems, in which technique is rarely effective, and
on problems in which the nonlinearity must be tackled head on.

\subsection{\label{sec:Typical-Parameters}Parameters for Experimental Evaluations}

We use parameter values that are typical in GNSS, except for the distances
to the references. More specifically, our baseline values are 
\begin{eqnarray*}
\lambda & = & 19\,\text{cm}\\
\sigma(\epsilon_{i}) & = & 10\,\text{m}\\
\sigma(\delta_{i}) & = & 0.1^{\circ}=19/3600\,\text{cm}\\
\sigma\left|\mathring{\ell}-\ell_{0}\right|_{j} & = & 10\,\text{m}\\
\left\Vert \mathring{\ell}-\rho_{i}\right\Vert  & \approx & 10\,\text{m}\text{ to }20,000\,k\text{m}
\end{eqnarray*}

We test the algorithms in 2D and in 3D. In our simulations, we select
$\mathring{\ell}-\ell_{0}$ at random from a Gaussian distribution
of each coordinate. We select the $\rho_{i}$s by selecting a uniform
random azimuth from $\ell$, selecting a point on a circular orbit
at that azimuth, and then perturbing the coordinates by adding random
Gaussian noise to each coordinate of $\rho_{i}$. The standard deviation
of these perturbations is 10\% of the nominal range.

\section{\label{sec:Deriving-Linear-Constraints}Deriving Linear Constraints}

Our observations~(\ref{eq:range-observations}) and~(\ref{eq:phase-observations})
are nonlinear in $\mathring{\ell}$ and they include both continuous
parameters, the coordinates of $\mathring{\ell}$, and integer parameters,
the $\mathring{n}_{i}$s. We present in this section several techniques
to derive from these observations constraints that are linear in both
$\mathring{\ell}$ and in a vector of $m$ discrete parameters. In
the multi-epoch setting, the discrete parameters are simply the $\mathring{n}_{i}$s,
which are integers, but in the single-epoch setting they are not integers.

Our derived constraints are all mathematical equalities with error
terms. The error terms are denoted clearly and we quantify them in
Section~\ref{sec:Quantifying-Error-Terms} below. Error terms that
are functions of $\delta_{i}$ are denoted $\delta_{i}^{(\cdots)},$where
letters within the parentheses specify the error expression, error
terms that are functions of $\epsilon_{i}$ are denoted $\epsilon_{i}^{(\cdots)}$,
and error terms that depend on neither $\delta_{i}$ nor $\epsilon_{i}$
are denoted $\eta_{i}$ or $\eta_{i}^{(\cdots)}$.

\subsection{\label{sec:Squared-Phase-Observations}Squared Phase Observations}

Squaring Equation~(\ref{eq:phase-observations})
\[
\lambda\left(\mathring{n}_{i}+\varphi_{i}\right)=\left\Vert \mathring{\ell}-\rho_{i}\right\Vert _{2}+\delta_{i}\;.
\]
we obtain
\begin{equation}
\lambda^{2}\left(\mathring{n}_{i}+\varphi_{i}\right)^{2}=\left\Vert \mathring{\ell}-\rho_{i}\right\Vert _{2}^{2}+2\delta_{i}\left\Vert \mathring{\ell}-\rho_{i}\right\Vert _{2}+\delta_{i}^{2}\;.\label{eq:squared_carrier_phase_with_error-1}
\end{equation}
We treat $2\delta_{i}\|\mathring{\ell}-\rho_{i}\|$ as an error term
denoted $\delta_{i}^{(\text{s})}$, the letter 's' denoting the squaring
operation. We use this constraint as is in multi-epoch problems, as
a building block for further transformations:
\begin{equation}
\lambda^{2}\left(\mathring{n}_{i}+\varphi_{i}\right)^{2}=\left\Vert \mathring{\ell}-\rho_{i}\right\Vert _{2}^{2}+\delta_{i}^{(\text{s})}+\delta_{i}^{2}\;.\label{eq:squared_carrier_phase_with_error-1-2}
\end{equation}
In single-epoch settings, we change the discrete variables. We substitute
$\mathring{s}_{i}=(\mathring{n}_{i}+\varphi_{i})^{2}$ to obtain
\begin{equation}
\lambda^{2}\mathring{s}_{i}=\left\Vert \mathring{\ell}-\rho_{i}\right\Vert _{2}^{2}+\delta_{i}^{(\text{s})}+\delta_{i}^{2}\;.\label{eq:squared_carrier_phase_with_error-1-1}
\end{equation}

\subsection{\label{sec:Squares-of-Range-Observations}Squared Range Observations}

We can also square the range observations~(\ref{eq:phase-observations})
\[
r_{i}=\left\Vert \mathring{\ell}-\rho_{i}\right\Vert _{2}+\epsilon_{i}
\]
to obtain 
\begin{eqnarray}
r_{i}^{2} & = & \left\Vert \mathring{\ell}-\rho_{i}\right\Vert _{2}^{2}+2\epsilon_{i}\left\Vert \mathring{\ell}-\rho_{i}\right\Vert _{2}+\epsilon_{i}^{2}\label{eq:squared-range-observation}\\
 & = & \left\Vert \mathring{\ell}-\rho_{i}\right\Vert _{2}^{2}+\epsilon_{i}^{(\text{s})}+\epsilon_{i}^{2}\;,\nonumber 
\end{eqnarray}
where $\epsilon_{i}^{(\text{s})}=2\epsilon_{i}\|\mathring{\ell}-\rho_{i}\|$.

\subsection{\label{sec:Differencing-Squared-Observations}Differencing Squared
Observations by Reference}

Subtracting squared observations associated with different references
eliminates the nonlinearity in $\mathring{\ell}$. Subtracting the
squared phase constraint of reference $1$ from that of reference
$i$, we obtain
\begin{equation}
\lambda^{2}\left(\left(\mathring{n}_{i}+\varphi_{i}\right)^{2}-\left(\mathring{n}_{1}+\varphi_{1}\right)^{2}\right)=\left\Vert \mathring{\ell}-\rho_{i}\right\Vert _{2}^{2}-\left\Vert \mathring{\ell}-\rho_{1}\right\Vert _{2}^{2}+\delta_{i}^{(\text{s})}+\delta_{i}^{2}-\delta_{1}^{(\text{s})}-\delta_{1}^{2}\;.\label{eq:subtracted_equation-1-1}
\end{equation}
Expanding the squared norms on the right-hand side gives

\begin{align}
\left\Vert \mathring{\ell}-\rho_{i}\right\Vert _{2}^{2}-\left\Vert \mathring{\ell}-\rho_{1}\right\Vert _{2}^{2} & =(\mathring{\ell}-\rho_{i})^{T}(\mathring{\ell}-\rho_{i})-(\mathring{\ell}-\rho_{1})^{T}(\mathring{\ell}-\rho_{1})\nonumber \\
 & =-2\mathring{\ell}^{T}(\rho_{i}-\rho_{1})+\rho_{i}^{T}\rho_{i}-\rho_{1}^{T}\rho_{1}\;.\label{eq:expanded_rhs-1-1}
\end{align}
The subtraction eliminated the squares $\mathring{\ell}^{T}\mathring{\ell}=x^{2}+y^{2}+z^{2}$
(or $\mathring{\ell}^{T}\mathring{\ell}=x^{2}+y^{2}$ in 2D), leaving
us with a expression that is a linear function of the unknown vector
$\mathring{\ell}$.

In the single-epoch setting this results in a constraint that is linear
in both $\mathring{\ell}$ and the discrete-valued parameters $\mathring{s}_{i}$,
\begin{equation}
\lambda^{2}\mathring{s}_{i}-\lambda^{2}\mathring{s}_{1}=-2\mathring{\ell}^{T}(\rho_{i}-\rho_{1})+\rho_{i}^{T}\rho_{i}-\rho_{1}^{T}\rho_{1}+\delta_{i}^{(\text{s})}+\delta_{i}^{2}-\delta_{1}^{(\text{s})}-\delta_{1}^{2}\label{eq:differenced-squared-phases_si}
\end{equation}
(the error terms $\delta_{i}^{(\text{s})}$ and $\delta_{1}^{(\text{s})}$
also depend on $\mathring{\ell}$; we analyze this dependence in Section~\ref{sec:epsilon-squared-delta-squared}
below). In the multi-epoch setting we stay with a constraint that
is still nonlinear in the $n_{i}$s,
\begin{equation}
\lambda^{2}\left(\left(\mathring{n}_{i}+\varphi_{i}\right)^{2}-\left(\mathring{n}_{1}+\varphi_{1}\right)^{2}\right)=-2\mathring{\ell}^{T}(\rho_{i}-\rho_{1})+\rho_{i}^{T}\rho_{i}-\rho_{1}^{T}\rho_{1}+\delta_{i}^{(\text{s})}+\delta_{i}^{2}-\delta_{1}^{(\text{s})}-\delta_{1}^{2}\;,\label{eq:differenced-squared-phases-ni}
\end{equation}
so we transform it again below to eliminate this nonlinearity.

Subtracting squared range observations by reference is also useful
and gives
\begin{eqnarray}
r_{i}^{2}-r_{1}^{2} & = & -2\mathring{\ell}^{T}(\rho_{i}-\rho_{1})+\rho_{i}^{T}\rho_{i}-\rho_{1}^{T}\rho_{1}+\epsilon_{i}^{(\text{s})}+\epsilon_{i}^{2}-\epsilon_{1}^{(\text{s})}-\epsilon_{1}^{2}\;.\label{eq:differenced-squared-ranges}
\end{eqnarray}

\subsection{\label{sec:Double-Differencing-by-Epoch}Double-Differencing by Epoch}

In multi-epoch settings, we eliminate the nonlinearity in the differenced
squared phase constraints by differencing again, this time by epoch.
Using the notation of Equation~(\ref{eq:phase-observations-1}) in
Equation~(\ref{eq:differenced-squared-phases-ni}), we have
\begin{eqnarray*}
\lambda^{2}\left(\left(\mathring{n}_{i}+\varphi_{i,j}\right)^{2}-\left(\mathring{n}_{1}+\varphi_{1,j}\right)^{2}\right) & = & -2\mathring{\ell}_{j}^{T}(\rho_{i,j}-\rho_{1,j})+\rho_{i,j}^{T}\rho_{i,j}-\rho_{1,j}^{T}\rho_{1,j}+\delta_{i,j}^{(\text{s})}+\delta_{i,j}^{2}-\delta_{1,j}^{(\text{s})}-\delta_{1,j}^{2}\;.
\end{eqnarray*}
Subtracting the corresponding equation for epoch $j'\neq j$ eliminates
the squared $\mathring{n}_{i}$s from the left-hand side. Consider
$(\mathring{n}_{i}+\varphi_{i,j})^{2}-(\mathring{n}_{i}+\varphi_{i,j'})^{2}$,
\begin{eqnarray*}
\left(\mathring{n}_{i}+\varphi_{i,j}\right)^{2}-\left(\mathring{n}_{i}+\varphi_{i,j'}\right)^{2} & = & \left(\varphi_{i,j}^{2}+2\varphi_{i,j}\mathring{n}_{i}+\mathring{n}_{i}^{2}\right)-\left(\varphi_{i,j'}^{2}+2\varphi_{i,j'}\mathring{n}_{i}+\mathring{n}_{i}^{2}\right)\\
 & = & \left(\varphi_{i,j}^{2}-\varphi_{i,j'}^{2}\right)+2\left(\varphi_{i,j}-\varphi_{i,j'}\right)\mathring{n}_{i}\;.
\end{eqnarray*}
The double-differenced constraints are
\begin{eqnarray*}
\left(\varphi_{i,j}^{2}-\varphi_{i,j'}^{2}\right)+2\left(\varphi_{i,j}-\varphi_{i,j'}\right)\mathring{n}_{i}\\
-\left(\varphi_{1,j}^{2}-\varphi_{1,j'}^{2}\right)-2\left(\varphi_{1,j}-\varphi_{1,j'}\right)\mathring{n}_{1} & = & -2\mathring{\ell}_{j}^{T}(\rho_{i,j}-\rho_{1,j})+\rho_{i,j}^{T}\rho_{i,j}-\rho_{1,j}^{T}\rho_{1,j}+\delta_{i,j}^{(\text{s})}+\delta_{i,j}^{2}-\delta_{1,j}^{(\text{s})}-\delta_{1,j}^{2}\\
 &  & +2\mathring{\ell}_{j'}^{T}(\rho_{i,j'}-\rho_{1,j'})-\rho_{i,j'}^{T}\rho_{i,j'}+\rho_{1,j'}^{T}\rho_{1,j'}-\delta_{i,j'}^{(\text{s})}-\delta_{i,j'}^{2}+\delta_{1,j'}^{(\text{s})}+\delta_{1,j'}^{2}\;.
\end{eqnarray*}
These constraints are linear in $\mathring{\ell}_{j}$, $\mathring{\ell}_{j'}$,
and the $\mathring{n}_{i}$s. 

When both the target and the references move between epochs, we use
these constraints as is. When the target and/or the references are
stationary, the constraints simplify. When the target is stationary
so $\mathring{\ell}_{j}=\mathring{\ell}_{j'}=\mathring{\ell}$ but
the references move, we substitute on the right-hand side
\[
-2\mathring{\ell}_{j}^{T}(\rho_{i,j}-\rho_{1,j})+2\mathring{\ell}_{j'}^{T}(\rho_{i,j'}-\rho_{1,j'})=-2\mathring{\ell}^{T}(\rho_{i,j}-\rho_{1,j}-\rho_{i,j'}+\rho_{1,j'})\;.
\]
When the target is moving but the references are stationary $\rho_{i,j}=\rho_{i,j'}$,
we substitute on the right-hand side
\[
-2\mathring{\ell}_{j}^{T}(\rho_{i,j}-\rho_{1,j})+\rho_{i,j}^{T}\rho_{i,j}-\rho_{1,j}^{T}\rho_{1,j}+2\mathring{\ell}_{j'}^{T}(\rho_{i,j'}-\rho_{1,j'})-\rho_{i,j'}^{T}\rho_{i,j'}+\rho_{1,j'}^{T}\rho_{1,j'}=-2(\mathring{\ell}_{j}-\mathring{\ell}_{j'})^{T}(\rho_{i}-\rho_{1})\;.
\]
Incorporate the resulting constraints into a minimization problem
allows us to estimate $\mathring{\ell}_{j}-\mathring{\ell}_{j'}$
but not $\mathring{\ell}_{j}$ or $\mathring{\ell}_{j'}$. We can
overcome this problem because the minimization problem does allow
us to estimate the $\mathring{n}_{i}$s, which we can then substitute
in the original phase constraints and estimate $\mathring{\ell}_{j}$
and $\mathring{\ell}_{j'}$ (separately).

When both the target and the references are stationary, the differencing
leaves only error terms,
\begin{eqnarray*}
\left(\varphi_{i,j}^{2}-\varphi_{i,j'}^{2}\right)+2\left(\varphi_{i,j}-\varphi_{i,j'}\right)\mathring{n}_{i}\\
-\left(\varphi_{1,j}^{2}-\varphi_{1,j'}^{2}\right)-2\left(\varphi_{1,j}-\varphi_{1,j'}\right)\mathring{n}_{1} & = & +\delta_{i,j}^{(\text{s})}+\delta_{i,j}^{2}-\delta_{1,j}^{(\text{s})}-\delta_{1,j}^{2}\\
 &  & -\delta_{i,j'}^{(\text{s})}-\delta_{i,j'}^{2}+\delta_{1,j'}^{(\text{s})}+\delta_{1,j'}^{2}\;.
\end{eqnarray*}
This is not a useful constraint; in this case we use the two single-epoch
constraints without subtracting them. 

\subsection{\label{sec:Linearizing-Euclidean-Distances}Linearizing Euclidean
Distances}

Finally, we can linearize the term $\|\ell-\rho_{i}\|_{2}$ that is
present in both the range and phase constraints around an approximate
location estimate $\ell_{0}$, which we can obtain by minimizing the
residual of the range constraints alone. The Taylor-series approximation
of the range near $\ell_{0}$ is
\begin{equation}
\left\Vert \ell-\rho_{i}\right\Vert _{2}=\left\Vert \ell_{0}-\rho_{i}\right\Vert _{2}+\frac{\partial\|\ell-\rho_{i}\|_{2}}{\partial\ell}\left(\ell-\ell_{0}\right)+\eta_{i}\;,\label{eq:linearized-range}
\end{equation}
where gradient $\partial\|\ell-\rho_{i}\|/\partial\ell$ is evaluated
at $\ell_{0}$ and where $\eta_{i}=O(\|\ell-\ell_{0}\|^{2})$ is a
truncation error. The symbolic form of the gradient is
\[
\frac{\partial\|\ell-\rho_{i}\|_{2}}{\partial\ell}=\frac{\left(\ell-\rho_{i}\right)^{T}}{\left\Vert \ell-\rho_{i}\right\Vert _{2}}
\]
so its value at $\ell_{0}$ is $(\ell_{0}-\rho_{i})^{T}/\|\ell_{0}-\rho_{i}\|$. 

The linear approximation gives us range and phase constraints that
are linear in both $\mathring{\ell}$ and $\mathring{n}_{i}$, but
with a large truncation error at short ranges:
\begin{eqnarray*}
r_{i} & = & \left\Vert \ell_{0}-\rho_{i}\right\Vert _{2}+\frac{\left(\ell_{0}-\rho_{i}\right)^{T}}{\left\Vert \ell_{0}-\rho_{i}\right\Vert _{2}}\left(\mathring{\ell}-\ell_{0}\right)+\eta_{i}+\epsilon_{i}\\
\lambda\varphi_{i}+\lambda\mathring{n}_{i} & = & \left\Vert \ell_{0}-\rho_{i}\right\Vert _{2}+\frac{\left(\ell_{0}-\rho_{i}\right)^{T}}{\left\Vert \ell_{0}-\rho_{i}\right\Vert _{2}}\left(\mathring{\ell}-\mathring{\ell}_{0}\right)+\eta_{i}+\delta_{i}\;.
\end{eqnarray*}

\subsection{\label{sec:Linearizing-Squared-Phase}Linearizing Squared Phase Constraints}

Finally, we also show how to linearize squared phase constraints,
in order to obtain one constraint that is linear in both $\mathring{\ell}$
and $\mathring{s}_{i}$. We start from Equation~(\ref{eq:squared_carrier_phase_with_error-1-1}),
\[
\lambda^{2}\mathring{s}_{i}=\left\Vert \mathring{\ell}-\rho_{i}\right\Vert _{2}^{2}+\delta_{i}^{(\text{s})}+\delta_{i}^{2}\;.
\]
In principle we can use a Taylor-series approximation again, but since
$\|\mathring{\ell}-\rho_{i}\|^{2}$ is a simple quadratic form in
$\mathring{\ell}$, we can use a simpler algebraic approach:
\[
\left\Vert \mathring{\ell}-\rho_{i}\right\Vert _{2}^{2}=\left\Vert \mathring{\ell}-\ell_{0}+\ell_{0}-\rho_{i}\right\Vert _{2}^{2}=\left\Vert \ell_{0}-\rho_{i}\right\Vert _{2}^{2}+2\left(\ell_{0}-\rho_{i}\right)^{T}\left(\mathring{\ell}-\ell_{0}\right)+\left\Vert \mathring{\ell}-\ell_{0}\right\Vert _{2}^{2}\;.
\]
We approximate the squared range using the first two terms on the
right hand side, and we treat the last term as a truncation error
$\eta^{(\text{s})}=\|\mathring{\ell}-\ell_{0}\|^{2}$, which is independent
of $i$. This gives us the linear constraint
\begin{equation}
\lambda^{2}\mathring{s}_{i}=\left\Vert \ell_{0}-\rho_{i}\right\Vert _{2}^{2}+2\left(\ell_{0}-\rho_{i}\right)^{T}\left(\mathring{\ell}-\ell_{0}\right)+\eta^{(\text{s})}+\delta_{i}^{(\text{s})}+\delta_{i}^{2}\;.\label{eq:linearized-squared-phase-si}
\end{equation}

\section{\label{sec:Assembling-Minimization-Problems}Assembling Minimization
Problems}

We estimate the $\mathring{n}_{i}$s and the location $\mathring{\ell}$
(or $\mathring{\ell}_{j}$s) by assembling constraints presented in
Section~\ref{sec:Deriving-Linear-Constraints} into residual minimization
problems.

\subsection{\label{subsec:Single-Epoch-Minimization-Problems}Single-Epoch Minimization
Problems}

In single-epoch settings, we assemble a minimization problem from
$m-1$ differenced squared phase constraints~(\ref{eq:differenced-squared-phases_si})
\[
\lambda^{2}\mathring{s}_{i}-\lambda^{2}\mathring{s}_{1}=-2\mathring{\ell}^{T}(\rho_{i}-\rho_{1})+\rho_{i}^{T}\rho_{i}-\rho_{1}^{T}\rho_{1}+\delta_{i}^{(\text{s})}+\delta_{i}^{2}-\delta_{1}^{(\text{s})}-\delta_{1}^{2}
\]
(recall that $m$ is the number of references). We also add the linearized
squared constraint for reference $1$ from Equation~(\ref{eq:linearized-squared-phase-si}),
\[
\lambda^{2}\mathring{s}_{1}=\left\Vert \ell_{0}-\rho_{1}\right\Vert _{2}^{2}+2\left(\ell_{0}-\rho_{1}\right)^{T}\left(\mathring{\ell}-\ell_{0}\right)+\eta^{(\text{s})}+\delta_{1}^{(\text{s})}+\delta_{1}^{2}\;.
\]
Next, we add $m-1$ differenced squared range constraints from Equation~(\ref{eq:differenced-squared-ranges}),
\begin{eqnarray*}
r_{i}^{2}-r_{1}^{2} & = & -2\mathring{\ell}^{T}(\rho_{i}-\rho_{1})+\rho_{i}^{T}\rho_{i}-\rho_{1}^{T}\rho_{1}+\epsilon_{i}^{(\text{s})}+\epsilon_{i}^{2}-\epsilon_{1}^{(\text{s})}-\epsilon_{1}^{2}
\end{eqnarray*}
as well as $m$ linearized range constraints
\[
r_{i}=\left\Vert \ell_{0}-\rho_{i}\right\Vert _{2}+\frac{\left(\ell_{0}-\rho_{i}\right)^{T}}{\left\Vert \ell_{0}-\rho_{i}\right\Vert _{2}}\left(\mathring{\ell}-\ell_{0}\right)+\eta_{i}+\epsilon_{i}\;.
\]
The last two families use each range constraint twice, once in each
family. We use both families because their error terms are completely
different.

\subsection{\label{subsec:Multiple-Epoch-Minimization-Problems}Multiple-Epoch
Minimization Problems}

In the multiple-epoch settings, we start from the double-differenced
squared phase constraints described in Section~\ref{sec:Double-Differencing-by-Epoch}
\begin{eqnarray*}
\left(\varphi_{i,j}^{2}-\varphi_{i,j'}^{2}\right)+2\left(\varphi_{i,j}-\varphi_{i,j'}\right)\mathring{n}_{i}\\
-\left(\varphi_{1,j}^{2}-\varphi_{1,j'}^{2}\right)-2\left(\varphi_{1,j}-\varphi_{1,j'}\right)\mathring{n}_{1} & = & -2\mathring{\ell}_{j}^{T}(\rho_{i,j}-\rho_{1,j})+\rho_{i,j}^{T}\rho_{i,j}-\rho_{1,j}^{T}\rho_{1,j}+\delta_{i,j}^{(\text{s})}+\delta_{i,j}^{2}-\delta_{1,j}^{(\text{s})}-\delta_{1,j}^{2}\\
 &  & +2\mathring{\ell}_{j'}^{T}(\rho_{i,j'}-\rho_{1,j'})-\rho_{i,j'}^{T}\rho_{i,j'}+\rho_{1,j'}^{T}\rho_{1,j'}-\delta_{i,j'}^{(\text{s})}-\delta_{i,j'}^{2}+\delta_{1,j'}^{(\text{s})}+\delta_{1,j'}^{2}\;,
\end{eqnarray*}
or their simplified variants when either the target or the references
are stationary. For the case of two epochs, this gives $m-1$ constraints.

We add the single-differenced range constraints from Equation~(\ref{eq:differenced-squared-ranges}),
\begin{eqnarray*}
r_{i,j}^{2}-r_{1,j}^{2} & = & -2\mathring{\ell}_{j}^{T}(\rho_{i,j}-\rho_{1,j})+\rho_{i,j}^{T}\rho_{i,j}-\rho_{1,j}^{T}\rho_{1,j}+\epsilon_{i,j}^{(\text{s})}+\epsilon_{i,j}^{2}-\epsilon_{1,j}^{(\text{s})}-\epsilon_{1,j}^{2}\;.
\end{eqnarray*}
In our implementation described below in Section~\ref{sec:Simulations-and-Evaluation},
we currently add differences of these constraints by epoch, but one
may consider adding them without double differencing. 

Finally, we add the linearized phase and range constraints described
in Section~\ref{sec:Linearizing-Euclidean-Distances}, 

\begin{eqnarray*}
r_{i,j} & = & \left\Vert \ell_{0,j}-\rho_{i,j}\right\Vert _{2}+\frac{\left(\ell_{0,j}-\rho_{i,j}\right)^{T}}{\left\Vert \ell_{0,j}-\rho_{i,j}\right\Vert _{2}}\left(\mathring{\ell}_{j}-\ell_{0,j}\right)+\eta_{i,j}+\epsilon_{i,j}\\
\lambda\varphi_{i,j}+\lambda\mathring{n}_{i} & = & \left\Vert \ell_{0,j}-\rho_{i,j}\right\Vert _{2}+\frac{\left(\ell_{0,j}-\rho_{i,j}\right)^{T}}{\left\Vert \ell_{0,j}-\rho_{i,j}\right\Vert _{2}}\left(\mathring{\ell}_{j}-\ell_{0,j}\right)+\eta_{i,j}+\delta_{i,j}
\end{eqnarray*}
(again for both epochs $j$ and $j'$).

\subsection{\label{subsec:Linearize-First-LAMBDA}Linearize-First Minimization
Problems (LAMBDA)}

For reference and to specify precisely the LAMBDA algorithm that we
used in the numerical experiments below, we specify the constraints
that are used in our implementation of the LAMBDA method. Specifically,
in this method we assemble the linearized phase and range constraints
described in Section~\ref{sec:Linearizing-Euclidean-Distances}, 

\begin{eqnarray*}
r_{i,j} & = & \left\Vert \ell_{0,j}-\rho_{i,j}\right\Vert _{2}+\frac{\left(\ell_{0,j}-\rho_{i,j}\right)^{T}}{\left\Vert \ell_{0,j}-\rho_{i,j}\right\Vert _{2}}\left(\mathring{\ell}_{j}-\ell_{0,j}\right)+\eta_{i,j}+\epsilon_{i,j}\\
\lambda\varphi_{i,j}+\lambda\mathring{n}_{i} & = & \left\Vert \ell_{0,j}-\rho_{i,j}\right\Vert _{2}+\frac{\left(\ell_{0,j}-\rho_{i,j}\right)^{T}}{\left\Vert \ell_{0,j}-\rho_{i,j}\right\Vert _{2}}\left(\mathring{\ell}_{j}-\ell_{0,j}\right)+\eta_{i,j}+\delta_{i,j}
\end{eqnarray*}
(for a single or multiple epochs, always with a single set of $n_{i}$s). 

\subsection{\label{subsec:From-Constraints-to-Generalized-Least-Squares}From
Sets of Constraints to Generalized Least Squares}

In the single-epoch case, we end up with a set of constraints of the
form
\[
A\mathring{\ell}+B\mathring{s}-d=e
\]
where $A$ is a known matrix, $B$ is a known matrix, $d$ is a known
vector, and $e$ is a vector of error terms. The unknown parameters
are the 2D or 3D location vector $\mathring{\ell}$ and the $m$-vector
$\mathring{s}$ of discrete parameters of the form $\mathring{s}_{i}=(\varphi_{i}+\mathring{n}_{i})^{2}$
where $\mathring{n}_{i}\in\mathbb{N}$. In the multi-epoch case, we
end up with a set of constraints of the form 
\[
A\mathring{\ell}+B\mathring{n}-d=e
\]
where $\mathring{n}\in\mathbb{N}^{m}$.

In both cases, some of the rows of $B$ are identically zero. These
rows corresponds to constraints that were derived from range observations
alone and are independent of any phase observations. 

Estimating the unknown vectors well requires a characterization of
the statistical and/or mathematical properties of the error vectors
$e$. Without loss of generality, one can assume that the expected
value of the $e_{i}$s is zero, because if some have a nonzero expectation
that can be computed or estimated, the expectation should be subtracted
from $d$, leaving an error vector with approximately zero expectation.
That is, we rearrange our equations as 
\[
A\mathring{\ell}+B\mathring{s}-\left(d+\text{expectation}(e)\right)=e-\text{expectation}(e)\;.
\]

We estimate $\mathring{\ell}$ and $\mathring{s}$ or $\mathring{n}$
by minimizing $\|W(e-\text{expectation}(e))\|_{2}$ where $W$ is
a weight matrix that is an inverse factor of an approximate covariance
matrix $C=(W^{T}W)^{-1}$. We stress that in general, the accuracy
of the estimate depends only weakly on the accuracy of $W$. Relative
accuracy to within 1\% or even just 10\% is usually sufficient, and
higher accuracy of $W$ does not improve much the accuracy of the
estimates.

We minimize the two-norm of $e-\text{expectation}(e)$ because this
is computationally easier than minimizing other norms, and because
when $e$ is distributed normally with a known expectation and a known
covariance matrix $C$, the two-norm minimizers are also the maximum-likelihood
estimators. 

However, in our case some elements of $e$ are not distributed normally,
not even approximately. This requires deeper insights on the choice
of $C$ and on the implications of this choice. Many years of experience
with LAMBDA applied to GNSS problems yielded the following insights,
which are also relevant to our problem~\parencite{LocationEstimationFromTheGroundUp}:
\begin{quote}
The set of phase observations have many local minima, including one
at the true $\mathring{n}$ and near $\mathring{\ell}$ , but typically
many more, especially when the number $m$ of references is small.
The residual at some of the other locations may be similar or even
smaller than at the true location, making it impossible to identify
the correct location from these constraints alone. The range (or ToA)
constraints help identify the correct solution. Far from the true
solution, these constraints have large residuals, which eliminates
these hypothetical solutions even if the residual of the phase constraints
alone there is tiny.
\end{quote}
Therefore, correctly weighing the phase constraints is key to the
accuracy of the estimates. If these are incorrectly weighed with respect
to one another, or too weakly weighed relative to the range/ToA observations,
the minimizer tends to remain near the true location, but its accuracy
suffers. On the other hand, if the range/ToA constraints are weighed
too weakly, the likelihood that the minimizer is not at all near the
true solution increases; but if the minimizer happens to be near the
true solution, the weak weighing of the range/ToA constraints has
little effect on the accuracy of the estimate.

These insights inform our analyses of the error terms in the next
section.

\section{\label{sec:Quantifying-Error-Terms}The Distribution of the Error
Terms and Approximations of Their Covariance Matrices}

We now analyze each error term introduced in Section~\ref{sec:Deriving-Linear-Constraints},
aiming to characterize their distribution. In particular, we aim to
estimate their expectation and their variance, or at least bound their
variance, in order to subtract the approximate expectation from $d$
and to correctly weigh the constraints.

We assume that the observation errors $\epsilon_{i}$ and $\delta_{i}$
are Gaussian, uncorrelated, and have zero mean and known standard
deviations specified in Section~\ref{sec:Typical-Parameters}.

\subsection{Squares of Observation Errors}

In constraints that are based on squaring, we have $\epsilon_{i}^{2}$
or $\delta_{i}^{2}$ terms. Their distribution is scaled $\chi^{2}(1)$
with expectation $\sigma^{2}(\epsilon_{i})$ or $\sigma^{2}(\delta_{i})$
and standard deviation $\smash{\sqrt{2}}\sigma^{2}(\epsilon_{i})$
or $\smash{\sqrt{2}}\sigma^{2}(\delta_{i})$ . 

We assume that these terms can be neglected. The $\delta_{i}^{2}$
errors are tiny, and the $\epsilon_{i}^{2}$ are also small and shadowed
by larger errors.

\subsection{\label{sec:epsilon-squared-delta-squared}The Errors $\epsilon_{i}^{(\text{s})}$
and $\delta_{i}^{(\text{s})}$}

The errors 
\begin{eqnarray*}
\delta_{i}^{(\text{s})} & = & 2\delta_{i}\|\mathring{\ell}-\rho_{i}\|\\
\epsilon_{i}^{(\text{s})} & = & 2\epsilon_{i}\|\mathring{\ell}-\rho_{i}\|
\end{eqnarray*}
are scaling of $\epsilon_{i}$ and $\delta_{i}$, since the norm $\|\mathring{\ell}-\rho_{i}\|$
is deterministic, even though it is not known. Therefore, their expectation
is zero, and their standard deviations are $2\sigma(\epsilon_{i})\|\mathring{\ell}-\rho_{i}\|$
and $2\sigma(\delta_{i})\|\mathring{\ell}-\rho_{i}\|$. We do not
know the precise value of $\|\mathring{\ell}-\rho_{i}\|$, but we
can approximate the the standard deviations with 
\begin{eqnarray*}
\bar{\sigma}\left(\delta_{i}^{(\text{s})}\right) & = & 2\delta_{i}\|\ell_{0}-\rho_{i}\|\\
\bar{\sigma}\left(\epsilon_{i}^{(\text{s})}\right) & = & 2\epsilon_{i}\|\ell_{0}-\rho_{i}\|\;.
\end{eqnarray*}
For the purpose of weighing the constraints, this is an excellent
approximation.

These errors are uncorrelated with each other, due to our assumption
that the $\epsilon_{i}$s and $\delta_{i}$s, but they are correlated
with some of the other error terms in the same constraints.

\subsection{Linearization Errors of Squared Ranges}

If we assume that $\mathring{\ell}-\ell_{0}$ is Gaussian, has zero
expectation, and has covariance matrix $C_{0}$, then the linearization
error $\eta^{(\text{s})}=\|\mathring{\ell}-\ell_{0}\|^{2}$ is a generalized
$\chi^{2}$ random variable whose expectation is $\text{trace}(C_{0})$
and whose variance is $2\|C_{0}\|_{F}^{2}$ (twice the Frobenius norm
of $C_{0}$). Note that 
\begin{equation}
\text{trace}(C_{0})/\sqrt{2}\leq\sqrt{2}\|C_{0}\|_{F}\leq\sqrt{2}\text{trace}(C_{0})\;.\label{eq:frobenius-norm-trace-bounds}
\end{equation}

We weigh the squared range constraint as if $\eta^{(\text{s})}$ is
distributed normally with zero expectation and with variance $\|C_{0}\|_{F}^{2}$.

We now prove~(\ref{eq:frobenius-norm-trace-bounds}).
\begin{lem}
Let $C_{0}$ be a 2-by-2 or 3-by-3 symmetric semidefinite matrix.
Then
\[
\text{trace}(C_{0})/\sqrt{2}\leq\sqrt{2}\|C_{0}\|_{F}\leq\sqrt{2}\text{trace}(C_{0})\;.
\]
\end{lem}
\begin{proof}
Let $0\leq v_{1},\ldots,v_{d}$ be the eigenvalues of $C_{0}$. We
have
\begin{eqnarray*}
\|C_{0}\|_{F} & = & \sqrt{{\scriptstyle {\textstyle \sum_{i}v_{i}^{2}}}}=\|v\|_{2}\\
\text{trace}\left(C_{0}\right) & = & {\textstyle \sum_{i}v_{i}}=\|v\|_{1}\;.
\end{eqnarray*}
The upper bound on $\smash{\sqrt{2}\|C_{0}\|_{F}}$ holds because
$\|v\|_{2}\leq\|v\|_{1}$ for any vector $v$. The lower bound holds
for dimensions $d=1,2,3,4$ because $\|v\|_{1}\leq\smash{\sqrt{d}}\|v\|_{2}$~\parencite[Fact~9.8.10]{bernstein2005matrix}.
\end{proof}

\subsection{\label{sec:Errors-in-Linearization-of-Ranges}Errors in Linearization
of Ranges}

We now analyze the error in the linearization of ranges. We have
\begin{eqnarray*}
\left\Vert \mathring{\ell}-\rho_{i}\right\Vert _{2} & = & \left\Vert \ell_{0}-\rho_{i}\right\Vert _{2}+\frac{\partial\|\ell-\rho_{i}\|_{2}}{\partial\ell}\left(\mathring{\ell}-\ell_{0}\right)+\eta_{i}\\
 & = & \left\Vert \ell_{0}-\rho_{i}\right\Vert _{2}+\frac{\left(\ell_{0}-\rho_{i}\right)^{T}}{\left\Vert \ell_{0}-\rho_{i}\right\Vert _{2}}\left(\mathring{\ell}-\ell_{0}\right)+\eta_{i}
\end{eqnarray*}
(recall that the derivative is evaluated at $\ell_{0}$). Therefore,
the error term that we need to analyze is
\begin{eqnarray*}
\eta_{i} & = & \left\Vert \mathring{\ell}-\rho_{i}\right\Vert _{2}-\left\Vert \ell_{0}-\rho_{i}\right\Vert _{2}-\frac{\left(\ell_{0}-\rho_{i}\right)^{T}}{\left\Vert \ell_{0}-\rho_{i}\right\Vert _{2}}\left(\mathring{\ell}-\ell_{0}\right)\\
 & = & \left\Vert \mathring{\ell}-\rho_{i}\right\Vert _{2}-\left\Vert \ell_{0}-\rho_{i}\right\Vert _{2}-\left\Vert \mathring{\ell}-\ell_{0}\right\Vert _{2}\frac{\left(\ell_{0}-\rho_{i}\right)^{T}}{\left\Vert \ell_{0}-\rho_{i}\right\Vert _{2}}\frac{\left(\mathring{\ell}-\ell_{0}\right)}{\left\Vert \mathring{\ell}-\ell_{0}\right\Vert _{2}}\\
 & = & \left\Vert \mathring{\ell}-\rho_{i}\right\Vert _{2}-\left\Vert \ell_{0}-\rho_{i}\right\Vert _{2}-\left\Vert \mathring{\ell}-\ell_{0}\right\Vert _{2}\cos\left(\pi-\alpha\right)\;.
\end{eqnarray*}
where $\alpha$ is the angle between the line segments $\ell_{0}\leftrightarrow\mathring{\ell}$
and $\ell_{0}\leftrightarrow\rho_{i}$, as shown in Figure~\ref{fig:linearization-errors}~(top).

\begin{figure}
\begin{centering}
\begin{tikzpicture}

  \clip (0,0.5) rectangle (12,4.25);

  \draw[thick,red] (0.5,4) circle (9.75cm);

  \fill[color=lightgray] (0.5,4) circle (1.25mm);
  \fill[color=lightgray] (7.5,2) circle (1.25mm);
  \fill[color=lightgray] (10.5,3.5) circle (1.25mm);

  \fill[color=lightgray] (intersection of 0.5,4--11.5,0.85714285714 and 10.5,3.5--9.5,0) circle (1.25mm);


  \draw[thick] (0.5,4) -- (intersection of 0.5,4--11.5,0.85714285714 and 10.5,3.5--9.5,0); 
  \draw[thick] (7.5,2) -- (10.5,3.5); 
  \draw[thick] (0.5,4) -- (10.5,3.5);
 \draw[thick] (10.5,3.5) -- (intersection of 0.5,4--11.5,0.85714285714 and 10.5,3.5--9.5,0);
  \draw[thin] (intersection of 0.5,4--11.5,0.85714285714 and 10.5,3.5--9.5,0) 
                   ++(-2.5mm,+0.71428571427mm) -- ++(0.71428571427mm,2.5mm) -- ++(+2.5mm,-0.71428571427mm);


  \draw[thin] (7.5,2) ++(-0.75,+0.21428571428) arc (164.06:-15.94:7.8mm); 

  \draw                       (8.25,2.05) node[right] {$\pi-\alpha$} ;
  \draw                       (7.25,2.45) node[right] {$\alpha$} ;
  \draw                       (7.5,2) node[below] {$\ell_0$} ;
  \draw                       (0.5,4) node[left] {$\rho_i$} ;
  \draw                       (10.5,3.5) node[right] {$\mathring{\ell}$} ;
 \draw                       (intersection of 0.5,4--11.5,0.85714285714 and 10.5,3.5--9.5,0) 
                              node[below=8pt,right=-3pt] {$p$} ;

\end{tikzpicture}

\vspace{1cm}

\begin{tikzpicture}
  \clip (0,-0.35) rectangle (12,4.25);

  \path[name path=arc1] (0.5,4) ++(9.3803,0) arc (0:-70:9.3803);
  \path[name path=arc2] (10.5,3.5) ++(0,-3.5) arc (-90:-120:3.5);
  
  \path[name intersections={of=arc1 and arc2, by={A}}];
 
  \draw[thick,red] (0.5,4) circle (9.3803cm);

  \fill[color=lightgray] (0.5,4) circle (1.25mm);
  \fill[color=lightgray] (A) circle (1.25mm);
  \fill[color=lightgray] (10.5,3.5) circle (1.25mm);

  \draw[thick] (0.5,4) -- (A); 
  \draw[thick] (A) -- (10.5,3.5); 
  \draw[thick] (0.5,4) -- (10.5,3.5);
  \draw[thin] (A) 
                   ++(-2.4mm,+0.9mm) -- ++(0.9mm,2.4mm) -- ++(+2.4mm,-0.9mm);



  \draw                       (A) node[right] {$\ell_0$} ;
  \draw                       (0.5,4) node[left] {$\rho_i$} ;
  \draw                       (10.5,3.5) node[right] {$\mathring{\ell}$} ;

\end{tikzpicture}
\par\end{centering}
\caption{\label{fig:linearization-errors}Top: The distance between $\rho_{i}$
and the red circle (equal to $\|\rho_{i}-p\|$) is the linear approximation
to $\|\rho_{i}-\mathring{\ell}\|$. Note that $\|\ell_{0}-p\|=\|\ell_{0}-\mathring{\ell}\|\cos(\pi-\alpha)$.
Bottom: the worst-case linearization error, when $\alpha=\pi/2$.
Note that the distances $\|\ell_{0}-\mathring{\ell}\|$ is the same
at the top and at the bottom, as are the coordinates of $\rho_{i}$
and $\mathring{\ell}$.}
\end{figure}

When $\rho_{i}$, $\ell_{0}$, and $\mathring{\ell}$ are collinear,
the approximation is exact; either $\alpha=\pi$ and $\cos(\pi-\alpha)=1$
and 
\[
\left\Vert \mathring{\ell}-\rho_{i}\right\Vert _{2}=\left\Vert \ell_{0}-\rho_{i}\right\Vert _{2}+\left\Vert \mathring{\ell}-\ell_{0}\right\Vert _{2}\;,
\]
or $\alpha=0$, $\cos(\pi-\alpha)=-1$, and
\[
\left\Vert \mathring{\ell}-\rho_{i}\right\Vert _{2}=\left\Vert \ell_{0}-\rho_{i}\right\Vert _{2}-\left\Vert \mathring{\ell}-\ell_{0}\right\Vert _{2}\;.
\]
In both cases we have $\eta_{i}=0$. The error is largest when $\alpha=\pi/2$
and $\cos(\pi-\alpha)=0$, so the approximation is $\left\Vert \ell_{0}-\rho_{i}\right\Vert _{2}$
with no correction, as shown in the bottom drawing in Figure~\ref{fig:linearization-errors}.
The magnitude of the linearization error in this case is 
\[
\left\Vert \mathring{\ell}-\rho_{i}\right\Vert _{2}-\left\Vert \ell_{0}-\rho_{i}\right\Vert _{2}\;,
\]
the difference between the hypotenuse of the right triangle $\rho_{i}\leftrightarrow\ell_{0}\leftrightarrow\mathring{\ell}$
and the leg $\rho_{i}\leftrightarrow\ell_{0}$ of that triangle. If
we denote the length long leg of the triangle $r=\left\Vert \ell_{0}-\rho_{i}\right\Vert $
and the short leg $d=\left\Vert \mathring{\ell}-\ell_{0}\right\Vert $,
the error is
\[
\sqrt{r^{2}+d^{2}}-r\;.
\]

We now estimate the magnitude of this expression for $d$ much smaller
than $r$. We define $f(d)=\smash{\sqrt{r^{2}+d^{2}}}$ and differentiate
\begin{eqnarray*}
\frac{\partial f}{\partial d} & = & \frac{d}{\sqrt{r^{2}+d^{2}}}\\
\frac{\partial^{2}f}{\partial d^{2}} & = & \frac{r^{2}}{\left(r^{2}+d^{2}\right)^{3/2}}\;.
\end{eqnarray*}
We approximate $f$ for small $d$ (relative to a large $r$) using
its Taylor series, evaluating the derivatives at $d=0$,
\begin{eqnarray*}
f\left(d\right) & = & \sqrt{r^{2}+d^{2}}\\
 & = & f(0)+\frac{\partial f}{\partial d}d+\frac{1}{2}\frac{\partial^{2}f}{\partial d^{2}}d^{2}+O(d^{3})\\
 & \approx & f(0)+\frac{\partial f}{\partial d}d+\frac{1}{2}\frac{\partial^{2}f}{\partial d^{2}}d^{2}\\
 & = & r+\frac{0}{r}d+\frac{1}{2}\frac{1}{r}d^{2}\\
 & = & r+\frac{1}{2}\frac{d^{2}}{r}\;.
\end{eqnarray*}
This shows that the linear approximation of the worst-case error,
for a large range $r$ and a small displacement $d$, behaves like
\[
\frac{1}{2}\frac{d^{2}}{r}\;.
\]
The denominator $r=\|\ell_{0}-\rho_{i}\|$ is deterministic, not random,
so the bound itself is a scaling of $d^{2}=\|\mathring{\ell}-\ell_{0}\|^{2}=\eta_{i}^{(\text{s})}$,
a generalized $\chi^{2}$ random variable whose expectation and variance
are $\text{trace}(C_{0})$ and $2\|C_{0}\|_{F}^{2}$. The scaling
factor is $1/(2r)=1/(2\|\ell_{0}-\rho_{i}\|)$.

The distribution of $\eta_{i}$, as opposed to this upper bound, also
depends on $\alpha$ and we do not attempt to fully characterize it.

\begin{figure}
\includegraphics[width=0.47\textwidth]{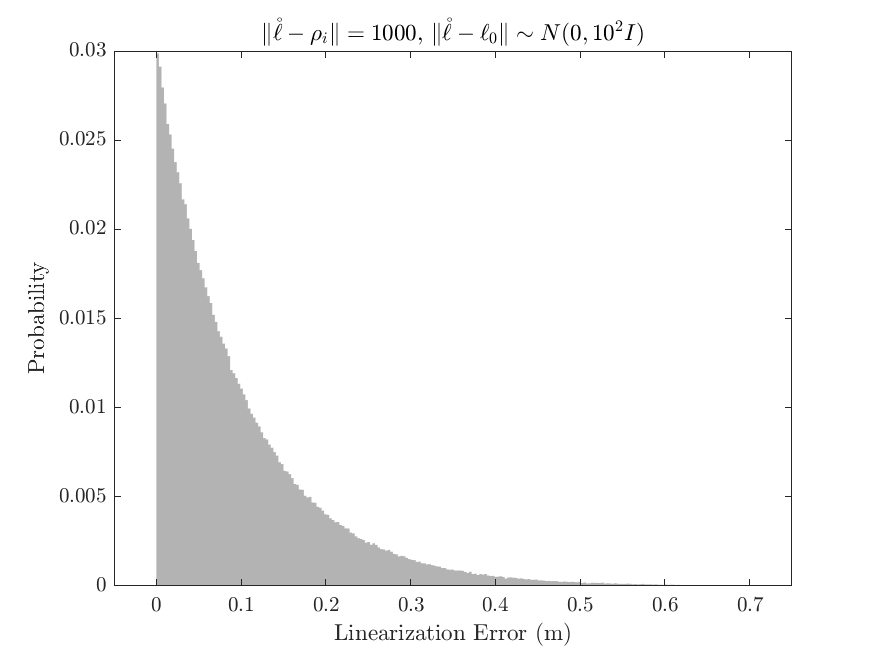}\hfill{}\includegraphics[width=0.47\textwidth]{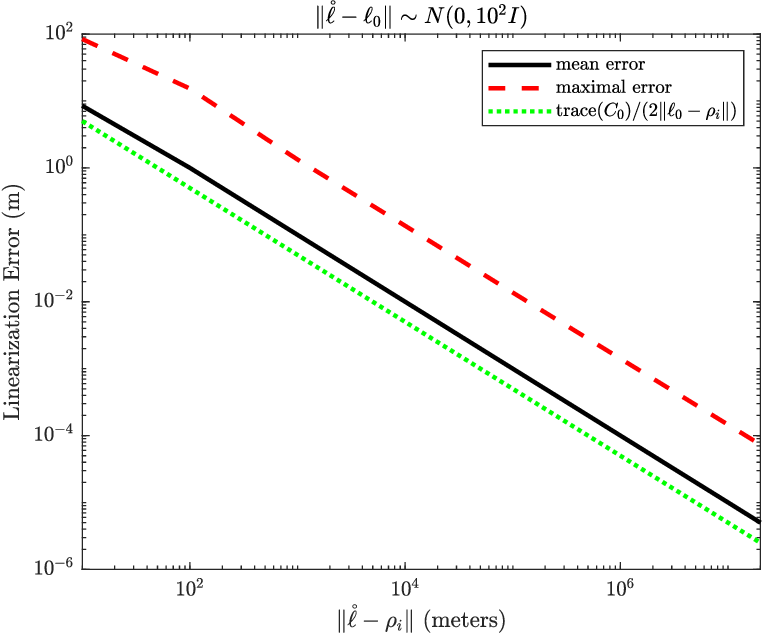}

\caption{\label{fig:linearization-error-stats-1}The distribution and mean
of linearization errors when $\ell-\ell_{0}$ is distributed normally
with zero expectation and covariance $10^{2}I$, in three dimensions.
The histogram on the left shows the distribution of errors in 1,000,000
experiments, when the target is a kilometer from the target, $\|\mathring{\ell}-\rho_{i}\|=1000$.
The value of $(1/2)\text{trace}\left(C_{0}\right)/\|\ell_{0}-\rho_{i}\|$
is about 0.05. The graph on the right shows the mean error, maximal
error, and the bound $(1/2)\text{trace}\left(C_{0}\right)/\|\ell_{0}-\rho_{i}\|$
for a whole spectrum of ranges $\|\mathring{\ell}-\rho_{i}\|$. }
\end{figure}

Figure~\ref{fig:linearization-error-stats-1} presents the statistics
of $\eta_{i}$ for a simple normal distribution of $\mathring{\ell}-\ell_{0}$.
We see that the mean error is about twice the value of 
\[
\frac{1}{2}\frac{\text{trace}\left(C_{0}\right)}{\left\Vert \ell_{0}-\rho_{i}\right\Vert }\;,
\]
and that the maximum error over one million experiments is much higher. 

In our algorithms, we weigh each linearized constraint 
\[
\left(\frac{1}{100}\frac{\text{trace}\left(C_{0}\right)}{\left\Vert \ell_{0}-\rho_{i}\right\Vert }\right)^{-1}\;,
\]
which works well in simulations, in spite of the somewhat high weights
that the $1/100$ factor induces. The $\eta_{i}$s are correlated
because of their shared dependence on $\|\mathring{\ell}-\ell_{0}\|$,
and because the relations between the $\alpha$s are a function of
the location of the references, but we do not model these correlations
in the weighing matrix.

\subsection{Overall Approximate Covariance Matrices}

\begin{table}
\begin{centering}
\begin{tabular}{llll}
\textbf{Constraint} & \textbf{Notation} & \textbf{Distribution} & \textbf{Approximation}\tabularnewline
\hline 
range observations (\ref{eq:range-observations}) & $\epsilon_{i}$ & $\mathcal{N}(0,\sigma^{2}(\epsilon_{i}))$ & \tabularnewline
carrier-phase observations (\ref{eq:phase-observations}) & $\delta_{i}$ & $\mathcal{N}(0,\sigma^{2}(\delta_{i}))$ & \tabularnewline
initial location estimate & $\mathring{\ell}-\ell_{0}$ & $\mathcal{N}(0,C_{0})$ & \tabularnewline
\hline 
squared range observations (\ref{eq:squared-range-observation}) & $\epsilon_{i}^{2}$ & $\chi_{1}^{2}(\sigma^{2}(\epsilon_{i}),2\sigma^{4}(\epsilon_{i}))$ & $0$\tabularnewline
squared range observations (\ref{eq:squared-range-observation}) & $\epsilon_{i}^{(\text{s})}=2\epsilon_{i}\|\mathring{\ell}-\rho_{i}\|$ & $\mathcal{N}(0,2\epsilon_{i}\|\mathring{\ell}-\rho_{i}\|)$ & $\mathcal{N}(0,2\epsilon_{i}\|\ell_{0}-\rho_{i}\|)$\tabularnewline
squared phase observations (\ref{eq:squared_carrier_phase_with_error-1}) & $\delta_{i}^{2}$ & $\chi_{1}^{2}(\sigma^{2}(\delta_{i}),2\sigma^{4}(\delta_{i}))$ & $0$\tabularnewline
squared phase observations (\ref{eq:squared_carrier_phase_with_error-1}) & $\epsilon_{i}^{(\text{s})}=2\delta_{i}\|\mathring{\ell}-\rho_{i}\|$ & $\mathcal{N}(0,2\delta_{i}\|\mathring{\ell}-\rho_{i}\|)$ & $\mathcal{N}(0,2\delta_{i}\|\ell_{0}-\rho_{i}\|)$\tabularnewline
linearized squared ranges (\ref{eq:linearized-squared-phase-si}) & $\eta^{(\text{s})}=\|\mathring{\ell}-\ell_{0}\|^{2}$ & $\chi_{1}^{2}(\text{trace}(C_{0}),2\|C_{0}\|_{F}^{2})$ & $\mathcal{N}(0,\text{trace}(C_{0}))$\tabularnewline
linearized ranges (\ref{eq:linearized-range}) & $\eta_{i}\lesssim\eta^{(\text{s})}/(2\|\ell_{0}-\rho_{i}\|)$ & --- & $\mathcal{N}(0,\text{trace}(C_{0})/100\|\ell_{0}-\rho_{i}\|)$\tabularnewline
\hline 
\end{tabular}
\par\end{centering}
\caption{\label{tab:error-distributions}Notation and distributions of the
error terms in the original constraints (top three rows) and in derived
constraints (bottom six). We have not fully characterized the distribution
of $\eta_{i}$ in the bottom row, so we show a bound on the error,
not a distribution; the $\lesssim$ signifies that the bound is a
second-order Taylor approximation, not an exact bound.}

\end{table}
Table~\ref{tab:error-distributions} lists all the error terms in
our constraints, our characterization of the distribution of each
error (or an upper bound for the one term for which we do not have
a complete characterization), and the approximations of the expectation
and variance of the errors that we use to weigh the constraints.

By our assumptions, the $\epsilon_{i}$s and the $\delta_{i}$s are
mutually independent, so they are uncorrelated. The squares $\epsilon_{i}^{2}$
and $\delta_{i}^{2}$ are statistically dependent on $\epsilon_{i}$
and $\delta_{i}$, respectively, but are uncorrelated with them. The
error in the initial location $\mathring{\ell}-\ell_{0}$ depends
on the $\epsilon_{i}$s; it is possible to approximate the covariance
of the error as a linear function of the $\epsilon_{i}$s, but our
implementation neglects this. The $\eta_{i}^{(\text{s})}$s and $\eta_{i}$s,
which are functions of $\mathring{\ell}-\ell_{0}$ also depends on
the $\epsilon_{i}$s; the dependence can be modeled but our implementation
neglects it.

We, therefore, approximate the covariance matrix of the individual
error terms
\[
e_{\text{raw}}=\left[\begin{array}{l}
\delta_{1}\\
\vdots\\
\delta_{m}\\
\hline \delta_{1}^{2}\\
\vdots\\
\delta_{m}^{2}\\
\hline \epsilon_{1}\\
\vdots\\
\epsilon_{m}\\
\hline \epsilon_{1}^{2}\\
\vdots\\
\epsilon_{m}^{2}\\
\hline \eta_{1}\\
\vdots\\
\eta_{m}\\
\hline \eta^{(\text{s})}
\end{array}\right]
\]
by
\[
C_{\text{raw}}=\left[\begin{array}{ccc|ccc|ccc|ccc|ccc|c}
\sigma^{2}(\delta_{1}) &  &  &  &  &  &  &  &  &  &  &  &  &  & \\
 & \ddots &  &  &  &  &  &  &  &  &  &  &  &  & \\
 &  & \sigma^{2}(\delta_{1}) &  &  &  &  &  &  &  &  &  &  &  & \\
\hline  &  &  & 0 &  &  &  &  &  &  &  &  &  &  & \\
 &  &  &  & \ddots &  &  &  &  &  &  &  &  &  & \\
 &  &  &  &  & 0 &  &  &  &  &  &  &  &  & \\
\hline  &  &  &  &  &  & \sigma^{2}(\epsilon_{1}) &  &  &  &  &  & ? & \cdots & ? & ?\\
 &  &  &  &  &  &  & \ddots &  &  &  &  & \vdots & \ddots & \vdots & \vdots\\
 &  &  &  &  &  &  &  & \sigma^{2}(\epsilon_{1}) &  &  &  & ? & \cdots & ? & ?\\
\hline  &  &  &  &  &  &  &  &  & 0 &  &  &  &  & \\
 &  &  &  &  &  &  &  &  &  & \ddots &  &  &  & \\
 &  &  &  &  &  &  &  &  &  &  & 0 &  &  & \\
\hline  &  &  &  &  &  & ? & \cdots & ? &  &  &  & \hat{\sigma}^{2}(\eta_{1}) & ? & \ddots & ?\\
 &  &  &  &  &  & \vdots & \ddots & \vdots &  &  &  & ? & \ddots & ? & \vdots\\
 &  &  &  &  &  & ? & \cdots & ? &  &  &  & \ddots & ? & \hat{\sigma}^{2}(\eta_{m}) & ?\\
\hline  &  &  &  &  &  & ? & \cdots & ? &  &  &  & ? & \cdots & ? & \hat{\sigma}^{2}(\eta^{(\text{s})})
\end{array}\right]\;,
\]
where $\hat{\sigma}^{2}(\eta_{i})$ and $\hat{\sigma}^{2}(\eta_{i}^{(\text{s})})$
denote our approximations of the variances of $\eta_{i}$ and $\eta_{i}^{(\text{s})}$.
Empty cells in this matrix denote covariance values that are known
to be $0$, and question marks denote covariance values that are,
in general, not zero, but which we do not model but treat as zero.
Note that correctly modeling the variance of the $\delta_{i}^{2}$s
and $\epsilon_{i}^{2}$ is easy.

We now show how to weigh the constraints in the single-epoch case.
The residual vector in this case is
\[
\text{residual}_{\text{single-epoch}}=\left[\begin{array}{ll}
\eta_{1}^{(\text{s})}+\delta_{1}^{(\text{s})}+\delta_{1}^{2} & \text{(linearized squared phase)}\\
\delta_{2}^{(\text{s})}+\delta_{2}^{2}-\delta_{1}^{(\text{s})}-\delta_{1}^{2} & (\text{differenced squared phases})\\
\delta_{3}^{(\text{s})}+\delta_{3}^{2}-\delta_{1}^{(\text{s})}-\delta_{1}^{2} & \vdots\\
\vdots\\
\delta_{m}^{(\text{s})}+\delta_{m}^{2}-\delta_{1}^{(\text{s})}-\delta_{1}^{2}\\
\epsilon_{2}^{(\text{s})}+\epsilon_{2}^{2}-\epsilon_{1}^{(\text{s})}-\epsilon_{1}^{2} & (\text{differenced squared ranges})\\
\vdots & \vdots\\
\epsilon_{m}^{(\text{s})}+\epsilon_{m}^{2}-\epsilon_{1}^{(\text{s})}-\epsilon_{1}^{2}\\
\eta_{1}+\epsilon_{1} & (\text{linearized ranges})\\
\vdots & \vdots\\
\eta_{m}+\epsilon_{m}
\end{array}\right]\;.
\]
We have
\[
\text{residual}_{\text{single-epoch}}={\scriptscriptstyle \left[\begin{array}{cccc|cccc|cccc|cccc|ccc|c}
\zeta_{1} &  &  &  & 1 &  &  &  &  &  &  &  &  &  &  &  &  &  &  & 1\\
\hline -\zeta_{1} & \zeta_{2} &  &  & -1 & 1 &  &  &  &  &  &  &  &  &  &  &  &  & \\
\vdots &  & \ddots &  & \vdots &  & \ddots &  &  &  &  &  &  &  &  &  &  &  & \\
-\zeta_{1} &  &  & \zeta_{m} & -1 &  &  & 1 &  &  &  &  &  &  &  &  &  &  & \\
\hline  &  &  &  &  &  &  &  & -\zeta_{1} & \zeta_{2} &  &  & -1 & 1 &  &  &  &  & \\
 &  &  &  &  &  &  &  & \vdots &  & \ddots &  & -1 &  & 1 &  &  &  & \\
 &  &  &  &  &  &  &  & -\zeta_{1} &  &  & \zeta_{m} & \vdots &  &  & \ddots &  &  & \\
\hline  &  &  &  &  &  &  &  & 1 &  &  &  &  &  &  &  & 1 &  & \\
 &  &  &  &  &  &  &  &  & \ddots &  &  &  &  &  &  &  & \ddots & \\
 &  &  &  &  &  &  &  &  &  &  & 1 &  &  &  &  &  &  & 1
\end{array}\right]\left[\begin{array}{l}
\delta_{1}\\
\vdots\\
\delta_{m}\\
\hline \delta_{1}^{2}\\
\vdots\\
\delta_{m}^{2}\\
\hline \epsilon_{1}\\
\vdots\\
\epsilon_{m}\\
\hline \epsilon_{1}^{2}\\
\vdots\\
\epsilon_{m}^{2}\\
\hline \eta_{1}\\
\vdots\\
\eta_{m}\\
\hline \eta_{1}^{(\text{s})}
\end{array}\right]}\;.
\]
where $\zeta_{i}=2\|\mathring{\ell}-\rho_{i}\|\approx2\|\ell_{0}-\rho_{i}\|$.
Denoting the marix in this expression by $G$, we conclude that we
can approximate the covariance matrix of the residual vector by 
\[
C_{\text{single-epoch}}=GC_{\text{raw}}G^{T}\;,
\]
and from this covariance matrix we compute the weight matrix $W_{\text{single-epoch}}$
such that $W_{\text{single-epoch}}^{T}W_{\text{single-epoch}}=C_{\text{single-epoch}}^{-1}$.

The process for the multi-epoch squaring and double-differencing is
similar. The structure of the residual vector for two epochs $j$
and $j'$ is
\[
\left[\begin{array}{ll}
\delta_{2,j}^{(\text{s})}+\delta_{2,j}^{2}-\delta_{1,j}^{(\text{s})}-\delta_{1,j}^{2}-\delta_{2,j'}^{(\text{s})}-\delta_{2,j'}^{2}+\delta_{1,j'}^{(\text{s})}+\delta_{1,j'}^{2} & (\text{double-differenced squared phases})\\
\delta_{3,j}^{(\text{s})}+\delta_{3,j}^{2}-\delta_{1,j}^{(\text{s})}-\delta_{1,j}^{2}-\delta_{3,j'}^{(\text{s})}-\delta_{3,j'}^{2}+\delta_{1,j'}^{(\text{s})}+\delta_{1,j'}^{2} & \vdots\\
\vdots\\
\delta_{m,j}^{(\text{s})}+\delta_{m,j}^{2}-\delta_{1,j}^{(\text{s})}-\delta_{1,j}^{2}-\delta_{m,j'}^{(\text{s})}-\delta_{m,j'}^{2}+\delta_{1,j'}^{(\text{s})}+\delta_{1,j'}^{2}\\
\epsilon_{1,j}^{(\text{s})}+\epsilon_{1,j}^{2}-\epsilon_{1,j}^{(\text{s})}-\epsilon_{1,j}^{2} & (\text{differenced squared ranges, epoch \ensuremath{j}})\\
\vdots & \vdots\\
\epsilon_{m,j}^{(\text{s})}+\epsilon_{m,j}^{2}-\epsilon_{1,j}^{(\text{s})}-\epsilon_{1,j}^{2}\\
\epsilon_{1,j'}^{(\text{s})}+\epsilon_{1,j'}^{2}-\epsilon_{1,j'}^{(\text{s})}-\epsilon_{1,j'}^{2} & (\text{differenced squared ranges, epoch \ensuremath{j'}})\\
\vdots & \vdots\\
\epsilon_{m,j'}^{(\text{s})}+\epsilon_{m,j'}^{2}-\epsilon_{1,j'}^{(\text{s})}-\epsilon_{1,j'}^{2}\\
\eta_{1,j}+\epsilon_{1,j} & (\text{linearized ranges, epoch \ensuremath{j}})\\
\vdots & \vdots\\
\eta_{m,j}+\epsilon_{im,j}\\
\eta_{1,j}+\delta_{1,j} & (\text{linearized phases, epoch \ensuremath{j'}})\\
\vdots & \vdots\\
\eta_{m,j}+\delta_{im,j}
\end{array}\right]\;.
\]

\section{\label{sec:Generalized-Schnorr-Euchner}The Generalized Schnorr-Euchner
Search Algorithm}

We have shown in Sections~\ref{subsec:Single-Epoch-Minimization-Problems}
and~\ref{subsec:Multiple-Epoch-Minimization-Problems} how to assemble
systems of constraints in the single-epoch and multi-epoch settings,
and we have shown in Sections~\ref{subsec:From-Constraints-to-Generalized-Least-Squares}
and~\ref{sec:Quantifying-Error-Terms} how to weigh the constraints.
In the single-epoch case, we end up with a generalized least squares
problem of the form
\[
\hat{\ell},\hat{s}=\arg\min_{\ell,s}\left\Vert W\left(A\ell+Bs-d\right)\right\Vert _{2},\quad s_{i}=\left(n_{i}+\varphi_{i}\right)^{2}\text{ where }n_{i}\in\mathbb{\mathbb{N}}\;,
\]
and in multi-epoch case, we end up with
\[
\hat{\ell},\hat{n}=\arg\min_{\ell,n}\left\Vert W\left(A\ell+Bn-d\right)\right\Vert _{2},\quad n_{i}\in\mathbb{\mathbb{N}}\;.
\]
In both cases, $W$, $A$, and $B$ are known matrices and $d$ is
a known vector.

In both cases the vector whose norm we minimize is linear in $\ell$
and $s$ (or $\ell$ and $n$) and in both cases we have both real
unknown parameters $\ell$ and discrete unknown parameters, $s$ or
$n$. In both cases the first step is identical: we use an orthogonal
projection to eliminate the real parameters $\ell$. Let $U$ be an
orthonormal basis for the column space of $WA$. We can compute $U$
using the singular value decomposition (SVD), or somewhat more cheaply,
using a $QR$ factorization. It is well known and easy to see that
\begin{eqnarray*}
\hat{s} & = & \arg\min_{\ell,s}\left\Vert W\left(A\ell+Bs-d\right)\right\Vert _{2}\\
 & = & \arg\min_{s}\left\Vert \left(I-UU^{T}\right)\left(Bs-d\right)\right\Vert _{2}\;.
\end{eqnarray*}
We now have a least-squares problem with discrete (or integer) parameters
only, typically over determined.

Next, we transform the problem so that the coefficient matrix is triangular
and square.

We compute the $QR$ factorization of $(I-UU^{T})B=QR$, where $Q$
is orthnormal with $m$ columns and and $R$ is square and upper triangular.
Multiplying the residual by $Q^{T}$ preserves its norm, so,
\begin{eqnarray}
\hat{s} & = & \arg\min_{s}\left\Vert \left(I-UU^{T}\right)\left(Bs-d\right)\right\Vert _{2}\nonumber \\
 & = & \arg\min_{s}\left\Vert Rs-Q^{T}\left(I-UU^{T}\right)d\right\Vert _{2}\;.\label{eq:square-upper-discrete-ls}
\end{eqnarray}

Least squares problems of the form~(\ref{eq:square-upper-discrete-ls})
with integer parameters are NP hard~\parencite{ArgellEtAl2002,CVPisNPHard,vanEmdeBoasCVP}
and hard to approximate~\parencite{DinurCVP}, but they are routinely
solved exactly in RTK using a three-phase approach. In the first phase,
the LLL algorithm~\parencite{doi:10.1007/BF01457454} or one of its variants~\parencite{ImprovedLLL,PivotedLLL}
applies a series of unimodular transformations to the columns of $R$,
transforming~(\ref{eq:square-upper-discrete-ls}) into
\[
\hat{s}=\arg\min_{s}\left\Vert \left(RG\right)\left(G^{-1}s\right)-Q^{T}\left(I-UU^{T}\right)d\right\Vert _{2}=G\arg\min_{\tilde{s}}\left\Vert \left(RG\right)\left(G^{-1}\tilde{s}\right)-Q^{T}\left(I-UU^{T}\right)d\right\Vert _{2}\;,
\]
where $RG$ retains the square upper-triangular form. In the second
phase, a specialized branch-and-bound algorithm, called the Schnorr-Euchner
search~\parencite{doi:10.1007/BF01581144} finds the minimizer of the
unimodularly-transformed problem. In the last phase, $G$ is applied
to the solution, to transform it into a minimizer of~(\ref{eq:square-upper-discrete-ls}).
The first phase is optional and its role is to heuristically reduce
the cost of the search phase.

In the next subsections we explain how to adapt both the first and
second phases to the case of discrete parameters that are not integers.
We start with the second phase, since understanding it makes the role
of the first easier to understand.

\subsection{Adapting the Schnorr-Euchner Search to Shifted Squares}

The Schnorr-Euchner search algorithm~\parencite{doi:10.1007/BF01581144}
performs a sequence of assignments to the unknown parameters, starting
from the last, $s_{m}$, progressing backwards to $s_{m-1}$, $s_{m-2}$
and up to $s_{1}$, but backtracking when necessary. Backtracking
allows the algorithm to try a different assignment to a parameter
$s_{i}$ that has already been assigned to. In the original algorithm,
designed for integer least-squares problems, integers are assigned
to the $s_{i}$s.

We modify the algorithm so that the set of values $s_{i,k}$ that
we assign to $s_{i}$ are not integers but rather numbers of the form
$s_{i,k}=(n_{i,k}+\varphi_{i})^{2}$ where $n_{i,k}\in\mathbb{N}$.

Backtracking in the Schorr-Euchner search algorithm defines an implicit
rooted tree, in which each level except for the root corresponds to
a parameter $s_{i}$, each vertex to an assignment of $s_{i}$ to
a particular value $s_{i,k}$, and each path from a vertex $s_{i,k}$
to the root represents a concrete assignment to $s_{i},\ldots,s_{m}$.
The abstract tree is infinite, with each vertex having an infinite
number of children, but the Schnorr-Euchner search always terminates,
so it explores a finite subtree. The algorithm explores the children
of $s_{i+1,k}$, which represent a specific assignment of $s_{i+1:m}$,
in a particular order. This order is critical both to the correctness
and to the efficiency of the algorithm. In particular, at $s_{i+1,k}$,
the contribution of row $i$ to the sum of squares is
\begin{equation}
\left(R_{i,i}s_{i}+R_{i,i+1:m}s_{i+1:m}-\left(Q^{T}\left(I-UU^{T}\right)d\right)_{i}\right)^{2}\;.\label{eq:contribution-of-row-i}
\end{equation}
For the algorithm to work correctly and efficiently, these contributions
must grow monotonically with each assignment to $s_{i}$. We denote
\[
z_{i}=-\frac{R_{i,i+1:m}s_{i+1:m}-\left(Q^{T}d\right)_{i}}{R_{i,i}}\;.
\]
When $s_{i}$ is constrained to the set $(n_{i}+\varphi_{i})^{2}$,
$n_{i}\in\mathbb{N}$, the contribution of row $i$ is proportional
to
\[
\left(\left(n_{i}+\varphi_{i}\right)^{2}-z_{i}\right)^{2}\;.
\]
The square is monotone, so this expression is minimized when $|(n_{i}+\varphi_{i})^{2}-z_{i}|$
is minimized. We test both $\lfloor\sqrt{z_{i}}-\varphi_{i}\rfloor$
and $\lceil\sqrt{z_{i}}-\varphi_{i}\rceil$ and assign to $n_{i,1}$
the value $n_{1}$ that yields the smaller contribution. This is the
equivalent of the so-called Babai rounding for this problem. Now suppose
that we already tested integer values $n_{1}-l$ to $n_{1}+r$. The
next largest contribution will be either at $n_{1}-l-1$ or $n_{1}+r+1$.
We again test both and take the one that yields the smaller contribution.

We note that this adaptation is valid for any set of discrete values
that we can enumerate in increasing distance from a real value $z_{i}$.

\subsection{\label{subsec:QR-column-ordering}QR Decompositions with Column Pivoting}

The Scnorr-Euchner search explores a smaller finite subtree of the
infinite abstract tree when diagonal elements of $R$ near the bottom
right corner have large absolute values. When the parameters are integer-valued,
applying the LLL algorithm or its variants to $R$ often increases
the absolute values of diagonal elements near the bottom-right, making
the subsequent search more efficient. LLL and its variants apply a
sequence of unimodular column operations to $R$ in a way that preserves
the upper-triangular structure. Unimodular transformations transforms
integer vectors to integer vectors, so the search can be applied to
the transformed $RG$, and the minimizing integer vector $G^{-1}\hat{s}$
can be transformed back to an integer minimizer $\hat{s}=GG^{-1}\hat{s}$
of~(\ref{eq:square-upper-discrete-ls}).

In the single-epoch case, we apply the Schnorr-Euchner search to sets
of shifted squares, so we need $G$ to preserve this property. Unimodular
matrices do not necessarily preserve it. Therefore, we utilize a more
restricted class of matrices, namely permutation matrices.

We explore two strategies for generating the permutation. Neither
of them is new. The simpler strategy incorporates greedy column pivoting
into the $QR$ factorization that produces $R$, aiming to populate
the diagonal of $RG$ with large entries near the bottom. The pivoting
selects the column with the smallest norm to be eliminated next. This
is the exact opposite of the policy of column pivoting in numerical
libraries, which has a different aim, to identify rank deficiency.

We also explore a more sophisticated strategy called V-BLAST~\parencite{doi:10.1109/TIT.2003.817444,doi:10.1016/j.apnum.2007.01.008}.
Initially, algorithms to produce this permutation were asymptotically
about as expensive as LLL~\parencite{doi:10.1109/TIT.2003.817444}, but
variants that are asymptotically as efficient as the standard $QR$
factorization have been developed~\parencite{doi:10.1016/j.apnum.2007.01.008}.
The idea behind these algorithms is to greedily select columns from
last to first, always selecting next the column that maximizes the
next diagonal value in $RG$.

We report below on the efficacy of these optimizations.

\section{\label{sec:geometric-arrangements}Polynomial-Time Algorithms Using
Geometric Arrangements}

This section describes geometric algorithms to find the minimizer
of~(\ref{eq:nonlinear-generalized-ls}) that run in polynomial time.
We focus on single-epoch problems for simplicity. 

\subsection{Introduction to Arrangements}

The algorithms that we present rely on a data structure called a \emph{geometric
arrangement}~\parencite{HalperinSharir2017arrangements} to decompose
the 2- or 3-dimensional Euclidean space into cells, faces, and vertices.
Consider a set of $m$ error-free carrier-phase constraints
\[
\lambda\varphi_{i}+\lambda\mathring{n}_{i}=\left\Vert \mathring{\ell}-\rho_{i}\right\Vert _{2}
\]
where $\varphi_{i}\in[0,1)$. The target lies outside or on the circle
(or sphere in 3D) defined by all points $\ell$ satisfying
\[
\lambda\mathring{n}_{i}=\left\Vert \ell-\rho_{i}\right\Vert _{2}
\]
and strictly inside the circle (sphere)
\[
\lambda\left(\mathring{n}_{i}+1\right)=\left\Vert \ell-\rho_{i}\right\Vert _{2}\;.
\]
These two circles (spheres) already induce an arrangement that partitions
the space into three cells and two faces or edges that separate them.
The cells are the disk (or ball in 3D) $\left\Vert \ell-\rho_{i}\right\Vert _{2}<\lambda\mathring{n}_{i}$,
the annulus $\lambda\mathring{n}_{i}<\left\Vert \ell-\rho_{i}\right\Vert _{2}<\lambda(\mathring{n}_{i}+1)$,
and the unbounded region $\left\Vert \ell-\rho_{i}\right\Vert _{2}>\lambda(\mathring{n}_{i}+1)$.
The faces or edges are the two circles or spheres. Since the two circles/spheres
do not intersect, there are no vertices in this arrangement. 

Since we do not know the value of $\mathring{n}_{i}$, algorithmically
it makes sense to construct an arrangement induced by all the plausible
circles/spheres whose center is $\rho_{i}$. We can derive a set of
plausible $n_{i}$s from the range constraint~(\ref{eq:range-observations})
\[
r_{i}=\left\Vert \mathring{\ell}-\rho_{i}\right\Vert _{2}+\epsilon_{i}\;.
\]
We define
\begin{eqnarray*}
\overline{n}_{i} & = & \left\lceil \frac{\left\Vert \mathring{\ell}-\rho_{i}\right\Vert _{2}+c\sigma\left(\epsilon_{i}\right)}{\lambda}\right\rceil \approx\left\lceil \frac{r_{i}+c\sigma\left(\epsilon_{i}\right)}{\lambda}\right\rceil \\
\underline{n}{}_{i} & = & \left\lfloor \frac{\left\Vert \mathring{\ell}-\rho_{i}\right\Vert _{2}-c\sigma\left(\epsilon_{i}\right)}{\lambda}\right\rfloor \approx\left\lfloor \frac{r_{i}-c\sigma\left(\epsilon_{i}\right)}{\lambda}\right\rfloor 
\end{eqnarray*}
for some appropriate constant $c$, say $6$, such that $\lambda\underline{n}{}_{i}<\|\mathring{\ell}-\rho_{i}\|<\lambda\overline{n}_{i}$
with high probability. This arrangement has $E_{i}=\overline{n}_{i}-\underline{n}{}_{i}$
annuli cells. The target lies, with high probability, in one of them
or on the circles that separate them.

Now consider another constraint specifying to carrier phase with respect
to another reference $i'$. It will define $E_{i'}=\overline{n}_{i'}-\underline{n}{}_{i'}$
additional annuli defined by $\overline{n}_{i'}-\underline{n}{}_{i'}+1$
circles. Each cell in the resulting arrangement corresponds to a particular
value of $n_{i}$ and a particular value of $n_{i'}$.

With $m$ pairs of range and carrier-phase constraints, we have at
most $mE$ circles/spheres that define the annuli, where $E=\max_{i}E_{i}$.
How many cells can we have in the arrangement? It turns out that the
number is $O((mE)^{2})$ in 2D and $O((mE)^{3})$ in 3D and that the
arrangement data structure can be constructed and its cells, faces,
and vertices enumerated in time and space $O((mE)^{2})$ in 2D and
$O((mE)^{3})$ in 3D. We need to qualify the statement that arrangements
can be constructed efficiently. Algorithms that correctly enumerate
the cells of arrangements of circles and spheres are complex~\parencite{HalperinSharir2017arrangements}.
Currently, there is an implementation of an algorithm that enumerates
arrangements of circles~\parencite{doi:10.1007/978-3-642-17283-0} but
not of spheres, even though the algorithm is known.

Each cell of the arrangement corresponds to a particular integer value
for each unknown $n_{i}$. We refer to this integer vector as the
\emph{label} of the cell. Therefore, we can enumerate the cells, determine
the label $n$ of each one, and solve the real nonlinear least squares
problem 
\[
\hat{\ell}(n)=\arg\min_{\ell}\left\Vert W\left(M\left(\ell\right)-\begin{bmatrix}0\\
\lambda I
\end{bmatrix}n-\begin{bmatrix}r_{i}\\
\lambda\varphi_{i}
\end{bmatrix}\right)\right\Vert _{2}\;.
\]
The assignment with the smallest residual defines the global solution
to (\ref{eq:nonlinear-generalized-ls}), 
\[
\hat{\ell},\hat{n}=\arg\min_{\ell,n}\left\Vert W\left(M\left(\ell\right)-\begin{bmatrix}0\\
\lambda I
\end{bmatrix}n-\begin{bmatrix}r_{i}\\
\lambda\varphi_{i}
\end{bmatrix}\right)\right\Vert _{2}\;.
\]
(The cells in an arragement are traditionally considered to be open
sets, faces are semi-open, and vertices are closed; the assumption
$\varphi_{i}\in[0,1)$ essentially unifies each face and vertex with
a particular cell, so that these unified cells form a partition of
the admissible space.) 
\begin{algorithm}
Construct a geometric arrangement of spheres (circles) induced by
carrier-phase constraints

Enumerate the labels of the cells, faces, and vertices

For each integer-vector label $n$, solve
\[
\hat{\ell}(n)=\arg\min_{\ell}\left\Vert W\left(M\left(\ell\right)-\begin{bmatrix}0\\
\lambda I
\end{bmatrix}n-\begin{bmatrix}r_{i}\\
\lambda\varphi_{i}
\end{bmatrix}\right)\right\Vert _{2}
\]
and record the norm of the weighted residual.

Output the solution with the minimal weighted residual.

\caption{\label{alg:brute-force-frameworks}A framework for geometric solvers
of our generalized least-squares problems.}
\end{algorithm}
The surprising aspect of this approach is that there are only $O((mE)^{2})$
different integer assignments of $n$ that are consistent with hypothetical
target locations $\ell$, even though each $n_{i}$ can assume $E_{i}$
different values. Enumerating all the possible assignments of all
the $n_{i}$s would have generated $\Pi_{i=0}^{m}E_{i}\approx E^{m}$,
which is exponential in $m$. The arrangement of $mE$ circles shows
that there are only $O((mE)^{2})$ geometrically-consistent assignments
of $m$ and it allows us to enumerate them efficiently.

In 3 dimensions, an arrangement of at most $mE$ spheres define the
geometrically-consistent assignments of $n$. The complexity of the
arrangement and the time and space to construct is are $O((mE)^{3})$. 

So far, we have ignored an important aspect of our problem: our carrier-phase
constraints are not exact but contain small additive errors $\delta_{i}$.
The next section shows that we must construct the arrangement in a
way that tolerates these observation errors. We then describe several
ways to construct arrangements that allow us to enumerate all the
geometrically-consistent assignments of $n$ in the presence of small
errors in the carrier-phase constraints.

\subsection{Observation Errors and Arrangements of Spheres with Integer Radii }

\begin{figure}
\begin{centering}
\begin{tikzpicture}[scale=1]
    \useasboundingbox (0,0) rectangle (12,6);
    
    \begin{scope}
        \clip (0,0) rectangle (12,6);
    \end{scope}
        
    \begin{scope}
        \clip (0,0) rectangle (12,6);
        \draw[line width=2pt, cyan, opacity=0.33] (9.5,-5.5) circle[radius=13];
        \draw[line width=2pt, cyan, opacity=0.33] (9.5,-5.5) circle[radius=11];
        \draw[line width=2pt, cyan, opacity=0.33] (9.5,-5.5) circle[radius=9];
        \draw[line width=2pt, cyan, opacity=0.33] (9.5,-5.5) circle[radius=7];
 
        \draw[line width=2pt, red, opacity=0.33] (0,-5.5) circle[radius=13];
        \draw[line width=2pt, red, opacity=0.33] (0,-5.5) circle[radius=11];
        \draw[line width=2pt, red, opacity=0.33] (0,-5.5) circle[radius=9];
        \draw[line width=2pt, red, opacity=0.33] (0,-5.5) circle[radius=7];
 
        \draw[line width=2pt, green, opacity=0.33] (12.75,-3.5) circle[radius=13];
        \draw[line width=2pt, green, opacity=0.33] (12.75,-3.5) circle[radius=11];
        \draw[line width=2pt, green, opacity=0.33] (12.75,-3.5) circle[radius=9];
        \draw[line width=2pt, green, opacity=0.33] (12.75,-3.5) circle[radius=7];
    \end{scope}

    \begin{scope}
        \clip (0,0) rectangle (12,6);
    \end{scope}

    \fill[color=lightgray] (5,4) circle (3pt);
    \node[below] at (5,4) {$\mathring{\ell}$};
    \node[above] at (8.75,5.4) {$\| \ell -\rho_1 \|=11\lambda$};
    \node[above] at (1.25,5.4) {$\| \ell -\rho_2 \|=11\lambda$};
    \node[above] at (9.75,4.0) {$\| \ell -\rho_3 \|=10\lambda$};

\end{tikzpicture}
\par\end{centering}
\caption{\label{fig:integer-radii-inconsistency}An illustration that demonstrates
that the minimizing set of integers might not be the label of any
cell in the arrangement. Circles with center $\rho_{1}$ are depicted
in cyan, those with center $\rho_{2}$ in red, and those with $\rho_{3}$
in green. Suppose that $\delta_{2}\protect\leq0$ and $\delta_{3}\protect\leq0$,
so that $\lambda\varphi_{2}+\lambda\cdot10=\left\Vert \mathring{\ell}-\rho_{2}\right\Vert +\delta_{2}$
and $\lambda\varphi_{3}+\lambda\cdot10=\left\Vert \mathring{\ell}-\rho_{3}\right\Vert +\delta_{3}$
with $\varphi_{2}$ and $\varphi_{3}$ both close to 1, but that $\delta_{1}>0$
is sufficiently large so that $\lambda\varphi_{1}+\lambda\cdot11=\left\Vert \mathring{\ell}-\rho_{1}\right\Vert +\delta_{1}$
with $\varphi_{1}$ is close to 0. The label of the cell containing
$\mathring{\ell}$ is $\begin{bmatrix}10 & 10 & 10\end{bmatrix}$
but the minimizing integer vector is $\begin{bmatrix}11 & 10 & 10\end{bmatrix}$,
which is not the label of any cell.}
\end{figure}
Consider an arrangement of circles of the form
\[
\lambda n_{i}=\left\Vert \ell-\rho_{i}\right\Vert _{2}\;,
\]
which we refer to as an arrangement of spheres or circles with integer
radii. Suppose that the target $\mathring{\ell}$, which lies in a
cell labeled $\mathring{n}$, lies very close to its $i$th outer
sphere $\mathring{n}_{i}+1$ and for which $\delta_{i}>0$ is small
but larger than the distance between $\mathring{\ell}$ and that sphere,
as shown in Figure~\ref{fig:integer-radii-inconsistency}. We have
\[
\lambda\varphi_{i}+\lambda\left(\mathring{n}_{i}+1\right)=\left\Vert \mathring{\ell}-\rho_{i}\right\Vert _{2}+\delta_{i}
\]
for some small positive $\varphi_{i}$. There might not be any cell
with label 
\[
\{\mathring{n}_{1},\ldots,\mathring{n}_{i-1},\mathring{n}_{1}+1,\mathring{n}_{i+1},\ldots,\mathring{n}_{m}\}\;,
\]
as shown in the figure, and the residual
\[
\lambda\varphi_{i}+\lambda\mathring{n}_{i}-\left\Vert \mathring{\ell}-\rho_{i}\right\Vert _{2}
\]
will be large, around $\lambda$, much larger than $\sigma(\delta_{i})$.
This shows that this particular construction cannot always handle
observation constraints, even when they are guaranteed to be much
smaller than $\lambda$. Note that when $\|\mathring{\ell}-\rho_{i}\|$
is close to a multiple of $\lambda$, this failure can occur with
a significant probability. The next constructions that we propose
avoid this failure mode.

\subsection{\label{subsec:intersections-of-annuli}Intersections of Families
of Annuli}

The constraint 
\[
\lambda\varphi_{i}+\lambda\mathring{n}_{i}=\left\Vert \mathring{\ell}-\rho_{i}\right\Vert _{2}+\delta_{i}
\]
together with assumed bounds $|\delta_{i}|\leq D_{i}$ constrains
$\mathring{\ell}$ to a family of concentric annuli (that is, to the
union of the annuli in the family)
\[
\lambda\varphi_{i}+\lambda n_{i}-D_{i}\leq\left\Vert \ell-\rho_{i}\right\Vert _{2}\leq\lambda\varphi_{i}+\lambda n_{i}+D_{i}\;,
\]
where $\underline{n}_{i}\leq n_{i}\leq\overline{n}_{i}$. We can derive
a bound $D_{i}$ that holds with high probability from the known distribution
of $\delta_{i}$, say $D_{i}=6\sigma(\delta_{i})$.

The target $\mathring{\ell}$ lies not only in one of the annuli whose
center is $\rho_{i}$, but also in one annuli in each of the other
$m-1$ families induced by the other $m-1$ carrier-phase constraints.,
as shown in Figure~\ref{fig:annuli}

Importantly, assuming also that $D_{i}<\lambda/2$, the annuli in
each family are disjoint.

\begin{figure}
\begin{centering}
\begin{tikzpicture}[scale=1]
    \useasboundingbox (0,0) rectangle (12,6);
    
    \begin{scope}
        \clip (0,0) rectangle (12,6);
    \end{scope}
    
    \def\centerX{9.5}
    \def\centerY{-5.5}
    
    \begin{scope}
        \clip (0,0) rectangle (12,6);
        \draw[line width=12pt, cyan, opacity=0.33] (9.5,-5.5) circle[radius=11];
        \draw[line width=12pt, cyan, opacity=0.33] (9.5,-5.5) circle[radius=8];
        \draw[line width=12pt, cyan, opacity=0.33] (9.5,-5.5) circle[radius=5];
        \draw[line width=12pt, cyan, opacity=0.33] (9.5,-5.5) circle[radius=14];
 
        \draw[line width=12pt, red, opacity=0.33] (0,-5.5) circle[radius=11];
        \draw[line width=12pt, red, opacity=0.33] (0,-5.5) circle[radius=8];
        \draw[line width=12pt, red, opacity=0.33] (0,-5.5) circle[radius=5];
        \draw[line width=12pt, red, opacity=0.33] (0,-5.5) circle[radius=14];
 
        \draw[line width=12pt, green, opacity=0.33] (12.35,-3.5) circle[radius=11];
        \draw[line width=12pt, green, opacity=0.33] (12.35,-3.5) circle[radius=8];
        \draw[line width=12pt, green, opacity=0.33] (12.35,-3.5) circle[radius=5];
        \draw[line width=12pt, green, opacity=0.33] (12.35,-3.5) circle[radius=14];
    \end{scope}
        
    \fill[lightgray] (4.75,4.45) circle (3pt);
    \node[above] at (4.75,4.45) {$\mathring{\ell}$};
    \draw[line width=1pt, ->] (6.5,2.7) -- (7.3,2.7);

    \begin{scope}
        \clip (0,0) rectangle (12,6);
    \end{scope}

\end{tikzpicture}
\par\end{centering}
\caption{\label{fig:annuli}An illustration of 3 families of annuli, depicted
in red, green, and cyan. The arrangement has only one cell that is
an intersection of one annulus from each family, and in this case
it indeed contains the target $\mathring{\ell}$, represented by a
gray dot. The arrangement also contains one almost-intersection, indicated
by the arrow.}
\end{figure}
We can exploit these insights by building an arrangement of up the
$2mE$ spheres
\begin{eqnarray*}
\left\Vert \ell-\rho_{i}\right\Vert _{2} & = & \lambda\varphi_{i}+\lambda n_{i}+D_{i}\\
\left\Vert \ell-\rho_{i}\right\Vert _{2} & = & \lambda\varphi_{i}+\lambda n_{i}-D_{i}
\end{eqnarray*}
for $\underline{n}_{i}\leq n_{i}\leq\overline{n}_{i}$. We then enumerate
the cells of the arrangement. For each cell, we determine whether
it is part of an annulus of each family, or else lies in the space
between two annuli associated with some $n_{i,j}$ and $n_{i,j}+1$.
If the cell is an intersection of $m$ annuli, we find the minimizer
$\hat{\ell}(n)$ where $n$ is the label of the cell. When the enumeration
ends, we output the minimizer with the smallest residual.

\subsection{\label{subsec:Near-Intersections-of-Annuli}Near Intersections of
Families of Annuli}

\begin{figure}
\begin{centering}
\begin{tikzpicture}[scale=1]
    \useasboundingbox (2,1) rectangle (9,7);
    
    
    \def\centerX{9.5}
    \def\centerY{-5.5}
    
    \begin{scope}
        \clip (2,1) rectangle (9,7);
        \draw[line width=12pt, cyan, opacity=0.33] (9.5,-5.5) circle[radius=14];
        \draw[line width=12pt, cyan, opacity=0.33] (9.5,-5.5) circle[radius=11];
        \draw[line width=12pt, cyan, opacity=0.33] (9.5,-5.5) circle[radius=8];
        \draw[line width=12pt, cyan, opacity=0.33] (9.5,-5.5) circle[radius=5];
        \draw[line width=1pt, cyan] (9.5,-5.5) circle[radius=11];
        \draw[line width=1pt, cyan] (9.5,-5.5) circle[radius=11];
        \draw[line width=1pt, cyan] (9.5,-5.5) circle[radius=8];
        \draw[line width=1pt, cyan] (9.5,-5.5) circle[radius=5];
        \draw[line width=1pt, cyan] (9.5,-5.5) circle[radius=14];
 
        \draw[line width=12pt, red, opacity=0.33] (0,-5.5) circle[radius=11];
        \draw[name path=red11, line width=1pt, red] (0,-5.5) circle[radius=11];
        \draw[line width=12pt, red, opacity=0.33] (0,-5.5) circle[radius=8];
        \draw[name path=red8, line width=1pt, red] (0,-5.5) circle[radius=8];
        \draw[line width=12pt, red, opacity=0.33] (0,-5.5) circle[radius=14];
        \draw[name path=red14, line width=1pt, red] (0,-5.5) circle[radius=14];
 
        \draw[line width=12pt, green, opacity=0.33] (12.35,-3.5) circle[radius=11];
        \draw[name path=green11, line width=1pt, green] (12.35,-3.5) circle[radius=11];

    \path[name intersections={of=green11 and red14, by={E,F}}];
    \path[name intersections={of=green11 and red11, by={A,B}}];
    \path[name intersections={of=green11 and red8, by={C,D}}];

    \draw[black, line width=0.5pt] (A) circle[radius=0.39];
    \draw[black, line width=0.5pt] (B) circle[radius=0.39];
 
    \draw[black, line width=0.5pt] (C) circle[radius=0.31];
    \draw[black, line width=0.5pt] (D) circle[radius=0.31];
 
    \draw[black, line width=0.5pt] (E) circle[radius=0.45];
    \draw[black, line width=0.5pt] (F) circle[radius=0.45];
    \end{scope}
        


\end{tikzpicture}
\par\end{centering}
\caption{\label{fig:annuli-approximation}The approximation to an arrangement
of annuli. The circle from the green family, centered at $\rho_{3}$,
intersects three circles from the red family, centered at $\rho_{2}$.
The blue family disqualifies the bottom-left and top-right intersections,
because no annulus from the blue family intersects the disks centered
at the intersections. Note that the closer the intersection angle
to $\pi/2$, the smaller the covering disk.}
\end{figure}
We now present an effective approximation of the algorithm from Section~\ref{subsec:intersections-of-annuli}
that is useful in the typical case in which $D_{i}\ll\lambda$. 

We start with just two or three families (in 2D or 3D, respectively).
Let us consider the 2D case for concreteness, denoting the two selected
families by $i'$ and $i''$. We compute all the intersections of
a circle $\lambda\varphi_{i'}+\lambda n_{i',j'}=\left\Vert \mathring{\ell}-\rho_{i'}\right\Vert _{2}$
and a circle $\lambda\varphi_{i''}+\lambda n_{i'',j''}=\left\Vert \mathring{\ell}-\rho_{i''}\right\Vert _{2}$,
for all admissible integers $n_{i',j'}$ and $n_{i'',j''}$ (intersections
of three spheres in 3D). There are $O(E^{2})$ such points, $O(E^{3}$)
in 3D. Each such point $p$ is in the intersection of an annulus from
family $i'$ and an annulus from $i''$. We now determine whether
the intersection of these two annuli contains a cell that is the intersection
of one annulus from each of the $m$ families. Our rule to determine
this is approximate, but only produces false positives. That is, when
our rule indicates that intersection of the two annuli does not intersects
all the other annuli, that determination is always correct. When our
rule indicates that the intersection also intersects all the other
annuli, that might be false in rare cases.

The intersection of two annuli in the plane is not necessarily a shape
that resembles a generalized parallelogram, so we begin with an easy-to-check
admissibility criterion.
\begin{defn}
\begin{figure}
\begin{centering}
\begin{tikzpicture}[scale=0.45]
    \coordinate (CL) at (2.5, 5.5);
    \coordinate (CR) at (8, 5.5);
    \def\rLL{5}
    \def\rLS{3}
    \def\rRL{4.5}
    \def\rRS{2.75}

    \begin{scope}
        \clip (-3,0.45) rectangle (16,11.05);

\begin{scope}
  \clip (CL) circle (\rLL);
  \clip (CR) circle (\rRL);
  \fill[green!50] (-3,0.45) rectangle (16,11.05);
  \fill[white] (CL) circle (\rLS);
  \fill[white] (CR) circle (\rRS);
\end{scope}
    \draw[line width=1pt] (CL) circle (\rLL);

    \draw[line width=1pt] (CL) circle (\rLS);

    \fill[black, opacity=0.33] (CL) circle (3pt);
    \node[below] at (CL) {$\rho_{i'}$};

    \draw[line width=1pt] (CR) circle (\rRL);

    \draw[line width=1pt] (CR) circle (\rRS);

    \fill[black, opacity=0.33] (CR) circle (3pt);
    \node[below] at (CR) {$\rho_i$};

    \draw[line width=0.4pt] (-3,5.5) -- (13,5.5);
\end{scope}
\end{tikzpicture}
\hfill{}\begin{tikzpicture}[scale=0.45]
    \coordinate (CL) at (2.5, 5.5);
    \coordinate (CR) at (8, 5.5);
    \def\rLL{8.00}
    \def\rLS{3.00}
    \def\rRL{4.5}
    \def\rRS{2.75}

    \begin{scope}
        \clip (-3,0.45) rectangle (16,11.05);

\begin{scope}
  \clip (CL) circle (\rLL);
  \clip (CR) circle (\rRL);
  \fill[green!50] (-3,0.45) rectangle (16,11.05);
  \fill[white] (CL) circle (\rLS);
  \fill[white] (CR) circle (\rRS);
\end{scope}
    \draw[line width=1pt] (CL) circle (\rLL);

    \draw[line width=1pt] (CL) circle (\rLS);

    \fill[black, opacity=0.33] (CL) circle (3pt);
    \node[below] at (CL) {$\rho_{i'}$};

    \draw[line width=1pt] (CR) circle (\rRL);

    \draw[line width=1pt] (CR) circle (\rRS);

    \fill[black, opacity=0.33] (CR) circle (3pt);
    \node[below] at (CR) {$\rho_i$};

    \draw[line width=0.4pt] (-3,5.5) -- (13,5.5);
\end{scope}
\end{tikzpicture}
\\
~\\
~\\
\begin{tikzpicture}[scale=0.45]
    \coordinate (CL) at (2.5, 5.5);
    \coordinate (CR) at (8, 5.5);
    \def\rLL{8.25}
    \def\rLS{2.75}
    \def\rRL{4.5}
    \def\rRS{2.75}

    \begin{scope}
        \clip (-3,0.45) rectangle (16,11.05);

\begin{scope}
  \clip (CL) circle (\rLL);
  \clip (CR) circle (\rRL);
  \fill[red!50] (-3,0.45) rectangle (16,11.05);
  \fill[white] (CL) circle (\rLS);
  \fill[white] (CR) circle (\rRS);
\end{scope}
    \draw[line width=1pt] (CL) circle (\rLL);

    \draw[line width=1pt] (CL) circle (\rLS);

    \fill[black, opacity=0.33] (CL) circle (3pt);
    \node[below] at (CL) {$\rho_{i'}$};

    \draw[line width=1pt] (CR) circle (\rRL);

    \draw[line width=1pt] (CR) circle (\rRS);

    \fill[black, opacity=0.33] (CR) circle (3pt);
    \node[below] at (CR) {$\rho_i$};

    \draw[line width=0.4pt] (-3,5.5) -- (13,5.5);
\end{scope}
\end{tikzpicture}
\hfill{}\begin{tikzpicture}[scale=0.45]
    \coordinate (CL) at (2.5, 5.5);
    \coordinate (CR) at (8, 5.5);
    \def\rLL{8.5}
    \def\rLS{2.5}
    \def\rRL{4.5}
    \def\rRS{2.75}

    \begin{scope}
        \clip (-3,0.45) rectangle (16,11.05);

\begin{scope}
  \clip (CL) circle (\rLL);
  \clip (CR) circle (\rRL);
  \fill[red!50] (-3,0.45) rectangle (16,11.05);
  \fill[white] (CL) circle (\rLS);
  \fill[white] (CR) circle (\rRS);
\end{scope}
    \draw[line width=1pt] (CL) circle (\rLL);

    \draw[line width=1pt] (CL) circle (\rLS);

    \fill[black, opacity=0.33] (CL) circle (3pt);
    \node[below] at (CL) {$\rho_{i'}$};

    \draw[line width=1pt] (CR) circle (\rRL);

    \draw[line width=1pt] (CR) circle (\rRS);

    \fill[black, opacity=0.33] (CR) circle (3pt);
    \node[below] at (CR) {$\rho_i$};

    \draw[line width=0.4pt] (-3,5.5) -- (13,5.5);
\end{scope}
\end{tikzpicture}
\par\end{centering}
\caption{\label{fig:admissible-generalized-parallelograms}Diagrams that illustrate
the definition of generalized parallelograms. Top: the circles that
define two annuli intersect at 8 points so the intersection of annuli
forms two generalized parallelograms, shown in green. Bottom: the
circles intersect at only 6 or 4 points, so the intersections of the
two annuli, shown in red, are not considered generalized parallelograms.}
\end{figure}
The intersections of two annuli in the plane
\begin{eqnarray*}
\{\ell & | & r_{i}-D_{i}\leq\|\ell-\rho_{i}\|\leq r_{i}+D_{i}\}\\
\{\ell & | & r_{i'}-D_{i'}\leq\|\ell-\rho_{i'}\|\leq r_{i'}+D_{i'}\}
\end{eqnarray*}
are called \emph{generalized parallelograms} if the two pairs of concentric
circles bound them
\begin{eqnarray*}
\|\ell-\rho_{i}\| & = & r_{i}\pm D_{i}\\
\|\ell-\rho_{i'}\| & = & r_{i'}\pm D_{i'}
\end{eqnarray*}
intersect at exactly 8 points.
\end{defn}
Figure~\ref{fig:admissible-generalized-parallelograms} illustrates
two cases in which the intersections of two annuli are admissible
and form two generalized parallelograms (top), as well as inadmissible
cases in which the four circles intersect at only 6 points (bottom
left) or only 4 points (bottom right).
\begin{defn}
The \emph{center }of a generalized parallelogram defined by the four
circles 
\begin{eqnarray*}
\|\ell-\rho_{i}\| & = & r_{i}\pm D_{i}\\
\|\ell-\rho_{i'}\| & = & r_{i'}\pm D_{i'}
\end{eqnarray*}

is the point that satisfies the two equations 
\begin{eqnarray*}
\|\ell-\rho_{i}\| & = & r_{i}\\
\|\ell-\rho_{i'}\| & = & r_{i'}\;.
\end{eqnarray*}
\end{defn}
There are two points that satisfy the two equations, one in each generalized
parallelogram. It is easy to see (see \ref{fig:admissible-generalized-parallelograms})
that the distance between the center of a generalized parallelogram
and $\rho_{i}$ and $\rho_{i'}$ can be either greater or smaller
than the distance between the center and a corner of the generalized
parallelogram. (For the former case, imagine a tiny parallelogram
formed by two very thin annuli; for the latter, imagine a long and
thin parallelogram that is almost a half circle.)

Our algorithm works as follows. We compute the four or six corners
of the generalized parallelogram or parallelpiped and determine the
maximum distance $\delta_{i',i''}$ from its center to a corner. We
prove below that under very mild conditions, a ball of radius $\delta_{i',i''}$
centered at the center of the generalized parallelogram/parallelpiped
covers the parallelogram/parallelpiped, as shown in Figure~\ref{fig:annuli-approximation}.
We now iterate over all the other families, computing the nearest
annulus from each family to $p$. If that annulus does not intersect
the ball centered at $p$, we discard the intersection $p$ because
we know that there is no intersection of all $m$ annuli near $p$.
If one annulus from each family intersects the ball, we declare $p$
to be a \emph{near intersection} and we evaluate $\hat{\ell(n)}$
. This algorithm enumerates all the intersections of $m$ annuli,
and perhaps some spurious vectors $n$ that do not correspond to an
actual intersection cell. The complexity of generating the candidate
integer vectors is $O(mE^{2})$ or $O(mE^{3}$).

We now prove the claim that the disk or ball cover the generalized
parallelogram or parallelpiped. We start with a technical lemma that
shows that the central angles defined by sides of generalized parallelograms
are smaller than $\pi$.
\begin{lem}
\label{lem:central-angles-of-sides-of-generalized-parallelograms}The
central angles defined by sides of a generalized parallelogram are
smaller than $\pi$.
\end{lem}
\begin{proof}
The 8 intersections of the two pairs of circles are symmetric about
the line incident on the two centers (shows as a thin horizontal line
in Figure~\ref{fig:admissible-generalized-parallelograms}). We claim
that a generalized parallelogram must be contained in a half plane
defined by this line. 

The assumption that the 4 bounding circles intersect at 8 points implies
that the two open disks $\|\ell-\rho_{i}\|<r_{i}+D_{i}$ and $\|\ell-\rho_{i'}\|<r_{i'}+D_{i'}$
must intersect, which means that the line segment between $\rho_{i}$
and $\rho_{i'}$ is not in the intersection of annuli. 

The same assumption also implies that $r_{i}+D_{i}<\|\rho_{i}-\rho_{i'}\|+r_{i'}-D_{i'}$,
otherwise the circle $\|\ell-\rho_{i}\|=r_{i}+D_{i}$ does not intersect
the circle $\|\ell-\rho_{i'}\|<r_{i'}-D_{i'}$; similarly $r_{i'}+D_{i'}<\|\rho_{i}-\rho_{i'}\|+r_{i}-D_{i}$
(this case is illustrated in the bottom right diagram in Figure~\ref{fig:admissible-generalized-parallelograms}).
Therefore, the rays that emanate outward from $\rho_{i}$ and $\rho_{i'}$
along the line connecting them are also not in the intersection of
annuli. 

Therefore, the entire line connecting the two centers is outside the
intersection of the two annuli, which implies that the intersection
occupies disjoint areas in the two half planes defined by this line.
This proves the claim.
\end{proof}
We now state and prove a second technical lemma.
\begin{lem}
\label{lem:short-arc-is-in-O}Let $K$ be a closed disk. Let $s$
and $t$ be points on a circle $O$ with center $\rho\not\in K$.
Then if $s\in K$ and $t\in K$, then the shorter of the two arcs
$\overline{st}$ that form $O$ is also in $K$.
\end{lem}
\begin{proof}
Figure~\ref{fig:proof-of-lemma} (left) illustrates the notation
used in the proof. Suppose for contradiction that there is a point
$q'$ on the short arc $\overline{st}$ of $O$. Since $s,t$ are
in $K$ and $q'$ is not, $O$ and the circle $C$ that bounds $K$
must intersect at exactly two points, and these points must be on
this shorter arc of $O$. Therefore, the longer arc of $O$ bounded
by $s$ and $t$, is inside the disk, and so is the chord $s\leftrightarrow t$.
The shape bounded by this long arc and the cord is convex. Its boundary
is inside $K$ so its interior is also in $K$. This interior includes
the center $\rho$ of $O$. We conclude that $\rho$ is in $K$, which
contradicts our assumptions. 
\end{proof}
Next, we state and prove the theorem that guarantees that the disk
that our algorithm constructs covers the generalized parallelogram
in two dimensions. We provide two proofs. One is intuitive but requires
a very mild condition. The second is more technical but does not require
the additional condition. Note that both proofs state a slightly more
general claim, which allows $p$ to be any point in the generalized
parallelogram. 
\begin{thm}
\label{lem:generalized-parallelogram-in-a-circle}Let $A$ be a closed
generalized parallelogram  formed by the intersection of two annuli,
\begin{eqnarray*}
r_{i}-D_{i}\leq & \|\ell-\rho_{i}\| & \leq r_{i}+D_{i}\\
r_{i'}-D_{i'}\leq & \|\ell-\rho_{i'}\| & \leq r_{i'}+D_{i'}
\end{eqnarray*}
for some $0<D_{i},D_{i'}$. Let $p$ be some point in the interior
of $A$ and let $\bar{p}$ be the furthest corner of the generalized
parallelogram from $p$. That is, $\bar{p}$ lies on both $\|\ell-\rho_{i}\|=r_{i}\pm D_{i}$
and $\|\ell-\rho_{i'}\|=r_{i'}\pm D_{i'}$ and it maximizes $\|p-\bar{p}\|$.

Let $K$ be the closed disk $K=\{\ell|\|\ell-p\|\leq\|\bar{p}-p\|\}$
and let $C=\partial K$ be the circle that is the boundary of $K$.

The first proof of the theorem also requires that $\rho_{i}$ and
$\rho_{i'}$ are outside $K$.

The disk $K$ contains the generalized parallelogram $A$, 
\[
A\subseteq K\;.
\]
\end{thm}
\begin{proof}
\begin{figure}
\begin{centering}
\begin{tikzpicture}[scale=1]
    \useasboundingbox (0,0) rectangle (7,6);
    
    
    \def\centerX{9.5}
    \def\centerY{-5.5}
    
    \begin{scope}
        \clip (0,0) rectangle (7,6);

        \draw[->, white, line width=1pt] (3.5,0) -- (3.5,5.9);
        
        \draw[line width=1pt, green, name path=O] (3.0,3.5) circle[radius=2];
        \fill[black, opacity=0.33] (3.0,3.5)   circle (3pt);
        \node[left] at (3.0,3.5) {$\rho$};
        \coordinate (rho) at (3.0,3.5);
        
        \draw[line width=1pt, black, name path=C] (3.5,-12.5) circle[radius=17];

        \path[name intersections={of=O and C, by={x1,x2}}];

        \fill[red, opacity=0.33] (rho) ++(15:2)   circle (3pt) node[right, black, opacity=1] {$t$};
        \fill[red, opacity=0.33] (x2) circle (3pt) node[above left, black, opacity=1] {$s$};

        \draw[line width=1pt, red, opacity=0.33] (rho) ++(15:2) -- (x2) ;

        \fill[black, opacity=0.33] (3.5,5.45) circle (3pt);
        \node[right] at (3.5,5.45) {$q'$};
        \node[right] at (5.0,3.5) {$O$};
        \node at (6,4.6) {$C=\partial K$};

    \end{scope}
\end{tikzpicture}\hfill{}\begin{tikzpicture}[scale=1]
    \useasboundingbox (0,0) rectangle (7,6);
    
    
    \def\centerX{9.5}
    \def\centerY{-5.5}
    
    \begin{scope}
        \clip (0,0) rectangle (7,6);

        \draw[->, gray, line width=1pt] (3.5,0) -- (3.5,5.9);
        
        \draw[line width=1pt, green, name path=Oi] (3.0,3.5) circle[radius=2];
        \coordinate (rho_i) at (3.0,3.5);
        
        \draw[line width=1pt, black, name path=C] (3.5,-12.5) circle[radius=17];

        \path[name intersections={of=Oi and C, by={x1,x2}}];

        \fill[red, opacity=0.33] (rho_i) ++(15:2)   circle (3pt) node[right, black, opacity=1] {$t$};
        \fill[red, opacity=0.33] (x2) circle (3pt) node[above left, black, opacity=1] {$s$};


        \fill[black, opacity=0.33] (3.5,5)   circle (3pt);
        \node[right] at (3.5,5) {$q$};
        \fill[black, opacity=0.33] (3.5,5.45) circle (3pt);
        \node[right] at (3.5,5.45) {$q'$};
        \node[right] at (5.0,3.5) {$O_i$};
        \node at (6,4.6) {$C=\partial K$};
        \node[right] at (3.5,0.7) {ray emanating from $p$};

    \end{scope}
\end{tikzpicture}
\par\end{centering}
\caption{\label{fig:proof-of-lemma}Diagram to illustrate the proofs of Lemma~\ref{lem:short-arc-is-in-O}
(left) and Theorem~\ref{lem:generalized-parallelogram-in-a-circle}
(right).}
\end{figure}
Figure~\ref{fig:proof-of-lemma} (right) illustrates the notation
that is used in the lemma. Suppose for contradiction that there is
a point $q\in A$ that is not in the disk $K$. We first claim that
there must also be a point $q'\in A$, $q'\not\in K$ that lies on
one of the four circles $\|\ell-\rho_{i}\|=r_{i}\pm D_{i}$, $\|\ell-\rho_{i'}\|=r_{i'}\pm D_{i'}$.
If $q$ already lies on one of the four circles, there is nothing
to prove. Otherwise, consider the ray emanating from $p$ and passing
through $q$. The ray starts at $p$ (the center of $K$), which is
both in $A$ and in $K$. When the ray reaches $q$, it is in $A$
(by definition of $q$) but outside $K$. Points even further from
$q$ along the ray are all outside $K$. Since $A$ is bounded, the
part of the ray that starts at $q$ must intersect the boundary of
$A$. This intersection point satisfies all the conditions that define
$q'$, so we mark it as $q'$.

Suppose without loss of generality that $q'$ lies on the circle $O_{i}=\{\ell|\|\ell-\rho_{i}\|=r_{i}+D_{i}\}$
and that the corners of $A$ on $O_{i}$ are $s$ and $t$. By Lemma~\ref{lem:central-angles-of-sides-of-generalized-parallelograms},
the angle $\angle s\rho_{i}t$ is smaller than $\pi$. Therefore,
the point $q'$ lies on the shorter arc of $O_{i}$ bounded by $s$
and $t$. By Lemma~\ref{lem:short-arc-is-in-O}, $q'$ is in $K$,
which contradicts our assumptions.  
\end{proof}

Appendix~A presents an alternative proof that does not require the
hypothesis that $\rho_{i}$ and $\rho_{i'}$ are outside $K$. We
now prove the containment result in 3 dimensions. Our proof requires
an assumption on the central angles, as well as an assumption that
the centers $\rho_{i}$, $\rho_{i'}$, and $\rho_{i''}$ and the line
segments connecting them are outside the ball that the algorithm constructs.
\begin{thm}
\label{lem:generalized-parallelpiped-in-a-sphere}Let $A$ be a closed
generalized parallelepiped formed by the intersection of three annuli
(spherical shells) in 3D,
\begin{eqnarray*}
r_{i}-D_{i}\leq & \|\ell-\rho_{i}\| & \leq r_{i}+D_{i}\\
r_{i'}-D_{i'}\leq & \|\ell-\rho_{i'}\| & \leq r_{i'}+D_{i'}\\
r_{i''}-D_{i''}\leq & \|\ell-\rho_{i''}\| & \leq r_{i''}+D_{i''}
\end{eqnarray*}
for some $0<D_{i},D_{i'},D_{i''}$. Let $p$ be a point in the interior
of $A$ and let $\bar{p}$ be the furthest corner of the generalized
parallelepiped from $p$. That is, $\bar{p}$ lies on three spheres:
$\|\ell-\rho_{i}\|=r_{i}\pm D_{i}$, $\|\ell-\rho_{i'}\|=r_{i'}\pm D_{i'}$,
and $\|\ell-\rho_{i''}\|=r_{i''}\pm D_{i''}$ and it maximizes $\|p-\bar{p}\|$.

Let ${\cal K}$ be the closed ball ${\cal K}=\{\ell|\|\ell-p\|\leq\|\bar{p}-p\|\}$
and let ${\cal C}=\partial{\cal K}$ be the sphere that is the boundary
of ${\cal K}$.

Suppose further that $\rho_{i}$, $\rho_{i'}$ and $\rho_{i''}$ are
outside ${\cal K}$, that the 3 line segments that connect $\rho_{i}$,
$\rho_{i'}$ and $\rho_{i''}$ are also outside ${\cal K}$, and that
for any point $\rho$ on one of these segments and any two points
$s,t\in A$, the angle $\angle s\rho t$ is smaller than $\pi$. 

Then
\[
A\subseteq{\cal K}\;.
\]
\end{thm}
\begin{proof}
We prove the lemma in two steps, using reductions to Lemma~\ref{lem:generalized-parallelogram-in-a-circle}.
In the first step, we claim that the edges of the parallelepiped are
in ${\cal K}$. These edges are arcs of circles, where the circles
are the intersections of two spheres out of three, say $\|\ell-\rho_{i}\|=r_{i}+D_{i}$
and $\|\ell-\rho_{i'}\|=r_{i'}-D_{i'}$. 

Let $s$ and $t$ be the corners of the parallelepiped at the ends
of a particular edge, say the edge that belongs to the two spheres
$\|\ell-\rho_{i}\|=r_{i}+D_{i}$ and $\|\ell-\rho_{i'}\|=r_{i'}-D_{i'}$.
Consider the restriction of our problem to the plane $f$ that passes
through the circle $O$ defined by the intersection of these two spheres.
The center $\rho$ of $O$ lies on the line segment that connects
the centers of the two spheres whose intersection defines $O$, $\rho_{i}$
and $\rho_{i'}$. This line segment is perpendicular to $f$. By the
hypotheses of the lemma, the line segment is not in ${\cal K}$ and
therefore $\rho\not\in{\cal K}$. The restriction of ${\cal K}$ to
our plane is a disk $K={\cal K}\cap f$. By definition, $s,t\in{\cal K}\cap f$,
and by our hypotheses, $\angle s\rho t$ is smaller than $\pi$. By
Lemma~\ref{lem:short-arc-is-in-O}, the edge that connects $s$ and
$t$ is also in ${\cal K}\cap f$ and hence in ${\cal K}$. We have
established our first claim.

Next, we claim that the faces of the parallelepiped are in $K$. Let
$q'$ be a point on a face of the parallelepiped but not on one of
its edges (if it is on an edge, our first claim shows that $q'\in K$).
Without loss of generality assume that $q'$ lies on the face that
is part of the sphere $\|\ell-\rho_{i}\|=r_{i}+D_{i}$. We now consider
the plane $f'$ that passes through $\rho_{i}$, $q'$, and $p$.
The restriction $K\cap f'$ is again a circle whose center is $p\in K\cap f'$.
The intersection of sphere $\|\ell-\rho_{i}\|=r_{i}+D_{i}$ that contains
the face with $f'$ is also a circle whose center is $\rho_{i}\in K\cap f'$.
The intersection of the face with $f'$ is an arc of this circle.
The endpoints $s$ and $t$ of this arc are points that are part of
edges of the parallelepiped, which by the first claim is in $K$.
The angle $\angle s\rho_{i}t$ sia smaller than $\pi$, so Lemma~\ref{lem:short-arc-is-in-O}
guarantees that the arc is in $K$ and that, therefore, $q'$ is in
$K$. 

Finally, we claim that any point $q$ in the interior of the parallelepiped
must also be in $K$. Let $L$ be an arbitrary line through $q$,
and let $S$ denote the maximal line segment of $L\cap A$ that contains
$q$. By the arguments so far the endpoints of $S$ are in $K$, hence,
by convexity of $K$, $q$ is in $K$ as well. 
\end{proof}

\subsection{\label{subsec:arrangements-of-spheres-shifted-radii}Arrangements
of Spheres with Shifted Integer Radii}

Another way to construct an arrangement with a cell labeled with the
correct integer vector is using the families of spheres
\[
\lambda n_{i}+\lambda\varphi_{i}+\frac{\lambda}{2}=\left\Vert \ell-\rho_{i}\right\Vert _{2}
\]
for $\underline{n}{}_{i}-1\leq n_{i}\leq\overline{n}_{i}+1$. We refer
to the circle (or sphere)
\[
\lambda n_{i}+\lambda\varphi_{i}+\frac{\lambda}{2}=\left\Vert \ell-\rho_{i}\right\Vert _{2}
\]
as the $i$th \emph{outer} circle (sphere) and to 
\[
\lambda n_{i}+\lambda\varphi_{i}-\frac{\lambda}{2}=\lambda\left(n_{i}-1\right)+\lambda\varphi_{i}+\frac{\lambda}{2}=\left\Vert \ell-\rho_{i}\right\Vert _{2}
\]
as the $i$th \emph{inner} circle or sphere.

\begin{figure}
\begin{centering}
\begin{tikzpicture}[scale=1]
    \useasboundingbox (0,0) rectangle (12,6);
    
    \begin{scope}
        \clip (0,0) rectangle (12,6);
    \end{scope}
    
    \def\centerX{9.5}
    \def\centerY{-5.5}
    
    \begin{scope}
        \clip (0,0) rectangle (12,6);
        \draw[line width=2pt, cyan, opacity=0.33] (9.5,-5.5) circle[radius=11];
        \draw[line width=2pt, cyan, opacity=0.33] (9.5,-5.5) circle[radius=9];
        \draw[line width=1pt, cyan, opacity=0.33] (9.5,-5.5) circle[radius=7];
        \draw[line width=1pt, cyan, opacity=0.33] (9.5,-5.5) circle[radius=13];
    \end{scope}
    
        \fill[lightgray] (5,3.5) circle (0.85cm);
    
    \fill[gray] (5,3.5) circle (3pt);
    \node[above] at (5,3.5) {$\mathring{\ell}$};

    \begin{scope}
        \clip (0,0) rectangle (12,6);
        \draw[line width=0.4pt] (5,3.5) -- (9.5,-5.5);
    \end{scope}

\end{tikzpicture}
\par\end{centering}
\caption{The distance between the target and its inner and outer spheres, depicted
using thick lines, bounded from below in Lemma~\ref{lemma:distance-target-spheres}.
The distance from $\rho_{i}$ to $\mathring{\ell}$ is $\lambda\mathring{n}_{i}+\lambda\varphi_{i}$,
to within $\pm\max_{i}|\delta_{i}|$, and the radii of the inner and
outer spheres are $\lambda\mathring{n}_{i}-\lambda\varphi_{i}-\lambda/2$
and $\lambda\mathring{n}_{i}-\lambda\varphi_{i}-\lambda/2$.}

\end{figure}

\begin{lem}
\label{lemma:distance-target-spheres}Let $D\geq\max_{i}|\delta_{i}|$
and assume that $D<\lambda/2$, and let the target $\mathring{\ell}$
lie in the cell labeled $\mathring{n}$. Then the distance between
$\mathring{\ell}$ and the boundary of its cell is at least $\lambda/2-D>0$.
\end{lem}
\begin{proof}
The radius of the $i$th inner sphere is $\lambda\mathring{n}_{i}+\lambda\varphi_{i}-\lambda/2$
and that of the outer sphere is $\lambda\mathring{n}_{i}+\lambda\varphi_{i}+\lambda/2$.
The range to the target satisfies the equation 
\[
\left\Vert \mathring{\ell}-\rho_{i}\right\Vert _{2}=\lambda\mathring{n}_{i}+\lambda\varphi_{i}-\delta_{i}\;.
\]
Therefore, the ranges satisfy the inequalities
\begin{eqnarray*}
\left\Vert \mathring{\ell}-\rho_{i}\right\Vert _{2}-\left(\frac{\lambda}{2}-D\right) & = & \lambda\mathring{n}_{i}+\lambda\varphi_{i}-\delta_{i}-\left(\frac{\lambda}{2}-D\right)\\
 & \geq & \lambda\mathring{n}_{i}+\lambda\varphi_{i}-\frac{\lambda}{2}
\end{eqnarray*}
and
\begin{eqnarray*}
\left\Vert \mathring{\ell}-\rho_{i}\right\Vert _{2}+\left(\frac{\lambda}{2}-D\right) & = & \lambda\mathring{n}_{i}+\lambda\varphi_{i}-\delta_{i}+\left(\frac{\lambda}{2}-D\right)\\
 & \leq & \lambda\mathring{n}_{i}+\lambda\varphi_{i}+\frac{\lambda}{2}\;.
\end{eqnarray*}
Therefore, an open ball of radius $\lambda/2-D$ centered at $\mathring{\ell}$
is contained in the annulus that the $i$th inner and outer spheres
define. This implies that the ball is contained in the cell labeled
$\mathring{n}$.
\end{proof}
Once we construct this arrangement, we can enumerate its cells, provably
including the one labeled $\mathring{n}$, and minimize $\hat{\ell}(n)$
in each cell. (Correctness depends, of course, on the validity of
$|\delta_{i}|<\lambda/2$, which implies that there is a $D<\lambda/2$
such that $D\geq\max_{i}|\delta_{i}|$; note that the construction
does not depend on $D$.)

\subsection{\label{subsec:Comparison-of-the-Algorithms}Comparison of the Algorithms}

We presented in this section three algorithms that can enumerate,
with high probability with respect to the distribution of observation
errors, a set of integer vectors that contains $\mathring{n}$. Each
has advantages and disadvantages:
\begin{itemize}
\item The \emph{intersection of annuli} algorithm constructs the largest
arrangement, consisting of up to $2mE$ circles or spheres, so it
will need to enumerate the largest number of integer vectors (cells
in the arrangement). However, most of these cells are not intersections
of $m$ annuli, so the continuous nonlinear least-squares minimization
to determine $\hat{\ell}(n)$ needs to be carried out just a few times.
The theoretical complexity of this algorithm is $O((mE)^{2})$ in
2D and $O((mE)^{3})$ in 3D; it can be implemented fairly easily in
2D, but not currently in 3D (since there is no existing implementation
of the algorithm that constructs an arrangement of spheres).
\item The \emph{intersection of circles} or spheres algorithm approximates
the previous one. It is easy to implement in both 2D and 3D, its theoretical
complexity is lower, $O(mE^{2})$ or $O(mE^{3})$, but it might run
the continuous least-squares minimization on some cells that are not
actually intersections of $m$ annuli.
\item The \emph{arrangement of circles with shifted integer radii }algorithm
constructs a smaller arrangement of at most $mE$ circles. Its theoretical
asymptotic complexity is the same as that of the\emph{ intersection-of-annuli
}algorithm, $O((mE)^{2})$ or $O((mE)^{3})$, but the smaller arrangement
reduces the time and space required to construct it. It can be implemented
in 2D but not in 3D. It runs the nonlinear least-squares minimization
on all the cells that it enumerates; it has no way to disqualify most
of them like the first two methods.
\end{itemize}

\section{\label{sec:Simulations-and-Evaluation}Simulations and Evaluation }

We conducted extensive analyses of the algorithms on simulated data.
The simulations used the parameters shown in Section~\ref{sec:Typical-Parameters}
unless noted otherwise. We conducted simulations in both 2D and 3D.
Simulation data was generated in Python.

\subsection{\label{subsec:results-lambda-squaring-differencing}Performance of
Mixed-Discrete Least-Squares Minimization Algorithms}

We implemented the mixed-discrete least-squares minimization algorithms
described in Sections~\ref{sec:Assembling-Minimization-Problems},
\ref{sec:Quantifying-Error-Terms} and~\ref{sec:Generalized-Schnorr-Euchner}
in Python and MATLAB. The parts in MATLAB are mostly the integer or
discrete least-squares search algorithms. We use a software package
called MILES~\parencite{doi:10.1007/s10291-007-0063-y}, implemented in
MATLAB, to solve mixed-integer least-squares minimization problems.
In other words, the implementations of the LLL algorithm and the Schnorr-Euchner
search are the ones in MILES. We note that in early experiments, MATLAB
implemenations of these algorithms turned out to be much faster in
MATLAB than in Python, probably due to the fact that MATLAB uses a
just-in-time-compiler. 

\begin{figure}
\includegraphics[width=0.47\textwidth]{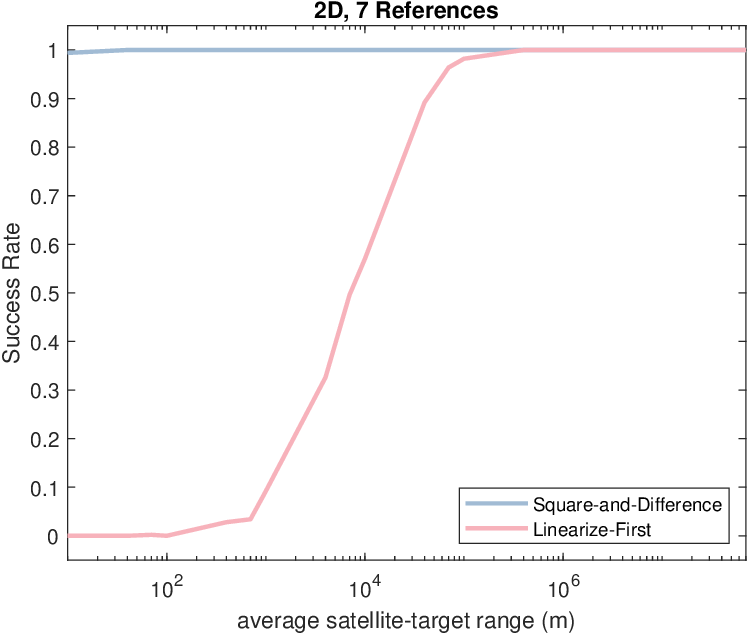}\hfill{}\includegraphics[width=0.47\textwidth]{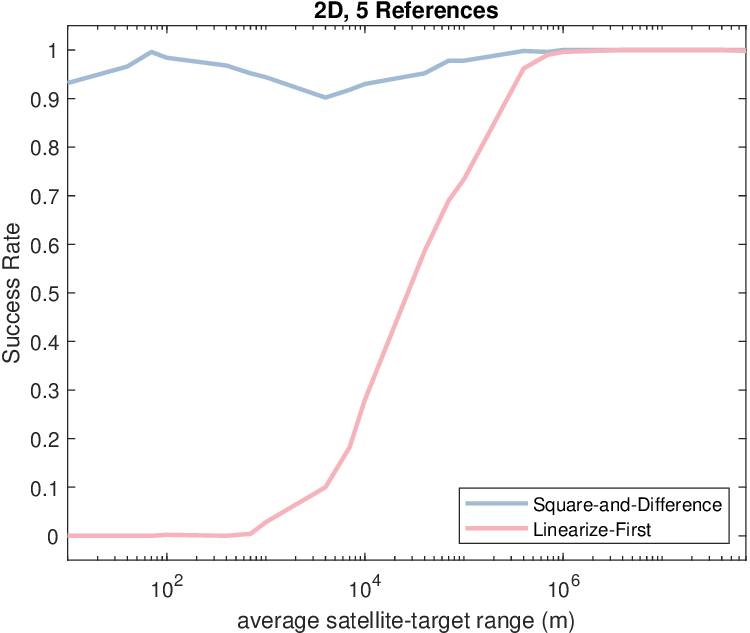}

\caption{\label{fig:-success-rates-2d}Success rates our new square-and-difference
algorithm in two dimensions, compared to that of the linearize-first
(LAMBDA) approach. The graphs plot the success rates for problems
with 7 and 5 references (left and right). }

\end{figure}
Figure~\ref{fig:-success-rates-2d} compares the success rates of
our new square-and-difference algorithm relative to those of the linearize-first
approach (i.e., LAMBDA) on problems in two dimensions with single-epoch
data (one observation of each reference). We define here success as
the correct recovery of all the integers, which leads to very accurate
localizations. 

The data shows that with 7 references, the square-and-difference algorithms
returns correct results down to very short ranges, 40~m. Even at
10~m the algorithm succeeds more than 99\% of the time. The linearization-first
approach, on the other hand, starts to fail with 7 references at ranges
of about 40~km, and fails badly at 10~km and below. With only 5
references, the square-and-difference algorithm succeeds about 90\%
of the time or more, but its performance is certainly worse than with
7 references. The results with 6 references are in between. The performance
of the linearization-first approach is also worse than with 7 references;
it fails at even longer ranges. We stress that failures in the linearization-first
approach are caused by the linearization; the linear mixed-integer
least-squares solver returns the optimal solution, but the optimal
solution of the linearized problem differs from that of the original
nonlinear problem.

\begin{figure}
\includegraphics[width=0.47\textwidth]{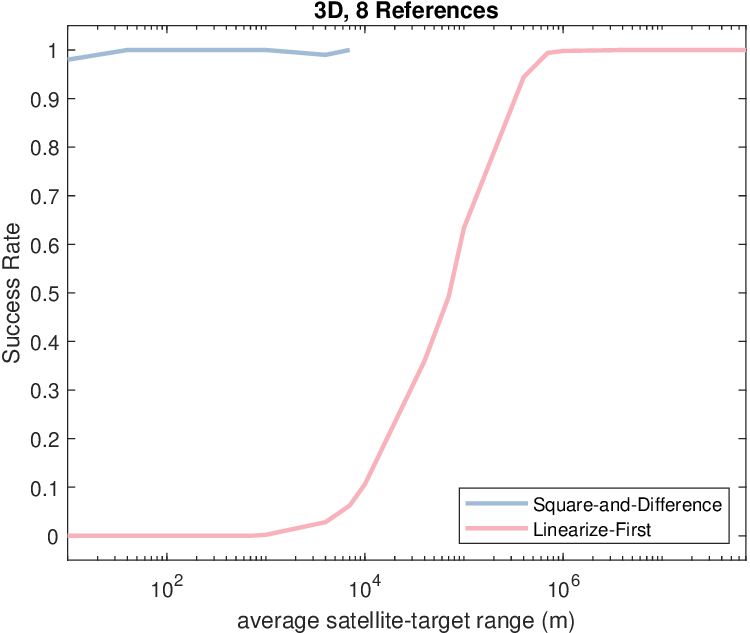}\hfill{}\includegraphics[width=0.47\textwidth]{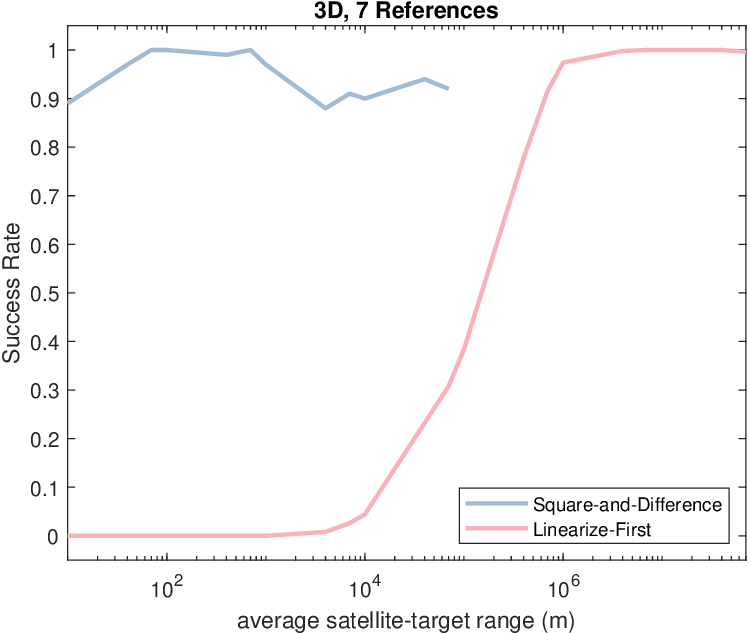}

\caption{\label{fig:-success-rates-3d}Success rates our new square-and-difference
algorithm in three dimensions, compared to that of the linearize-first
approach. The graphs plot the success rates for problems with 8 and
7 references (left and right).}
\end{figure}

The results in three dimensions are qualitatively similar, but more
references are required. Our algorithm gets quite slow in 3D at large
ranges, so we present its performance at ranges of up to 10~km. Figure~\ref{fig:-success-rates-3d}
shows that with 8 references, our algorithm almost always succeeds
in 3D at short ranges, whereas MILES fails completely at the same
ranges. With only 7 references, our algorithm succeeds 90\% of the
time or more, but its performance is clearly worse than with 8 references. 

\begin{figure}
\includegraphics[width=0.47\textwidth]{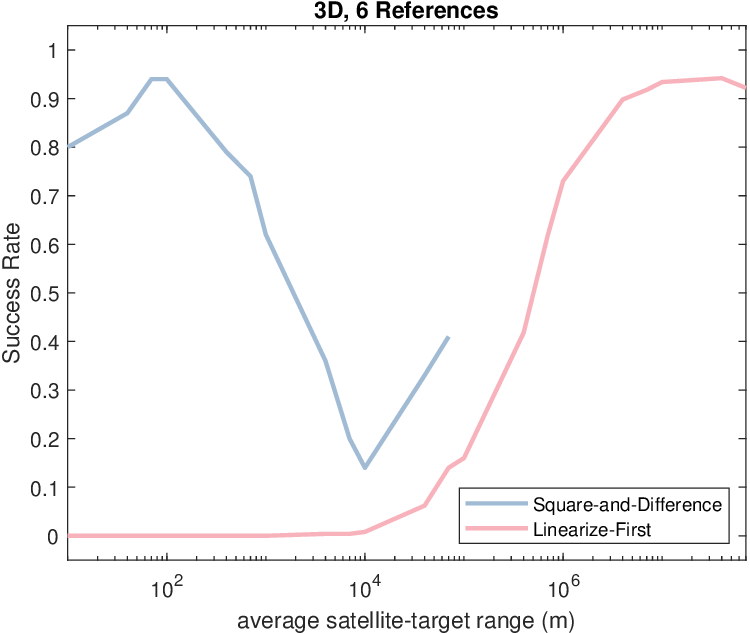}\hfill{}\includegraphics[width=0.47\textwidth]{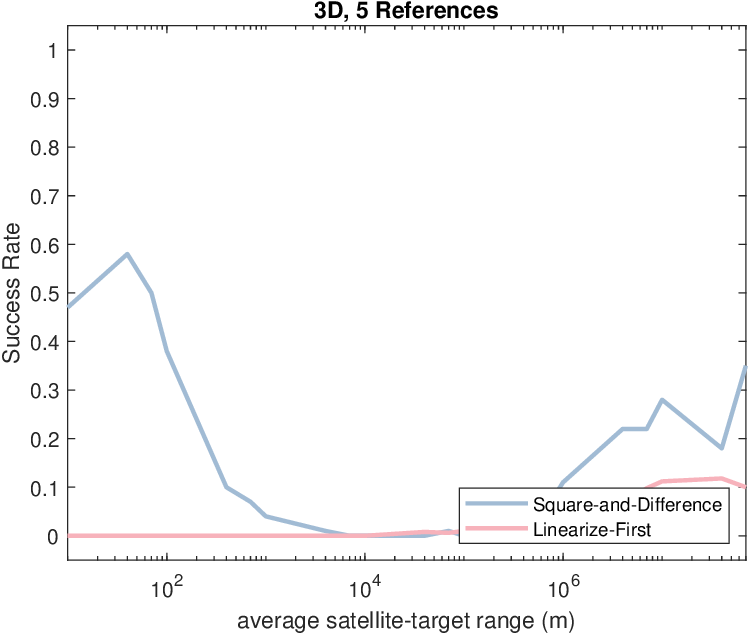}

\caption{\label{fig:-success-rates-too-few}Success rates with too few references.}
\end{figure}

When there are too few constraints, both algorithms fail, as shown
in Figure~\ref{fig:-success-rates-too-few}. The graphs in the figure
show that with 6 references both algorithms achieve a success rate
of about 80-90\% at either short ranges (our algorithm) or long ranges
(LAMBDA) but both fail quite badly at other ranges. With only 5 references,
both algorithms fail miserably.

\begin{figure}
\includegraphics[width=0.47\textwidth]{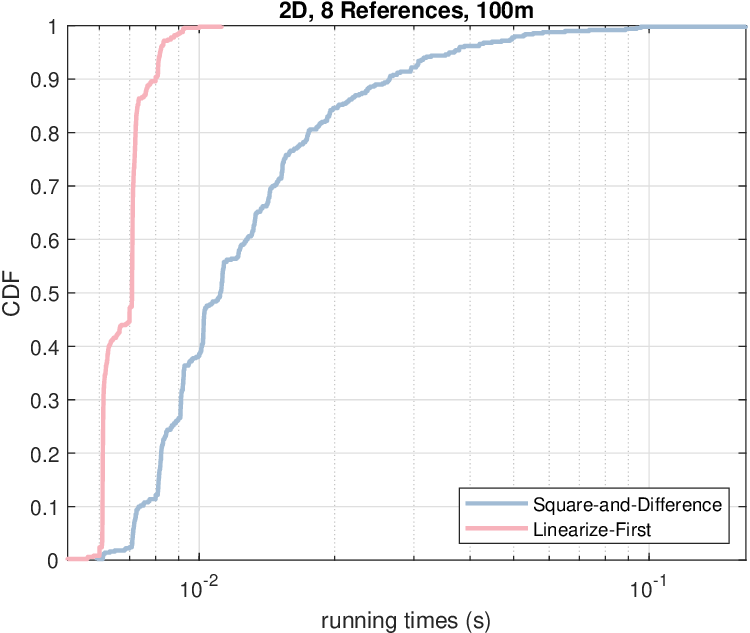}\hfill{}\includegraphics[width=0.47\textwidth]{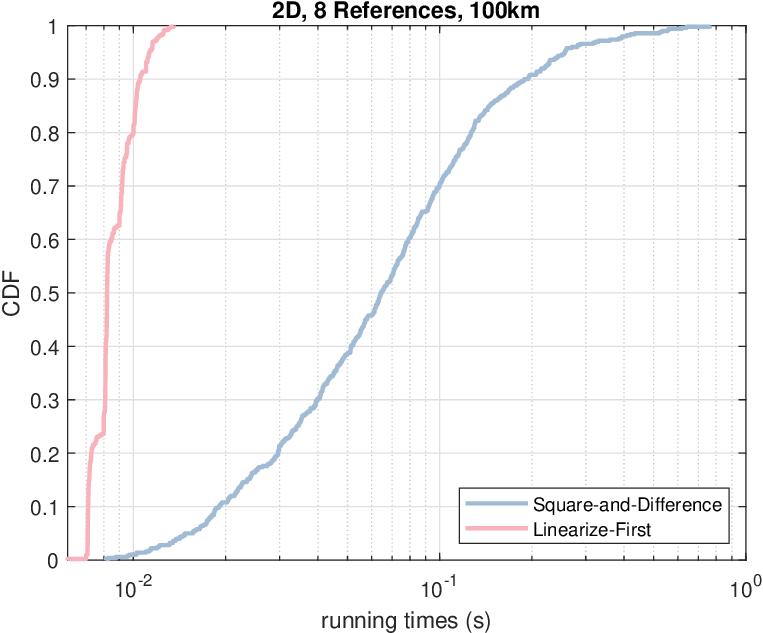}\\
~\\
\includegraphics[width=0.47\textwidth]{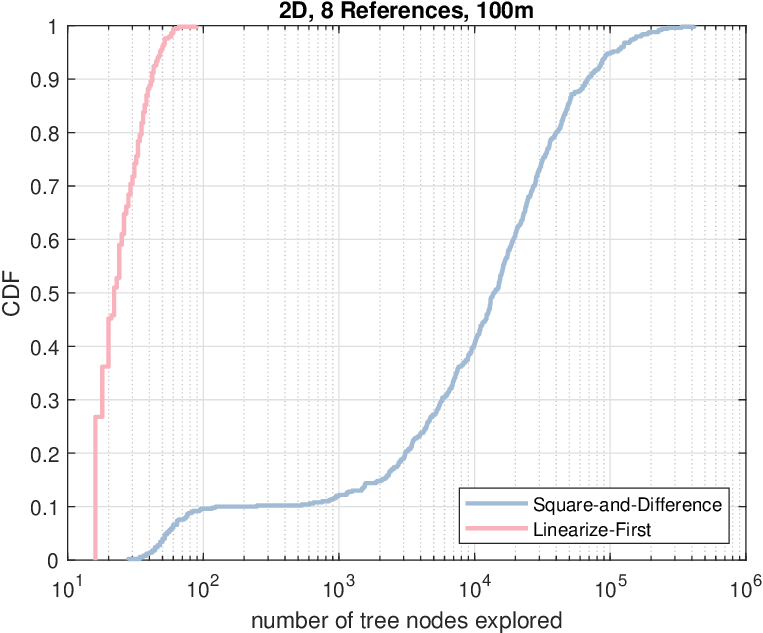}\hfill{}\includegraphics[width=0.47\textwidth]{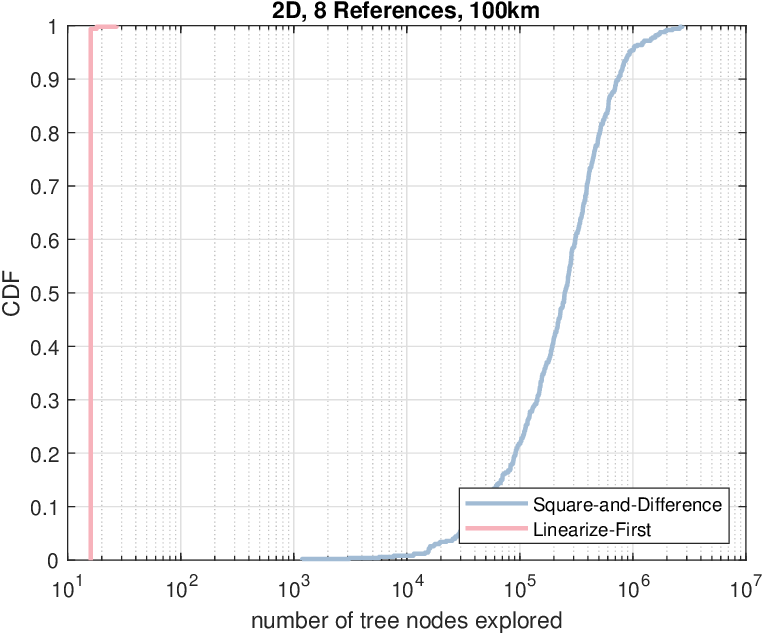}

\caption{\label{fig:running-times-2D}The running times (top row) of the algorithms
and the number of tree nodes they explore (bottom row) in two dimensions
with 8 references, on ranges of 100~m or 100~km between the target
and the references.}
\end{figure}
The two algorithms also behave very differently in terms of their
computational efficiency, which determines the running times. Figure~\ref{fig:running-times-2D}
presents the running times of the algorithms on 2D problems with 8
references at target-reference ranges of 100~m and 100~km. Running
times were measured on a computer with an Intel i7-7500U processor
running at 2.7~GHz running Windows and MATLAB version 2024a. At short
ranges, the running times of the square-and-difference and the linearize-first
algorithms are quite similar, with a median running times around 10~ms.
Virtually all the runs completed within 100~ms. At ranges of 100~km,
in which the linearize-first approach is effective, its running times
are short and stable, taking less than 10~ms. The square-and-difference
approach is much slower at these large ranges, with a median of over
60~ms and with 10\% of the runs taking longer than 200~ms. 

The degradation of the performance of the square-and-difference algorithm
at high ranges is caused by the $2\delta_{i}\left\Vert \mathring{\ell}-\rho_{i}\right\Vert _{2}$
term in Equation~\ref{eq:squared_carrier_phase_with_error-1}. The
algorithm drops this term and compensates with an appropriately-large
variance, which forces the search algorithm to consider many more
discrete solutions.

The graphs in the bottom row of Figure~\ref{fig:running-times-2D}
present the number of tree nodes that the two algorithms explore in
the discrete least-squares search. The number of tree nodes is the
number of assignments to discrete variables during the Schorr-Euchner
search (original or generalized version). The data reveals that at
short ranges (100~m), the median number of nodes explored is only
40 in the linearize-first approach, versus more than 6,800 in the
square-and-difference approach. The wall-clock running times are similar
at this range due to the cost of the LLL basis-reduction step, which
is used in the linearize-first approach, which searches over the integers,
but not in the square-and-difference approach, which searches over
squares of shifted integers.

At ranges of 100~km the linearize-first-approach searches even fewer
integer assignments, almost always 16, but the square-and-difference
approach performs even more assignments, median of more than 143,000.
This causes this algorithm an appreciable slowdown relative to shorter
ranges.\lyxadded{stoledo}{Sun Sep 28 12:32:10 2025}{¶}

\begin{figure}
\includegraphics[width=0.47\textwidth]{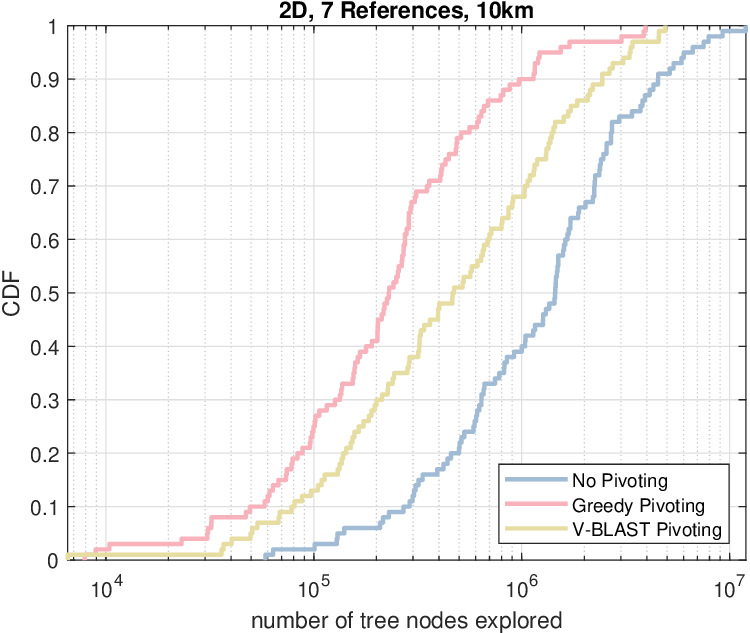}\hfill{}\includegraphics[width=0.47\textwidth]{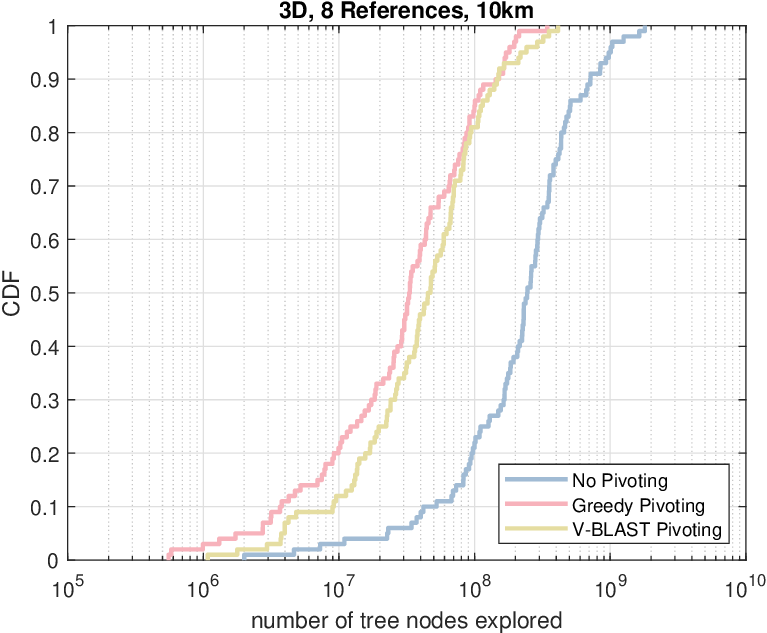}\caption{\label{fig:nodes-column-ordering}The number of nodes that the square-and-difference
algorithm explores during the Schnorr-Euchner search with three different
column-ordering strategy: no pivoting, greedy column pivoting, and
V-BLAST.}
\end{figure}

Figure~\ref{fig:nodes-column-ordering} shows the effect of different
column ordering strategies, discussed in Section~\ref{subsec:QR-column-ordering},
on the number of nodes explored during the Schnorr-Euchner search.
The figure presents results for ranges of 10~km in both 2D and 3D;
the results are qualitatively similar for other ranges and numbers
of references. We see that the greedy column pivoting strategy, which
selects columns from first to last by minimizing the Euclidean norm
among remaining columns, reduces the number of nodes searched by about
a factor of 10. The more sophisticated V-BLAST strategy, which selects
columns from last to first by maximizing diagonal values from last
to first, actually performs a little worse than greedy pivoting.

\begin{figure}
\includegraphics[width=0.47\textwidth]{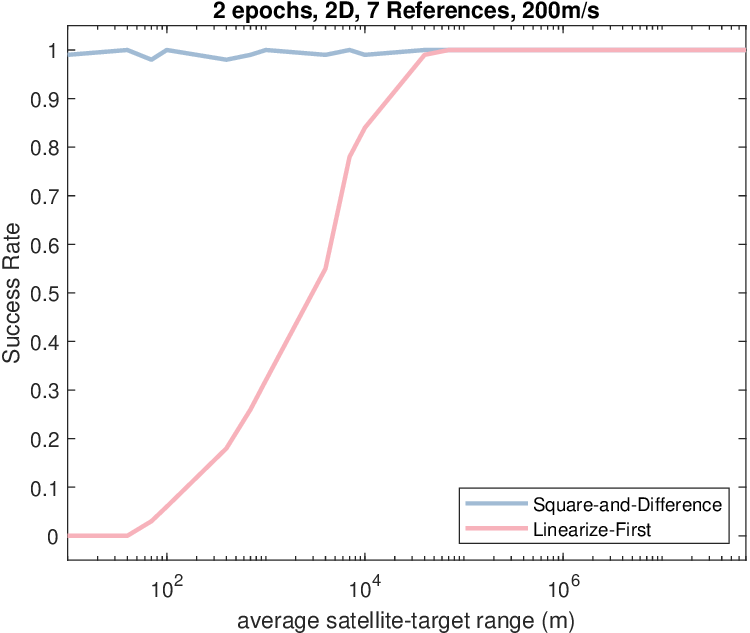}\hfill{}\includegraphics[width=0.47\textwidth]{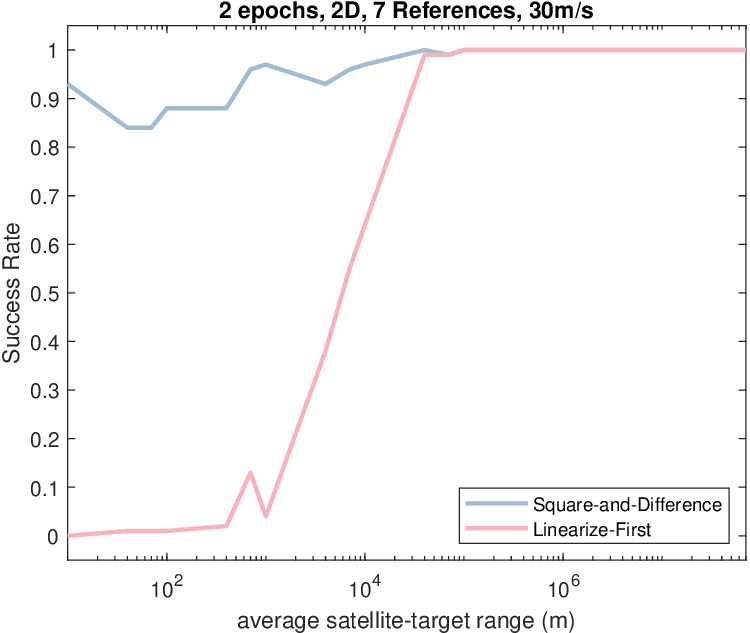}

\caption{\label{fig:-success-rates-2D-2epochs}Success rates our new square-and-difference
algorithm in two dimensions, compared to that of the linearize-first
approach. The graphs plot the success rates for problems 7 references
moving at velocities of 200~m/s or 30~m/s (left and right).}
\end{figure}

Figure~\ref{fig:-success-rates-2D-2epochs} presents the success
rates of the multi-epoch square-and-difference algorithm in 2D using
7 references that move in circular orbits, simulating satellites.
The graph on the left shows that when the references move at high
speed, 200~m/s, our algorithm solves for the correct integers almost
100\% of the time, even at short ranges. The linearize first approach
works well at large ranges but fails at short one, just like with
one epoch. When the references move more slowly, at 30~m/s, the estimation
problem becomes harder and our algorithm recovers the integers correctly
only about 85\% of the time at short ranges.

\begin{figure}
\includegraphics[width=0.47\textwidth]{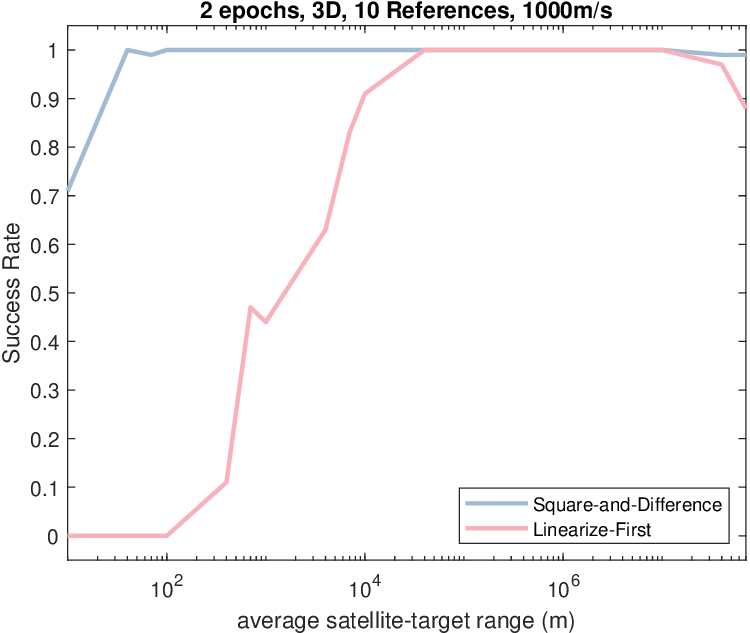}\hfill{}\includegraphics[width=0.47\textwidth]{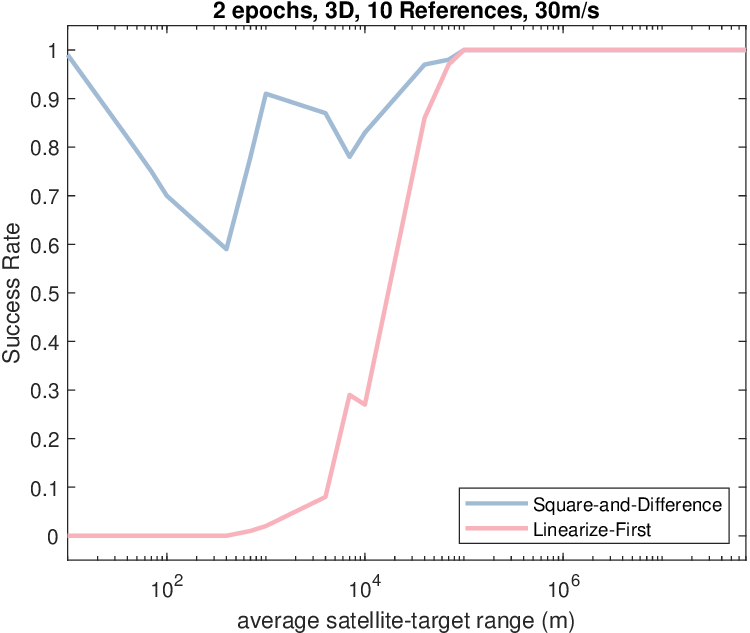}

\caption{\label{fig:-success-rates-3D-2epochs}Success rates our new square-and-difference
algorithm in three dimensions, compared to that of the linearize-first
approach. The graphs plot the success rates for problems with 10 references
moving at velocities of 1000~m/s or 30~m/s (left and right).}
\end{figure}

Figure~\ref{fig:-success-rates-3D-2epochs} shows that the behavior
of the algorithms in 3D is similar. With 10 high-velocity references,
moving at 1000~m/s, our algorithm succeeds nearly 100\% of the time
down to about 40~m. When the references move at only 30~m/s, performance
degrades; our algorithm recovers the integers correctly only 60--90\%
of the time at distances smaller than about 10~km.

Note that the failures of the squaring and double-differencing method
at slow reference velocities can be avoided by running the singe-epoch
method on each epoch separately. As shown in Figure~\ref{fig:-success-rates-3d},
this approach works correctly even with 8 references. However, the
single-epoch method is slower; it does not benefit from the LLL phase
so it explores many more assignments of the discrete parameters.

\begin{figure}
\includegraphics[width=0.47\textwidth]{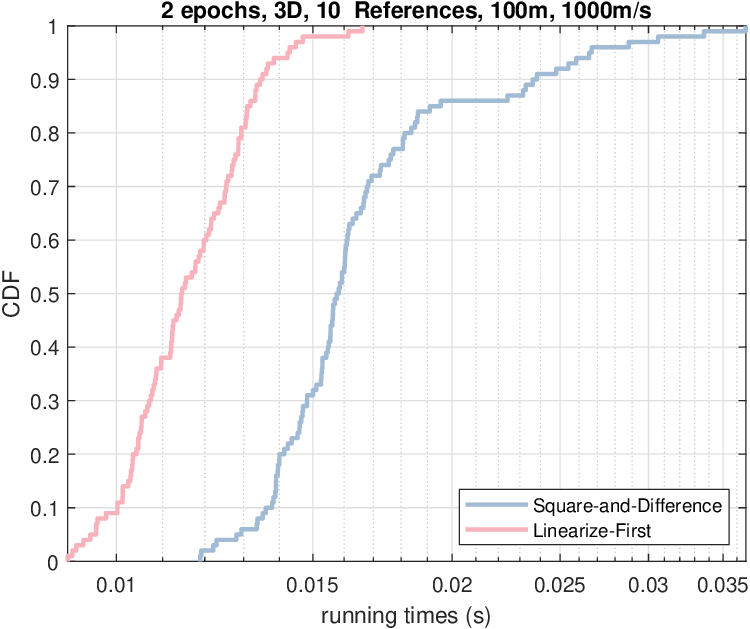}\hfill{}\includegraphics[width=0.47\textwidth]{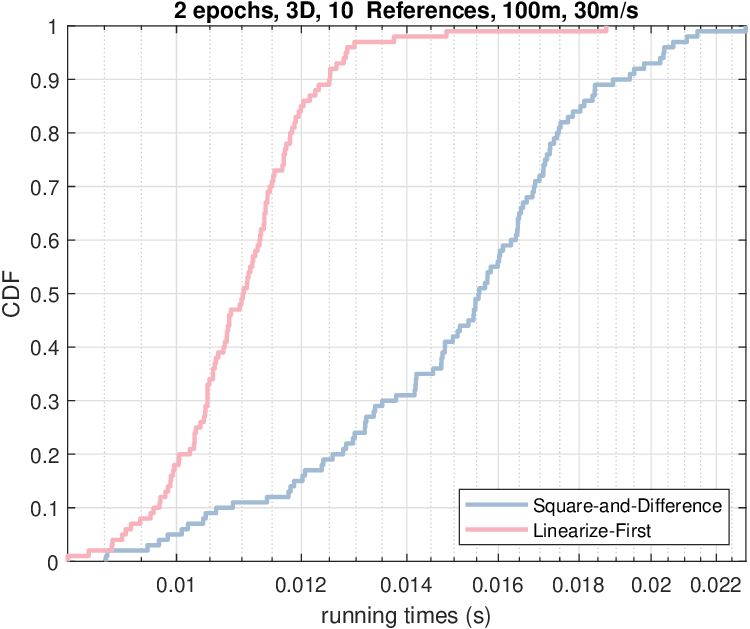}\\
~\\
\includegraphics[width=0.47\textwidth]{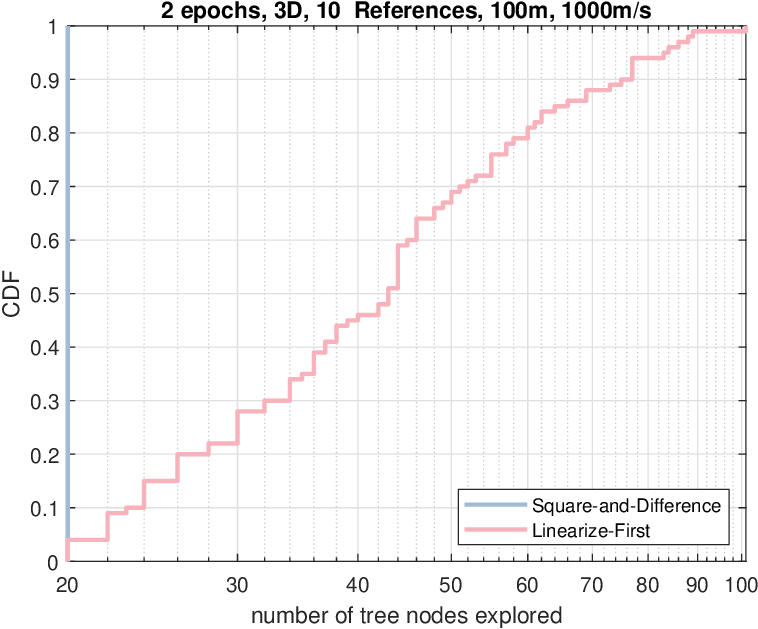}\hfill{}\includegraphics[width=0.47\textwidth]{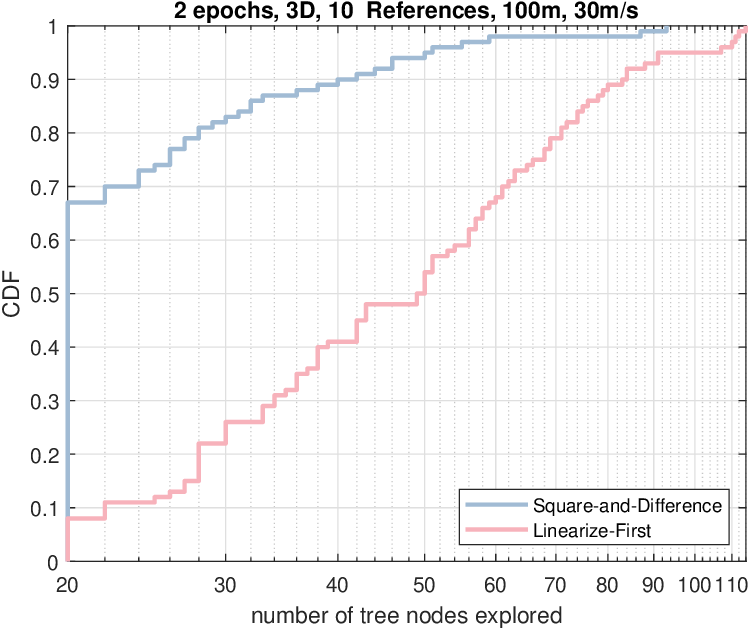}

\caption{\label{fig:running-times-3D-2epochs}The running times (top row) of
the algorithms and the number of tree nodes they explore (bottom row)
in 3D and two epochs, with 10 references at a range of~100~m moving
at velocities of 1000 or 30~m/s.}
\end{figure}

Figure~\ref{fig:running-times-3D-2epochs} presents the running times
and the number of nodes explored in the two-epoch 3D case, using 10
references moving at 1000 or 30~m/s. With fast-moving references,
the median running time of our algorithm is 16~ms. Most of the time
is spent in the LLL phase; the number of nodes explored is small,
with a median of 44. The linearize-first algorithm is a little faster,
but not by much: its median running time is over 11~ms. The performance,
in terms of running times an d number of nodes explored, are not influenced
significantly by the velocity of references; the medians at 30~m/s
are similar.

\subsection{\label{subsec:results-geometric-qpproaches}Performance of Geometric
Approaches}

We implemented the three geometric approaches described in Section~\ref{sec:geometric-arrangements}:
\begin{itemize}
\item The method that enumerates intersections and near-intersections of
annuli (spherical rings in 3D), described in Section~\ref{subsec:Near-Intersections-of-Annuli}.
We implemented this method in Python using 64-bit floating-point arithmetic.
In this method, we choose two or three references arbitrarily (in
2D and 3D, respectively), compute all the intersections of one circle
or sphere centered at each reference, and checking that a circle or
sphere centered at each other reference $\rho_{i}$ passes within
distance $D_{i}$ of the intersection. The radii of the circles/spheres
is shifted by $\lambda\varphi_{i}$ from an integer multiple of the
wavelength $\lambda$. We normally set $D_{i}=100\sigma(\delta_{i})\approx5\,\text{mm}$.
This implementation is much faster than the others, both because of
its lower asymptotic complexity and because of the use of floating-point
arithmetic. We refer to it as the \emph{intersection of circles}. 
\item The intersection-of-annuli method, described in Section~\ref{subsec:intersections-of-annuli}.
This method is implemented using the CGAL C++ computational-geometry
library~\parencite{cgal:6.0.1}, with a Python interface. CGAL uses a
combination of floating-point arithmetic and symbolic computation
to ensure that floating-point errors do not cause its algorithms to
fail. This method is only implemented in 2D because CGAL currently
supports arrangements of circles but not arrangements of spheres.
We set the width of the annuli to $1000\sigma(\delta_{i})\approx50\,\text{mm}$.
We refer to this implementation as the \emph{intersection of annuli}.
\item A simplified version of the method that computes the geometric arrangement
of circles with shifted integer radii, described in Section~\ref{subsec:arrangements-of-spheres-shifted-radii}.
Our implementation shifts the circles by $\lambda\varphi_{i}$, not
by $(\lambda/2)\varphi_{i}$. The algorithm constructs the arrangement
and then enumerates its vertices, checking the residual norm in each
of them. Normally, one or a few have very low residual norms, including
one near the target. The cost of constructing this arrangement is
essentially the same as that of the method proposed in Section~\ref{subsec:arrangements-of-spheres-shifted-radii},
with the $(\lambda/2)\varphi_{i}$ shift. This method also uses CGAL
through a Python interface, and it is only implemented in 2D, due
to the same limitation of CGAL. We refer to this method as the \emph{arrangement
of circles} method.
\end{itemize}
All the implementations enumerate a set of vertices or cells, and
when they enumerate cells, they also identify a point in the cell.
Each vertex or cell maps to a set of integers. The implementations
than solve the continuous nonlinear least squares (with the integers
resolved) starting at the vertex or point in the cell to find the
least-squares optimal solutions.

All the implementations filter out candidates that are more than $3\sigma(\epsilon_{i})$
from the minimizer of the range constraints (equivalent to the code-phase
solution in GNSS). This is a little aggressive and is likely to cause
failures in about 1\% of the tests. Using a more conservative bound,
say $6\sigma(\epsilon_{i})$, would reduce the failure rate but would
increase the running time because this would generate a factor of
4 or 8 (in 2D or 3D) more candidate locations.

All the test problems use a mean reference range of just 100~m, carrier-phase
errors with standard deviation $\sigma(\delta_{i})=0.1^{\circ}=19/3600\,\text{cm}$,
as specified in Section~\ref{sec:Typical-Parameters}, and range-measurement
errors with standard deviation $\sigma(\epsilon_{i})=10\,\text{m}$,
as specified in Section~\ref{sec:Typical-Parameters}, or a tighter
$\sigma(\epsilon_{i})=1\,\text{m}$. The smaller range-measurement
errors lead to fewer candidates locations, speeds up the algorithm,
and allows us to explore more scenarios. 

Each scenario was repeated 100 times using random data. 

\begin{figure}
\includegraphics[width=0.47\textwidth]{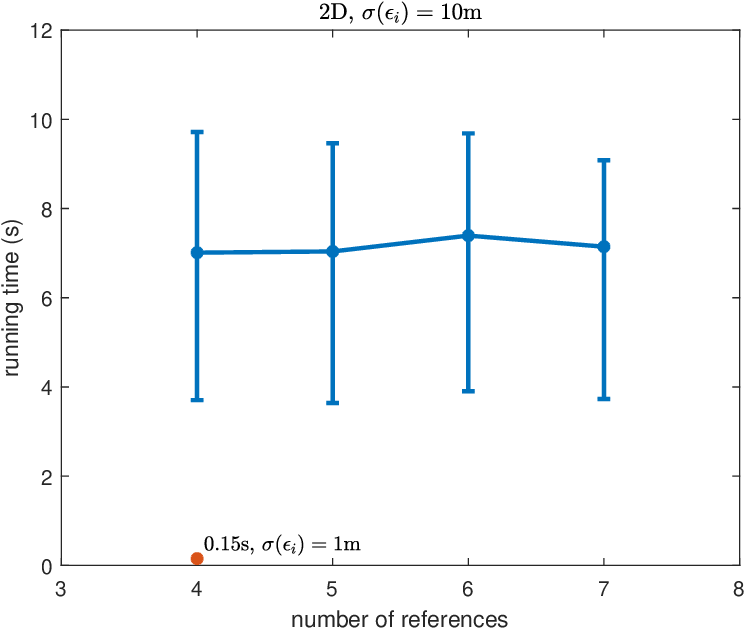}\hfill{}\includegraphics[width=0.47\textwidth]{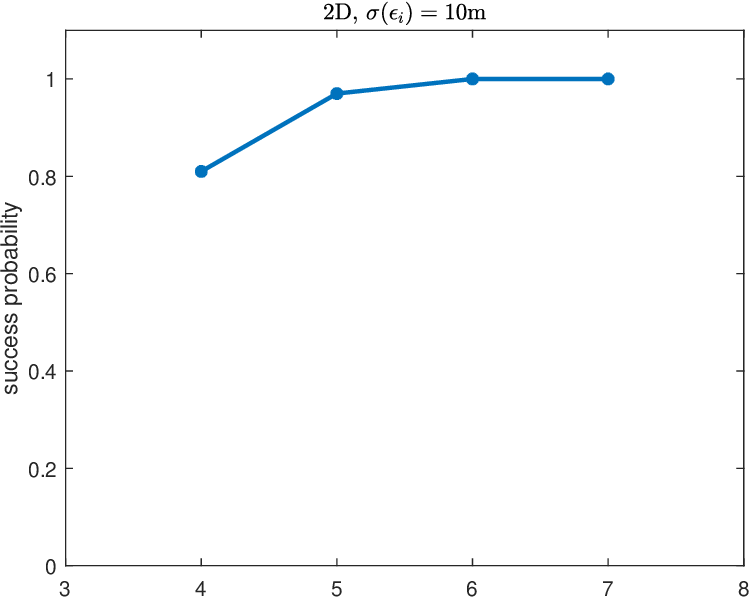}

\caption{\label{fig:intersection-of-circles-nsats}Running times and success
rates of the intersection-of-circles as a function of the number of
references (left and right). The reported running times (circular
dots) are medians and the error bars indicate the 5\% and 95\% percentiles.
The red dot near the bottom of the graph on the left shows the median
running time when $\sigma(\epsilon_{i})$ is only 1~m, not 10~m. }
\end{figure}

Figure~\ref{fig:intersection-of-circles-nsats} shows the performance
of the intersection-of-circles algorithm. On our standard test setup,
with $\sigma(\epsilon_{i})=10\,\text{m}$, the code takes about 4-10~s
to run, essentially independently of the number of references. With
6 or 7 references, the algorithm always filters all the intersections
but one, correctly resolving all the $N_{i}$s. With 4 and 5 references
the algorithm sometimes identifies more than one admissible intersection,
leading to correction resolution of the $N_{i}s$ in 81\% and 97\%
of the cases, respectively. The number of intersections within $3\sigma(\epsilon_{i})$
from the minimizer of the range constraints ranges from about 2.4~k
to about 93~k. The running times are highly correlated with the number
of intersections.

Reducing $\sigma(\epsilon_{i})$ to 1~m reduces the running time
by about a factor of 50, from a median of 7~s to a median of 1.5~s.
The reduction results from a factor-of-100 shrinkage in the area within
the radius-$3\sigma(\epsilon_{i})$ disk.

\begin{figure}
\includegraphics[width=0.47\textwidth]{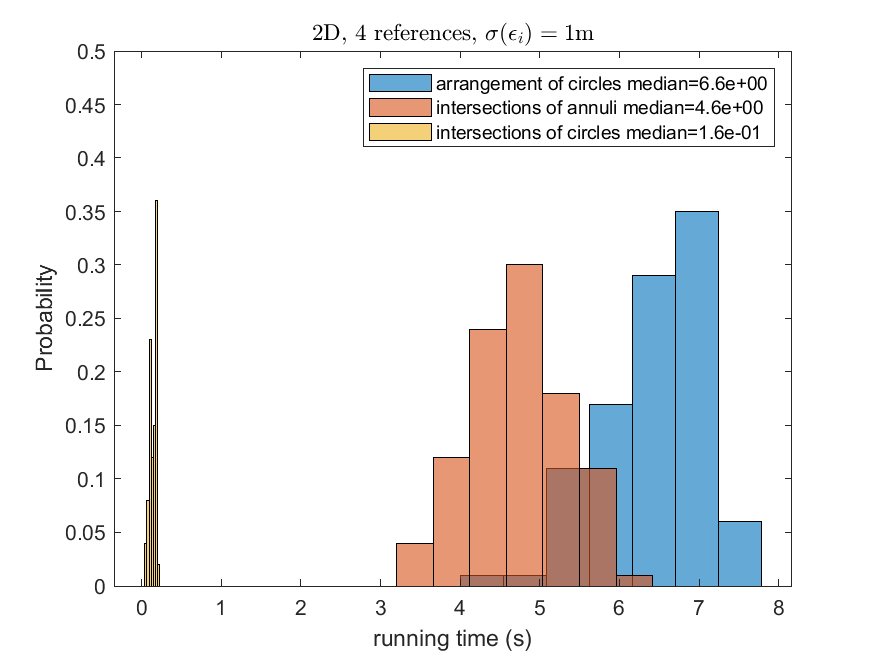}\hfill{}\includegraphics[width=0.47\textwidth]{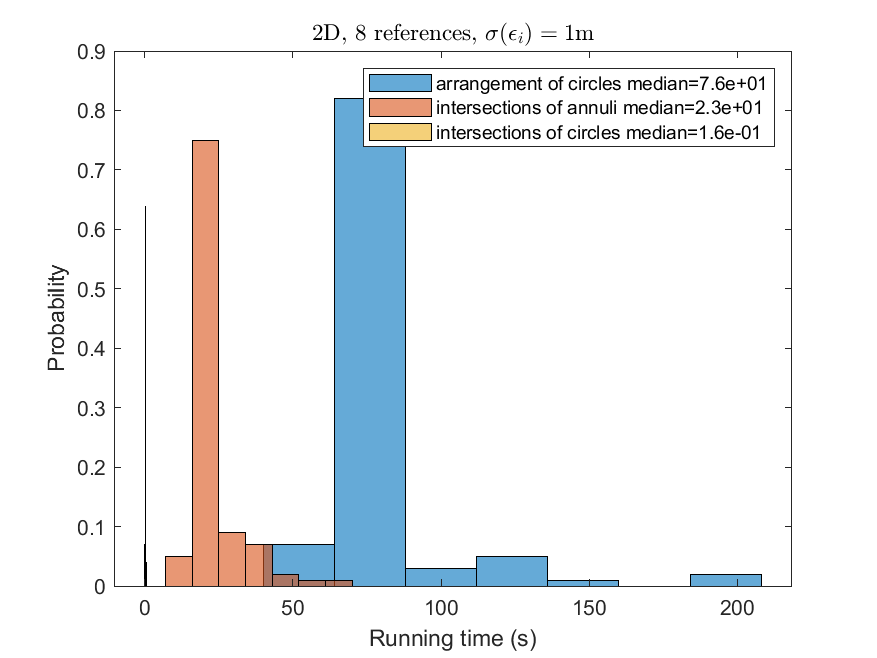}\\
~\\
\includegraphics[width=0.47\textwidth]{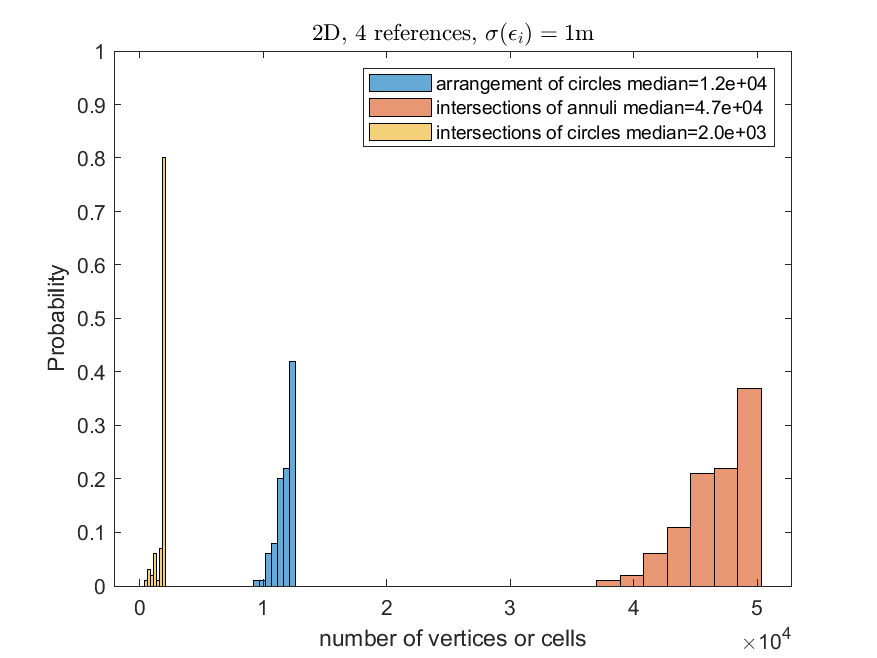}\hfill{}\includegraphics[width=0.47\textwidth]{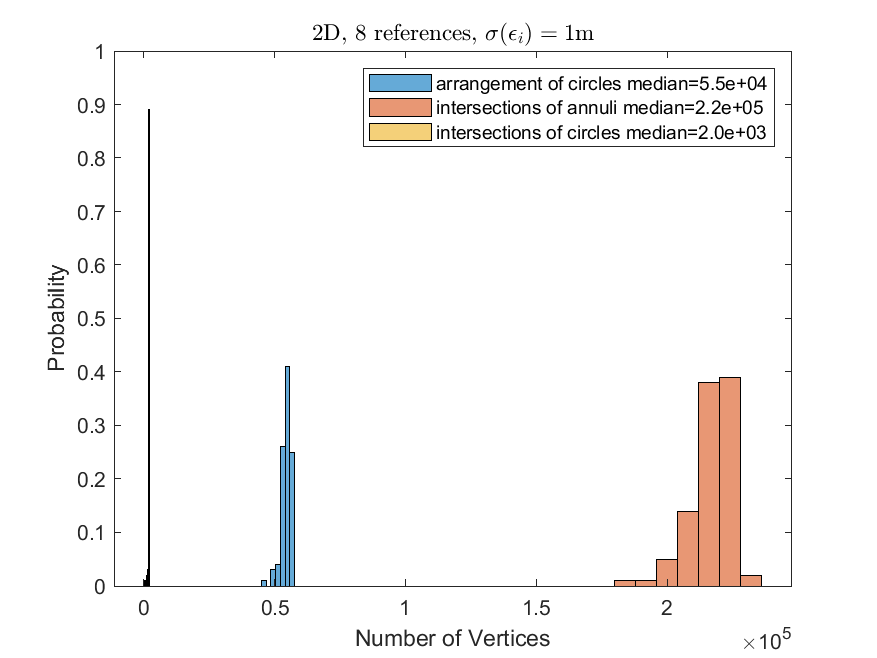}\\
~\\
\includegraphics[width=0.47\textwidth]{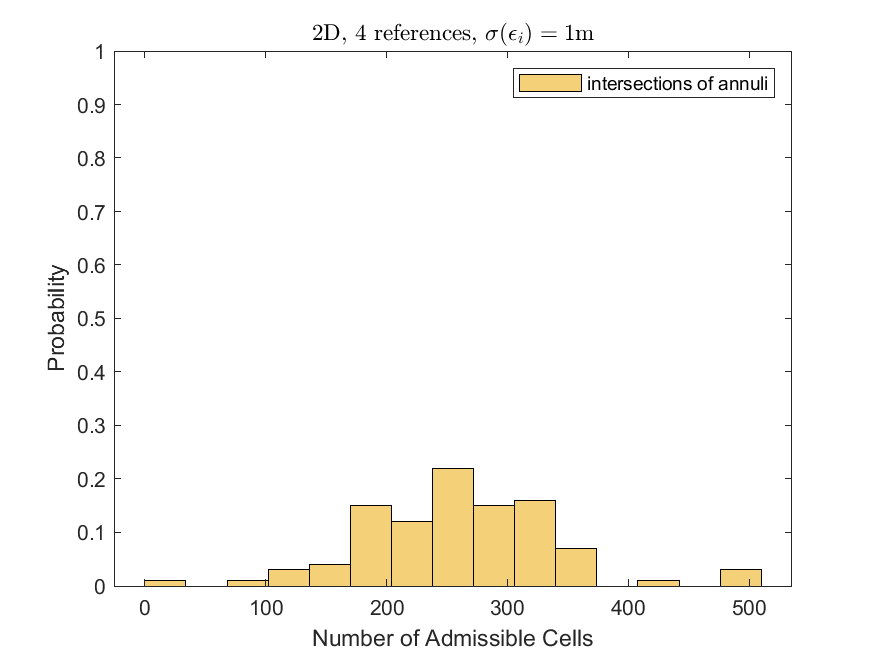}\hfill{}\includegraphics[width=0.47\textwidth]{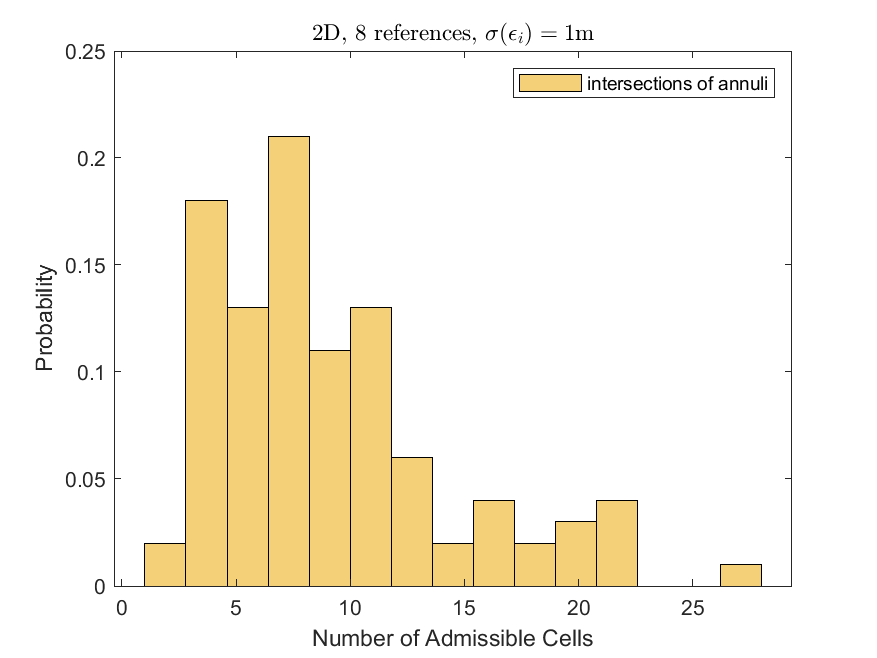}

\caption{\label{fig:geometric-all-2d-4-8}The running times (top row) and
number of candidates tested (middle) in all three implementations,
as well as the number of intersection cells (bottom) in the intersection-of-annuli
implementation. }
\end{figure}

Figure~\ref{fig:geometric-all-2d-4-8} describes the results of experiments
in 2 dimensions with all three methods, with both 4 and 8 references.
We performed these experiments with $\sigma(\epsilon_{i})=1\,\text{m}$
in order to achieve more reasonable running. The running times of
the intersection-of-circles are much faster than those of the methods
that rely on constructing arrangements. The running times of the intersection-of-circle
are also insensitive to the number of references, as we have seen
already in Figure~\ref{fig:intersection-of-circles-nsats}, whereas
the running times of the arrangement-constructing methods depend strongly
on the number of references. This is expected, since their theoretical
complexity scales quadratically with the number of references. These
scaling behaviors are also evident in the graphs that describe the
number of vertices, cells, or intersections that the methods inspect.

We also see, in the bottom row, that the number of admissible cells
shrink dramatically as the number of references increases. This phenomenon
makes more likely that the integers are resolved correctly.

We also verified that the intersection-of-circles method works correctly
in 3D, where each intersection is defined by three spheres. We only
ran the method on 5 random test cases, due to the long running times.
We used 7 references, $\sigma(\epsilon_{i})=10\,\text{m}$, and set
$D_{i}=1000\sigma(\delta_{i})\approx50\,\text{mm}$. The algorithm
correctly resolved the integers in all cases. The median running time
was a little over an hour (range of 3274--3810~s) and the median
number of intersections was about 5~M (range of 4.3--19.0~M).

\section{\label{sec:Conclusions}Conclusions}

The use of simplified system model, in which the system observes the
range of a target from a number of references both directly and in
the carrier-sense (remainder) sense, allowed us to demonstrate that
LAMBDA approaches are inappropriate for carrier-phase localization
at short ranges, and allowed us to develop a range of algorithms that
are effective at short ranges.

The failure of LAMBDA at short ranges stems from the large linearization
error. There is no doubt that the same failure also occurs in more
realistic system models with nuisance parameters, like the models
that are routinely used for carrier-phase localization in GNSS.

Our results also outline very clearly the research that still needs
to be conducted before we have effective algorithms for carrier-phase
localization at short ranges. We need to investigate which of the
methods that we proposed in this paper can be extended to more realistic
system models, and how.

Another challenge is performance, in the sense of running time and
parallelization. Our implementations, especially of the square-and-difference
algorithms, perform well in many cases, but in general they are more
expensive than the LAMBDA approach. However, even the fastest geometric
algorithm (intersections of circles) is much slower than the square-and-difference
algorithms. Additional work is required to develop more efficient
variants of our new algorithms, to optimize their implementations,
and to create implementations that can exploit multiple cores and/or
accelerators like GPUs.

\paragraph*{Acknowledgments}

We thank Xiao-Wen Chang for several insightful discussions and for
helpful comments on early versions of the manuscript and for providing
us an implementation of his V-BLAST algorithm. We thank Hector Rotstein
for many comments and suggestions that improved the article. This
research was supported in part by grants from the Blavatnik Computer
Science Research Fund. Work by D.H. and E.F. has also been supported
in part by the Israel Science Foundation (grant 3598/25).

\paragraph*{Competing Interests}

The authors declare none.

\printbibliography

\appendix

\section{Alternative Proofs of Geometric Lemmas}

This appendix presents an alternative proof of Theorem~\ref{lem:generalized-parallelogram-in-a-circle}
that does not require that $\rho_{i}$ and $\rho_{i}$ are outside
$K$. \setcounter{thm}{\numexpr\getrefnumber{lem:generalized-parallelogram-in-a-circle}-1\relax}
\begin{thm}
\label{lem:generalized-parallelogram-in-a-circle-alt} Let $A$ be
a closed generalized parallelogram  formed by the intersection of
two annuli,
\begin{eqnarray*}
r_{i}-D_{i}\leq & \|\ell-\rho_{i}\| & \leq r_{i}+D_{i}\\
r_{i'}-D_{i'}\leq & \|\ell-\rho_{i'}\| & \leq r_{i'}+D_{i'}
\end{eqnarray*}
for some $0<D_{i},D_{i'}$. Let $p$ be a point in the interior of
$A$ and let $\bar{p}$ be the furthest corner of the generalized
parallelogram from $p$. That is, $\bar{p}$ lies on both $\|\ell-\rho_{i}\|=r_{i}\pm D_{i}$
and $\|\ell-\rho_{i'}\|=r_{i'}\pm D_{i'}$ and it maximizes $\|p-\bar{p}\|$.

Let $K$ be the closed disk $K=\{\ell|\|\ell-p\|\leq\|\bar{p}-p\|\}$.

The disk $K$ contains the generalized parallelogram $A$, 
\[
A\subseteq K\;.
\]
\end{thm}
\begin{figure}
\begin{centering}
\begin{tikzpicture}[scale=1]

  \clip (-0.5,-0.5) rectangle (7.7,7.7);
   
  \draw[name path=a1] (0,0) circle[radius=3];
  \path[name path=a2] (0,0) circle[radius=5];
  \draw[name path=a3] (0,0) circle[radius=7];

  \draw[name path=b1] (6,-2) circle[radius=3.5];
  \path[name path=b2] (6,-2) circle[radius=5];
  \draw[name path=b3] (6,-2) circle[radius=6.6];

  \intersectAndLabel{a2}{b2}{2}{2}{p}{90}
  \intersectAndLabel{a3}{b1}{3}{1}{s}{90}
  \intersectAndLabel{a3}{b3}{3}{3}{t}{90}
  \intersectAndLabel{a1}{b1}{3}{3}{u}{0}
  \intersectAndLabel{a1}{b3}{3}{3}{v}{0}
 
  \coordinate (co) at (0,0);
  \node[point,label={[label distance=-1pt]270:{\textcolor{black}{$o$}}}] (po) at (co) {};

  \draw[blue] (pap)--(pas);
  \draw[blue] (pap)--(pat);
  \draw[blue] (pap)--(pau);
  \draw[blue] (pap)--(pav);

  \path[name path=opr] (co) -- ($(co)!10!(cap)$); 
  \path[name intersections={of=opr and a3, by=cpr}];
  \node[point,label={[label distance=-2pt]45:{\textcolor{black}{$p'$}}}] (ppr) at (cpr) {};
  \draw[gray] (po) -- (ppr) {};

  \coordinate (caq) at (5.75,2.55);
  \node[point,label={[label distance=-2pt]90:{\textcolor{black}{$q$}}}] (paq) at (caq) {};

  \path[name path=oqr] (co) -- ($(co)!10!(caq)$); 
  \path[name intersections={of=oqr and a3, by=cqr}];
  \node[point,label={[label distance=-1pt]0:{\textcolor{black}{$q'$}}}] (pqr) at (cqr) {};
  \draw[gray] (po) -- (pqr) {};

  \draw[gray] (po) -- (pas);
  \draw[orange] (pap) -- (pqr);
  \draw[purple] (pap) -- (paq);

  \path (caq) coordinate (sg);
  \path (cas) coordinate (st);
  \path (co) coordinate (o);
  \path (cap) coordinate (e);
  \drawarc[->,>=stealth,orange]{sg}{o}{e}{2.2cm};
  \drawarc[->,>=stealth,blue]{st}{o}{e}{2.0cm};


\end{tikzpicture}
\par\end{centering}
\caption{\label{fig:alternative-proofs}Notation for the proof of Lemma~\ref{lem:generalized-parallelogram-in-a-circle-alt}. }

\end{figure}

\begin{proof}
The notation for the proof is illustrated in Figure~\ref{fig:alternative-proofs}.
We denote the vertices of the generalized parallelogram $A$ by $s$,
$t$, $v$, and $u$. Without loss of generality, we assume that the
arcs $st$ and $tv$ are the convex boundary parts of $A$ and that
the arcs $vu$ and $us$ are the concave boundary parts of $A$. .

We prove a slightly more general claim, to simplify the proof. We
show that $A\subseteq A'\subseteq K$, where $A'$ is the convex hull
of $A$. 

The shape $A'$ is a union of four closed shapes, pairwise disjoint
in their interiors: the two generalized triangles $\triangle pst$
and $\triangle ptv$, and two straight-edge triangles $\triangle pvu$,
and $\triangle pus$. We show that $K$ contains $A'$ by showing
that $K$ contains all four shapes.

By definition, the three corners of each of the straight-edge triangles
$\triangle pvu$ and $\triangle pus$ are in $K$, so the two entire
triangles are in $K$, since $K$ is convex.

Showing that the two generalized triangles, $\triangle pst$ and $\triangle ptv$,
are contained in $K$ is more challenging. Without loss of generality,
we prove the claim for $\triangle pst$. Let $o$ be the center of
the circle that the arc $st$ belongs to. Consider the ray emanating
from $o$ and passing through $p$. The ray intersects the circle
that contains the arc $st$. We denote the intersection point by $p'$;
it may lie within the arc $st$ or outside it. We handle the two cases
separately.
\begin{itemize}
\item If $p'$ lies within the arc, then the line segment between $p$ and
$p'$ splits the generalized triangle $\triangle pst$ into two generalized
triangles, $\triangle ptp'$ and $\triangle psp'$. Consider a point
$q$ in one of them; without loss of generality, we assume that it
is $\triangle psp'$. We show below that $q\in K$ by showing $|pq|\leq|ps|$.
\item If $p'$ lies outside the arc, assume without loss of generality that
$t$ lies in the arc $sp'$. This implies that the generalized triangle
$\triangle psp'$ contains the generalized triangle $\triangle spt$.
Consider a point $q\in\triangle spt$, which implies $q\in\triangle psp'$.
We show below that $q\in K$ by showing $|pq|\leq|ps|$.
\end{itemize}
In both cases, the ray from $o$ through $q$ intersects the arc $sp'$.
We denote the intersection point by $q'$.

We denote distances from $p$ by capital letters:
\begin{eqnarray*}
Q & = & \left|pq\right|\\
Q' & = & \left|pq'\right|\\
P' & = & \left|pp'\right|\\
S & = & \left|ps\right|\\
O & = & \left|op\right|\;.
\end{eqnarray*}
Note that $O+P'=|op'|=|oq'|=|os|$. We also denote $D=|oq|-O$, so
that $|oq|=O+D$. We claim that $D\leq P'$. This is true since $P'=|pp'|=|op'|-O$
and since $|op'|\geq|oq|$.

We now use the law of cosines to the triangles $\triangle ops$, $\triangle opq$,
and $\triangle opq'$. Denoting the angle of $\triangle ops$ at $o$
by $\theta$ and the angle of both $\triangle opq$ and $\triangle opq'$
at $o$ by $\gamma$, the law of cosines gives:
\begin{eqnarray*}
S^{2} & = & O^{2}+\left(O+P'\right)^{2}-2O\left(O+P'\right)\cos\theta\\
Q^{2} & = & O^{2}+\left(O+D\right)^{2}-2O\left(O+D\right)\cos\gamma\\
Q'{}^{2} & = & O^{2}+\left(O+P'\right)^{2}-2O\left(O+P'\right)\cos\gamma\;.
\end{eqnarray*}
The inequality $\gamma\leq\theta\leq\pi$ implies $\cos\gamma\geq\cos\theta$.
We subtract the last equality from the first to obtain
\begin{eqnarray*}
S^{2}-Q'{}^{2} & = & -2O\left(R+P'\right)\cos\theta+2O\left(O+P'\right)\cos\gamma\\
 & = & 2O\left(R+P'\right)\left(\cos\gamma-\cos\theta\right)\\
 & \geq & 0\;,
\end{eqnarray*}
establishing $Q'\leq S$. We now subtract $Q'{}^{2}$ from $Q^{2}$:
\begin{eqnarray*}
Q'{}^{2}-Q{}^{2} & = & \left(O+P'\right)^{2}-2O\left(O+P'\right)\cos\gamma-\left(O+D\right)^{2}+2O\left(O+D\right)\cos\gamma\\
 & = & P'{}^{2}-D^{2}+2OP'-2OD-2P'\cos\gamma+2D\cos\gamma\\
 & = & P'{}^{2}-D^{2}+2O(P'-D)+2O\left(D-P'\right)\cos\gamma\\
 & = & \left(P'{}^{2}-D^{2}\right)+2O\left(P'-D\right)\left(1-\cos\gamma\right)\;.
\end{eqnarray*}
The first term is nonnegative since $P'\geq D$, as we have shown
above. The second term is also nonnegative because $O>0$, $P'-D\geq0$,
and $1-\cos\gamma\geq0$ (for any $\gamma$). We conclude that $Q\leq Q'$.
Therefore, $Q\leq Q'\leq S$.

The fact that $Q\leq S$ shows that $q$ is indeed in $K$, completing
the proof of the lemma.

\begin{figure}
\begin{centering}
\begin{tikzpicture}[scale=1]

  \clip (-0.5,-0.5) rectangle (7.7,7.7);
   
  \draw[name path=a1] (0,0) circle[radius=3];
  \path[name path=a2] (0,0) circle[radius=5];
  \draw[name path=a3] (0,0) circle[radius=7];

  \draw[name path=b1] (6,-2) circle[radius=4.7];
  \path[name path=b2] (6,-2) circle[radius=5];
  \draw[name path=b3] (6,-2) circle[radius=5.3];

  \intersectAndLabel{a2}{b2}{2}{2}{p}{180}
  \intersectAndLabel{a3}{b1}{3}{1}{s}{270}
  \intersectAndLabel{a3}{b3}{3}{3}{t}{90}
  \intersectAndLabel{a1}{b1}{3}{3}{u}{0}
  \intersectAndLabel{a1}{b3}{3}{3}{v}{0}
 
  \coordinate (co) at (0,0);
  \node[point,label={[label distance=-1pt]270:{\textcolor{black}{$o$}}}] (po) at (co) {};

  \draw[blue] (pap)--(pas);
  \draw[blue] (pap)--(pat);
  \draw[blue] (pap)--(pau);
  \draw[blue] (pap)--(pav);

  \path[name path=opr] (co) -- ($(co)!10!(cap)$); 
  \path[name intersections={of=opr and a3, by=cpr}];
  \node[point,label={[label distance=-2pt]180:{\textcolor{black}{$p'$}}}] (ppr) at (cpr) {};
  \draw[gray] (po) -- (ppr) {};

  \coordinate (caq) at (5.85,2.8);
  \node[point,label={[label distance=-2pt]90:{\textcolor{black}{$q$}}}] (paq) at (caq) {};

  \path[name path=oqr] (co) -- ($(co)!10!(caq)$); 
  \path[name intersections={of=oqr and a3, by=cqr}];
  \node[point,label={[label distance=-1pt]0:{\textcolor{black}{$q'$}}}] (pqr) at (cqr) {};
  \draw[gray] (po) -- (pqr) {};

  \draw[gray] (po) -- (pas);
  \draw[orange] (pap) -- (pqr);
  \draw[purple] (pap) -- (paq);

  \path (caq) coordinate (sg);
  \path (cas) coordinate (st);
  \path (co) coordinate (o);
  \path (cap) coordinate (e);
  \drawarc[->,>=stealth,orange]{sg}{o}{e}{2.2cm};
  \drawarc[->,>=stealth,blue]{st}{o}{e}{2.0cm};


\end{tikzpicture}\vspace{2cm}
\par\end{centering}
\begin{centering}
\begin{tikzpicture}[scale=4]

  \clip (4,2) rectangle (7,4);
   
  \draw[name path=a1] (0,0) circle[radius=3];
  \path[name path=a2] (0,0) circle[radius=5];
  \draw[name path=a3] (0,0) circle[radius=7];

  \draw[name path=b1] (6,-2) circle[radius=4.7];
  \path[name path=b2] (6,-2) circle[radius=5];
  \draw[name path=b3] (6,-2) circle[radius=5.3];

  \intersectAndLabel{a2}{b2}{2}{2}{p}{180}
  \intersectAndLabel{a3}{b1}{3}{1}{s}{270}
  \intersectAndLabel{a3}{b3}{3}{3}{t}{90}
  \intersectAndLabel{a1}{b1}{3}{3}{u}{0}
  \intersectAndLabel{a1}{b3}{3}{3}{v}{0}
 
  \coordinate (co) at (0,0);
  \node[point,label={[label distance=-1pt]270:{\textcolor{black}{$o$}}}] (po) at (co) {};

  \draw[blue] (pap)--(pas);
  \draw[blue] (pap)--(pat);
  \draw[blue] (pap)--(pau);
  \draw[blue] (pap)--(pav);

  \path[name path=opr] (co) -- ($(co)!10!(cap)$); 
  \path[name intersections={of=opr and a3, by=cpr}];
  \node[point,label={[label distance=-2pt]180:{\textcolor{black}{$p'$}}}] (ppr) at (cpr) {};
  \draw[gray] (po) -- (ppr) {};

  \coordinate (caq) at (5.85,2.8);
  \node[point,label={[label distance=-2pt]90:{\textcolor{black}{$q$}}}] (paq) at (caq) {};

  \path[name path=oqr] (co) -- ($(co)!10!(caq)$); 
  \path[name intersections={of=oqr and a3, by=cqr}];
  \node[point,label={[label distance=-1pt]0:{\textcolor{black}{$q'$}}}] (pqr) at (cqr) {};
  \draw[gray] (po) -- (pqr) {};

  \draw[gray] (po) -- (pas);
  \draw[orange] (pap) -- (pqr);
  \draw[purple] (pap) -- (paq);

  \path (caq) coordinate (sg);
  \path (cas) coordinate (st);
  \path (co) coordinate (o);
  \path (cap) coordinate (e);
  \drawarc[->,>=stealth,orange]{sg}{o}{e}{2.2cm};
  \drawarc[->,>=stealth,blue]{st}{o}{e}{2.0cm};


\end{tikzpicture}
\par\end{centering}
\caption{\label{fig:alternative-proofs-anomalies}An extreme case of a generalized
parallelogram, illustrating why the proof of Lemma~\ref{lem:generalized-parallelogram-in-a-circle-alt}
shows that the shape bounded by the convex hull of $A$ is in $K$:
The line segments $ps$ and $pu$ and the arc $us$ do not necessarily
form a generalized triangle, so we rely on the containment of the
straight-edge triangle $\triangle psu$ in $K$ instead. The illustration
at the top shows the entire generalized parallelogram and the one
at the bottom zooms in on the edge $ps$, which intersects the arc
$us$ in two points.}
\end{figure}
\end{proof}

\end{document}